\documentclass[12pt]{article}
\usepackage{amsfonts,epsf,latexsym}

\newcommand{\beq}{\begin{equation}}
\newcommand{\eeq}{\end{equation}}
\newcommand{\beqs}{\begin{eqnarray}}
\newcommand{\eeqs}{\end{eqnarray}}
\newcommand{\sgn}{\mathop{\rm sgn}\nolimits}

\catcode`@=11
\@addtoreset{equation}{section}
\@addtoreset{equation}{subsection}
\def\theequation{\ifnum\value{section}=0 \arabic{equation}\ignorespaces
\else \ifnum\value{section}=-1 A.\arabic{equation}\ignorespaces
\else \ifnum\value{subsection}=0 \thesection.\arabic{equation}\ignorespaces
\else \thesection.\arabic{subsection}.\arabic{equation}\ignorespaces
                           \fi
                      \fi
                 \fi}
\catcode`@=12

\begin{document}

\def\thefootnote{\fnsymbol{footnote}}

\baselineskip 6.0mm

\vspace{4mm}

\begin{center}

{\Large \bf Exact Potts Model Partition Functions for Strips of the 
Square Lattice} 

\vspace{8mm}

\setcounter{footnote}{0}
Shu-Chiuan Chang$^{(a)}$\footnote{email: shu-chiuan.chang@sunysb.edu}
Jes\'us Salas$^{(b)}$\footnote{email: salas@laurel.unizar.es} 
\setcounter{footnote}{6}
Robert Shrock$^{(a)}$\footnote{email: robert.shrock@sunysb.edu}

\vspace{6mm}

(a) \ C. N. Yang Institute for Theoretical Physics  \\
State University of New York       \\
Stony Brook, N. Y. 11794-3840  \\
USA

(b) \ Departamento de F\'{\i}sica Te\'orica \\
Facultad de Ciencias \\ 
Universidad de Zaragoza \\
Zaragoza 50009 \\
Spain

\vspace{10mm}

{\bf Abstract}
\end{center}

We present exact calculations of the Potts model partition function $Z(G,q,v)$
for arbitrary $q$ and temperature-like variable $v$ on $n$-vertex
square-lattice strip graphs $G$ for a variety of transverse widths $L_t$ and
for arbitrarily great length $L_\ell$, with free longitudinal boundary
conditions and free and periodic transverse boundary conditions.  These have
the form $Z(G,q,v)=\sum_{j=1}^{N_{Z,G,\lambda}}
c_{Z,G,j}(\lambda_{Z,G,j})^{L_\ell}$.  We give general formulas for $N_{Z,G,j}$
and its specialization to $v=-1$ for arbitrary $L_t$ for both types of boundary
conditions, as well as other general structural results on $Z$.  The free
energy is calculated exactly for the infinite-length limit of the graphs, and
the thermodynamics is discussed.  It is shown how the internal energy
calculated for the case of cylindrical boundary conditions is connected with
critical quantities for the Potts model on the infinite square lattice.
Considering the full generalization to arbitrary complex $q$ and $v$, we
determine the singular locus ${\cal B}$, arising as the accumulation set of
partition function zeros as $L_\ell \to \infty$, in the $q$ plane for fixed $v$
and in the $v$ plane for fixed $q$.

\vspace{16mm}

\pagestyle{empty}
\newpage

\pagestyle{plain}
\pagenumbering{arabic}
\renewcommand{\thefootnote}{\arabic{footnote}}
\setcounter{footnote}{0}

\section{Introduction}

The $q$-state Potts model has served as a valuable model for the study of phase
transitions and critical phenomena \cite{potts}-\cite{baxterbook}.  On a
lattice, or, more generally, on a (connected) graph $G$, at temperature $T$,
this model is defined by the partition function
\beq
Z(G,q,v) = \sum_{ \{ \sigma_n \} } e^{-\beta {\cal H}}
\label{zfun}
\eeq
with the (zero-field) Hamiltonian
\beq
{\cal H} = -J \sum_{\langle i j \rangle} \delta_{\sigma_i \sigma_j}
\label{ham}
\eeq
where $\sigma_i=1,...,q$ are the spin variables on each vertex $i \in G$;
$\beta = (k_BT)^{-1}$; and $\langle i j \rangle$ denotes pairs of adjacent
vertices.  The graph $G=G(V,E)$ is defined by its vertex set $V$ and its edge
set $E$; we denote the number of vertices of $G$ as $n=n(G)=|V|$ and the
number of edges of $G$ as $e(G)=|E|$.  We use the notation
\beq
K = \beta J \ , \quad a = e^K = u^{-1} \ , \quad v = a-1
\label{kdef}
\eeq
so that the physical ranges are (i) $v \ge 0$ corresponding to $\infty \ge T
\ge 0$ for the Potts ferromagnet, and (ii) $-1 \le v \le 0$, corresponding to
$0 \le T \le \infty$ for the Potts antiferromagnet.  One defines the (reduced)
free energy per site $f=-\beta F$, where $F$ is the actual free energy, via
\beq
f(G,q,v) = n^{-1}\ln [Z(G,q,v)]
\label{ffinite}
\eeq
and, in the $n \to \infty$ limit, 
\beq
f(\{G\},q,v) = \lim_{n \to \infty} n^{-1} \ln [Z(G,q,v)]
\label{f}
\eeq
where we use the symbol $\{G\}$ to denote $\lim_{n \to \infty}G$ for a given
family of graphs.

Let $G^\prime=(V,E^\prime)$ be a spanning subgraph of $G$, i.e. a subgraph
having the same vertex set $V$ and an edge set $E^\prime \subseteq E$. Then
$Z(G,q,v)$ can be written as the sum \cite{kf}
\beq
Z(G,q,v) = \sum_{G^\prime \subseteq G} q^{k(G^\prime)}v^{e(G^\prime)}
\label{cluster}
\label{zpol}
\eeq
where $k(G^\prime)$ and $e(G^\prime)$ denote respectively the number of 
connected components and the number of edges of $G^\prime$.
Since we only consider connected graphs $G$, we have $k(G)=1$. The formula
(\ref{cluster}) enables one to generalize $q$ from ${\mathbb Z}_+$ to ${\mathbb
R}_+$ and, indeed, to ${\mathbb C}$.  The Potts model partition function on a
graph $G$ is essentially equivalent to the Tutte polynomial
\cite{tutte1}-\cite{tutte3} and Whitney rank polynomial \cite{wurev},
\cite{bbook}-\cite{boll} for this graph.

\vspace{6mm}

In this paper we shall present exact calculations of the Potts model partition
function $Z(G,q,v)$, for arbitrary $q$ and $v$, on square-lattice strip graphs
$G$ of various widths $L_t$ and arbitrarily great length $L_\ell$, with
boundary conditions that are (i) free in both longitudinal and transverse
directions, denoted ``open'' and (ii) periodic in the transverse direction and
free in the longitudinal direction, denoted ``cylindrical''.  Specifically, we
shall consider strips with $L_t$ vertices along a transverse slice (hence $L_t$
edges on this slice in the case of cylindrical boundary conditions and $L_t-1$
edges in the case of free boundary conditions) and $L_\ell$ vertices along a
longitudinal slice.  We shall often denote the strip with open boundary
conditions as $(L_t)_{\rm F} \times (L_\ell)_{\rm F}$ and the strip with
cylindrical boundary conditions as $(L_t)_{\rm P} \times (L_\ell)_{\rm F}$.
Previous related calculations of $Z(G,q,v)$ for arbitrary $q$ and $v$ on
fixed-width, arbitrary-length lattice strip graphs are in \cite{bcc}-\cite{s3a}
for the square lattice, in \cite{ta} for the triangular lattice, in \cite{hca}
for the honeycomb lattice, and in \cite{ka} for the square lattice with
next-nearest-neighbor edges (bonds).  Some general properties are given in
\cite{ss00,cf}.  Calculations of $Z(G,q,v)$ for arbitrary $q$ and $v$ on finite
sections of lattices have been given in \cite{ks,kc} and used for studies of
zeros in the $q$ and $v$ plane (the latter being complex-temperature or Fisher
zeros \cite{fisher}); for fixed positive integral values of $q$, such
calculations have been given, e.g., in \cite{mm}-\cite{p2}.  Since the Ising
case $q=2$ is exactly solvable on two-dimensional lattices, one can calculate
exactly the complex-temperature phase boundaries that are the accumulation sets
of the complex-temperature zeros in the thermodynamic limit (e.g.,
\cite{steph}-\cite{wudensity} and references therein).

There are several motivations for the present work.  Clearly, new exact
calculations of Potts model partition functions are of value in their own
right.  Since the results apply for arbitrary length, one can take the limit of
infinite length and calculate the free energy and specific heat for the Potts
model with any $q$ and various widths.  In addition, it was shown \cite{a} that
these calculations can give insight into the complex-temperature phase diagram
of the 2D Potts model on the given lattice.  This is useful since the 2D Potts
model has never been solved except in the (zero-field) $q=2$ Ising case. 

Various special cases of the Potts model partition function are of interest.
One special case is the zero-temperature limit of the Potts antiferromagnet
(AF), i.e., $v=-1$. 
For sufficiently large $q$, on a given lattice or graph $G$, this
exhibits nonzero ground state entropy (without frustration).  Such nonzero
ground state entropy is an exception to the third law of thermodynamics
\cite{al,chowwu}.  It is equivalent to a ground state degeneracy per site
(vertex), $W > 1$, since $S_0 = k_B \ln W$.  The $T=0$ (i.e., $v=-1$) partition
function of the above-mentioned $q$-state Potts antiferromagnet (PAF) on $G$
satisfies
\beq 
Z(G,q,-1)=P(G,q)
\label{zp}
\eeq
where $P(G,q)$ is the chromatic polynomial (in $q$) expressing the number
of ways of coloring the vertices of the graph $G$ with $q$ colors such that no
two adjacent vertices have the same color \cite{bbook,rrev,rtrev}. The 
minimum number of colors necessary for this coloring is the chromatic number
of $G$, denoted $\chi(G)$.  We have 
\beq 
W(\{G\},q)= \lim_{n \to \infty}P(G,q)^{1/n} \ .
\label{w}
\eeq
Some previous related works on chromatic polynomials and the $W$ function are
\cite{lenard}-\cite{dg}.  Using the formula (\ref{cluster}) for $Z(G,q,v)$, one
can generalize $q$ from ${\mathbb Z}_+$ not just to ${\mathbb R}_+$ but to
${\mathbb C}$ and $a$ from its physical ferromagnetic and antiferromagnetic
ranges $0 \le v \le \infty$ and $-1 \le v \le 0$ to $v \in {\mathbb C}$.  A
subset of the zeros of $Z$ in the two-complex dimensional space ${\mathbb C}^2$
defined by the pair of variables $(q,v)$ can form an accumulation set in the $n
\to \infty$ limit, denoted ${\cal B}$, which is the continuous locus of points
where the free energy is nonanalytic.  This locus is determined as the solution
to a certain $\{G\}$-dependent equation \cite{bcc,a}.  For a given value of
$v$, one can consider this locus in the $q$ plane, and we denote it as ${\cal
B}_q(\{G\},v)$.  In the special case $v=-1$ where the partition function is
equal to the chromatic polynomial, the zeros in $q$ are the chromatic zeros,
and ${\cal B}_q(\{G\},v=-1)$ is their continuous accumulation set in the $n \to
\infty$ limit.  Previous works have given exact calculations of the chromatic
polynomials and nonanalytic loci ${\cal B}_q$ for various families of graphs.
With the exact Potts partition function, one can study ${\cal B}_q$ for
arbitrary temperature in both the antiferromagnetic and ferromagnetic cases
and, for a given value of $q$, one can study the continuous accumulation set of
the zeros of $Z(G,q,v)$ in the $v$ plane; this will be denoted ${\cal
B}_v(\{G\},q)$.  (Where the context is clear, we shall refer to these
accumulation sets in the $q$ and $v$ plane as just ${\cal B}$.)

Following the notation in \cite{w} and our other earlier works on ${\cal
B}_q(\{G\})$ for $v=-1$, we denote the maximal region in the complex $q$ plane
to which one can analytically continue the function $W(\{G\},q)$ from physical
values where there is nonzero ground state entropy as $R_1$.  The maximal
value of $q$ where ${\cal B}_q$ intersects the (positive) real axis was
labelled $q_c(\{G\})$.  Thus, region $R_1$ includes the positive real axis for
$q > q_c(\{G\})$.  Correspondingly, in the studies of complex-temperature phase
diagrams of spin models \cite{chisq,cmo} a nomenclature was used according to
which the complex-temperature extension(CTE) of the physical paramagnetic phase
was denoted as (CTE)PM, which will simply be labelled PM here, the extension
being understood, and similarly with ferromagnetic (FM) and antiferromagnetic
(AFM) phases; other complex-temperature phases, having no overlap with any
physical phase, were denoted $O_j$ (for ``other''), with $j$ indexing the
particular phase.  Here we shall continue to use this notation for
the respective slices of ${\cal B}$ in the $q$ and $v$ planes.

We record some special values of $Z(G,q,v)$ below, beginning with the $q=0$
special case
\beq
Z(G,0,v)=0
\label{zq0}
\eeq
which implies that $Z(G,q,v)$ has an overall factor of $q$, as is also clear
from (\ref{cluster}). In general (and for all the graphs considered here), this
is the only overall factor that it has.  We also have
\beq
Z(G,1,v)=\sum_{G^\prime \subseteq G} v^{e(G^\prime)} = a^{e(G)} \ .
\label{zq1}
\eeq
For temperature $T=\infty$, i.e., $v=0$, 
\beq
Z(G,q,0)=q^{n(G)} \ .
\label{za1}
\eeq
where $n(G)$ is the total number of sites of $G$. Further, 
\beq
Z(G,q,-1)=P(G,q)=\biggl [ \prod_{s=0}^{\chi(G)-1}(q-s) \biggr ] U(G,q)
\label{pchi}
\eeq
where $U(G,q)$ is a polynomial in $q$ of degree $n(G)-\chi(G)$. 

Another basic property, evident from eq. (\ref{cluster}), is that (i) the zeros
of $Z(G,q,v)$ in $q$ for real $v$ and hence also the continuous accumulation
set ${\cal B}_q$ are invariant under the complex conjugation $q \to q^*$; (ii)
the zeros of $Z(G,q,v)$ in $v$ or equivalently $a$ for real $q$ and hence also
the continuous accumulation set ${\cal B}_v$ are invariant under the complex
conjugation $v \to v^*$.  In certain cases one must take account of the 
noncommutativity \cite{w,a} 
\beq 
\lim_{n \to \infty} \lim_{q
\to q_s} Z(G,q,v)^{1/n} \ne \lim_{q \to q_s} \lim_{n \to \infty} Z(G,q,v)^{1/n}
\ .
\label{fnoncomm}
\eeq

A general form for the Potts model partition function for the strip graphs
considered here, or more generally, for recursively defined families of graphs
comprised of $m$ repeated subunits (e.g. the columns of squares of transverse
width $L_t$ vertices that are repeated $L_\ell$ times in the longitudinal
direction to form an $L_\ell \times L_t$ strip of a regular lattice with some
specified boundary conditions), is \cite{a}
\beq 
Z(G,q,v) = \sum_{j=1}^{N_{Z,G,\lambda}} c_{G,j}(\lambda_{G,j})^m
\label{zgsum}
\eeq
where the coefficients $c_{G,j}$ and corresponding terms $\lambda_{G,j}$, as
well as the total number $N_{Z,G,\lambda}$ of these terms, depend on the type
of recursive graph $G$ (width and boundary conditions) but not its length.  In
the special case $v=-1$ where $Z$ reduces to the chromatic polynomial
(zero-temperature Potts antiferromagnet), eq. (\ref{zgsum}) reduces to the form
\cite{bkw}
\beq
P(G,q) = \sum_{j=1}^{N_{P,G,\lambda}} c_{G,j}(\lambda_{P,G,j})^m \ .
\label{pgsum}
\eeq
For the lattice strips of interest here, we define the following explicit
notation.  Let $N_{Z,sq,BC_t \ BC_\ell,L_t,\lambda}$ denote the total number of
$\lambda$'s for the square-lattice strip with the transverse ($t$) and
longitudinal ($\ell$) boundary conditions $BC_t$ and $BC_\ell$ of width $L_t$.
Henceforth where no confusion will result, we shall suppress the $\lambda$
subscript.  The explicit labels are 
$N_{Z,sq,FF,L_t}$ and $N_{Z,tri,FF,L_t}$ for the strips of the square and
triangular lattices with free boundary conditions, and 
$N_{Z,sq,PF,L_t}$ and $N_{Z,tri,PF,L_t}$ for the strips of these respective
lattices with cylindrical boundary conditions.  In the special case $v=-1$, our
nomenclature for the corresponding total numbers of $\lambda$'s in the 
chromatic polynomials are listed below, together with the labels used for these
numbers in \cite{ss00}:
\beq
N_{P,sq,FF,L_t} \equiv \hbox{\rm SqFree}(L_t)
\label{npffsq}
\eeq
\beq
N_{P,tri,FF,L_t} \equiv \hbox{\rm TriFree}(L_t)
\label{npfftri}
\eeq
\beq
N_{P,sq,PF,L_t} \equiv \hbox{\rm SqCyl}(L_t)
\label{nppfsq}
\eeq
\beq
N_{P,tri,PF,L_t} \equiv \hbox{\rm TriCyl}(L_t) \ .
\label{nppftri}
\eeq

The Potts ferromagnet has a zero-temperature phase transition in the $L_\ell
\to \infty$ limit of the strip graphs considered here.  However, in contrast to
the infinite-length, finite-width lattice strips with periodic (or twisted
periodic) longitudinal boundary conditions, where this property was reflected
in the feature that ${\cal B}$ in the $u=1/a$ plane passes through the $T=0$
point $u=0$ and hence is noncompact in the $a$ plane, for infinite-length
strips with free longitudinal boundary conditions, ${\cal B}$ is generically
compact in the $a$ (equivalently, the $v$) plane.

\section{General Structural Properties} 

In this section we give some results for the numbers of $\lambda$'s that enter
in the respective chromatic polynomials and Potts model partition functions for
the strips of interest here.  Recall the definitions of the Catalan number
$C_n$ and Motzkin number $M_n$ \cite{motzkin}-\cite{stanley}:
\beq
C_n=\frac{1}{n+1}{2n \choose n}
\label{catalan}
\eeq
\beq
M_n =  \sum_{j=0}^n (-1)^j C_{n+1-j} {n \choose j}
\label{motzkin}
\eeq
These numbers occur in many
combinatoric applications \cite{motzkin}-\cite{sl}.  Among these is the
construction of non-intersecting chords on a circle; the number of ways of
connecting a subset of $n$ points on a circle by non-intersecting chords is
$M_n$, while the number of ways of completely connecting $2n$ points on the
circle by non-intersecting chords is $C_n$.  Summing over subsets of points
connected by chords, this yields the known relation
\beq
M_n=\sum_{k=0}^{[\frac{n}{2}]} {n \choose 2k} C_k  \ .
\label{cattomot}
\eeq
where $[\nu]$ denotes the integral part of $\nu$.
These numbers have the asymptotic behaviors
\beq
C_n \sim \pi^{-1/2} n^{-3/2} 4^n \bigg [ 1 + O(n^{-1}) \bigg ] \quad 
\rm{as} \ \ n \to \infty
\label{cnasymp}
\eeq 
\beq 
M_n \sim 3^{3/2}2^{-1}\pi^{-1/2}n^{-3/2} 3^n \bigg [ 1 + O(n^{-1}) \bigg ] 
\quad \rm{as} \ \ n \to \infty \ . 
\label{motzkinasymp}
\eeq

Ref. \cite{cf} derived expressions for the total number of
$\lambda$'s in $Z$ and $P$ with periodic boundary conditions in the 
longitudinal direction and free boundary conditions in the 
transverse direction. We denote such quantities respectively as 
$N_{Z,sqtri,FP,L_t}$ and $N_{P,sqtri,FP,L_t}$, where $sqtri$ means 
either square or triangular:
\beq
N_{Z,sqtri,FP,L_t} = {2L_t \choose L_t}
\label{nztotcyc}
\eeq
\beq
N_{P,sqtri,FP,L_t} = 2(L_t-1)! \ \sum_{j=0}^{[\frac{L_t}{2}]} 
\frac{(L_t-j)}{(j!)^2(L_t-2j)!} \ . 
\label{nptotcyc}
\eeq
These have the leading asymptotic behaviors
$N_{Z,sqtri,FP,L_t} \sim L_t^{-1/2} \ 4^{L_t}$ and 
$N_{P,sqtri,FP,L_t} \sim L_t^{-1/2} \ 3^{L_t}$ as $L_t \to \infty$. 

\subsection{$N_{P,G,BC,L_t}$}

\tiny

\begin{table}
\caption{\footnotesize{Numbers of $\lambda$'s for the chromatic polynomial
for the strips of the triangular and square lattices having free and
cylindrical boundary conditions and various widths $L_t$.}} 
\scriptsize
\begin{center}
\begin{tabular}{|c|c|c|c|c|c|c|c|c|}
\hline\hline $L_t$ & $N_{P,tri,FF,L_t}$ & $N_{P,sq,FF,L_t}$ &
$2N_{P,sq,FF,L_t}$ & $d_{L_t}$ & $N_{P,tri,PF,L_t}$ & $N_{P,sq,PF,L_t}$ &
$2N_{P,sq,PF,L_t}$ & $L_tN_{P,tri,PF,L_t}$ \\
 & & & $-N_{P,tri,FF,L_t}$ & & & & $-N_{P,tri,PF,L_t}$ & $-d_{L_t}$ \\
\hline\hline
 1  &    1 &    1 &    1 &    1 &   1 &   1 &   1 &    0  \\ \hline
 2  &    1 &    1 &    1 &    1 &   1 &   1 &   1 &    1  \\ \hline
 3  &    2 &    2 &    2 &    1 &   1 &   1 &   1 &    2  \\ \hline
 4  &    4 &    3 &    2 &    3 &   2 &   2 &   2 &    5  \\ \hline
 5  &    9 &    7 &    5 &    6 &   2 &   2 &   2 &    4  \\ \hline
 6  &   21 &   13 &    5 &   15 &   5 &   5 &   5 &   15  \\ \hline
 7  &   51 &   32 &   13 &   36 &   6 &   6 &   6 &    6  \\ \hline
 8  &  127 &   70 &   13 &   91 &  15 &  14 &  13 &   29  \\ \hline
 9  &  323 &  179 &   35 &  232 &  28 &  22 &  16 &   20  \\ \hline
10  &  835 &  435 &   35 &  603 &  67 &  51 &  35 &   67  \\ \hline
11  & 2188 & 1142 &   96 & 1585 & 145 &  95 &  45 &   10  \\ \hline
12  & 5798 & 2947 &   96 & 4213 & 368 & 232 &  96 &  203  \\ \hline
13  &15511 & 7889 &  267 &11298 & 870 & 498 & 126 &   12  \\ \hline
14  &41835 &21051 &  267 &30537 &2211 &1239 & 267 &  417  \\ \hline
\end{tabular}
\end{center}
\label{nptable}
\end{table} 

\normalsize

Recall that for the open strip of the triangular lattice of width $L_t$, 
Ref. \cite{ss00} obtained the result
\beq
N_{P,tri,FF,L_t} = M_{L_t-1} \ .
\label{nptottriff}
\eeq
The proof of this was based on the property that there is a one-one
correspondence between the number of distinct eigenvalues $\lambda$ of the
transfer matrix and the number of non-crossing, non-nearest-neighbor partitions
of the set $\{1,2,...,L_t\}$.  But the latter number is precisely the Motzkin
number $M_{L_t-1}$.  

Now the number of $\lambda$'s for the open square-lattice strip with a given
width $L_t$ is less than for the open triangular strip of this width because
some of the symmetry under reflection about the longitudinal axis, which
renders some of the partitions equivalent to each other.  We proceed to derive
this number.  Because of this reflection symmetry,
$2N_{P,sq,FF,L_t}-N_{P,tri,FF,L_t}$ gives the number of non-crossing
non-nearest-neighbor partitions for a transverse slice that is symmetric under
this reflection.

Using this observation, we proceed to derive the following result:

\medskip
          
{\bf Theorem 1} \quad 
\beq
2N_{P,sq,FF,L_t}-N_{P,tri,FF,L_t} = 
\frac{1}{2}N_{P,sqtri,FP,[\frac{L_t+1}{2}]} \ . 
\label{2npsqff-nptff}
\eeq

{\sl Proof}. \quad Let us denote $2N_{P,sq,FF,L_t}-N_{P,tri,FF,L_t}$ by
$X_{L_t}$ for short.  To show the general method of the proof, we first list,
for small $L_t$, the sets ${\bf P}_{X_{L_t}}$ of partitions having the 
above-mentioned reflection symmetry:  
${\bf P}_{X_1} = \{ 1\}$, ${\bf P}_{X_2} = \{ 1\}$, 
${\bf P}_{X_3} = \{ 1,\delta_{1,3} \}$, 
${\bf P}_{X_4} = \{ 1, \delta_{1,4} \}$, 
${\bf P}_{X_5} = \{1, \delta_{2,4}, \delta_{1,5}, \delta_{2,4}\delta_{1,5}, 
                 \delta_{1,3,5} \}$,
${\bf P}_{X_6} = \{ 1, \delta_{2,5}, \delta_{1,6}, \delta_{2,5}\delta_{1,6},
                       \delta_{1,3}\delta_{4,6} \}$. 
Here it can be seen that going from the partition
for an odd width $L_t$ to the partition for the next greater width, $L_t+1$,
one only has to split the central vertex into two vertices (one connected only
to vertices left of center, the other only to vertices right of center)
without otherwise changing the partitions.  Hence,
\beq X_{L_t+1} = X_{L_t} \quad {\rm for \ odd} \ L_t \ .
\label{xoddly}
\eeq
Going from the partition for an even width $L_t$ to the partition for the next
greater $L_t+1$, there are two possibilities. The first possibility is that
there is no delta function connecting any vertex in the set $\{1, 2, ..,
L_t/2\}$ and the corresponding vertex in the set $\{L_t/2+1, L_t/2+2, ..,
L_t\}$. Let us denote the number of these kinds of partitions by $X_{L_t,a}$,
which is equal to $N_{P,tri,FF,L_t/2}$.  In this case, one can add a vertex in
the middle.  The second possibility is that there is more than one delta
function connecting the two vertex sets mentioned above.  Let us denote the
number of these kinds of partitions by $X_{L_t, b}$.  Clearly, $X_{L_t,
a}+X_{L_t,b}=X_{L_t}$ for the case of even $L_t$ considered here.  In this
case, one can either add a vertex in the middle without any further delta
function or have one more delta function connecting this middle vertex and the
nearest vertex pair which are already connected by a delta function.

Another way to get a partition for $L_t+1$ vertices is to start with a basis
consisting of $L_t-1$ vertices.  Here one splits the central vertex for $L_t-1$
into three vertices with the upper and lower vertices connected by a delta
function. The number of these types of partition is $X_{L_t-1}$.  With
eqs. (\ref{nptottriff}) and (\ref{xoddly}), we obtain
\beqs 
X_{L_t+1} & = & X_{L_t, a} + 2 X_{L_t, b} + X_{L_t-1} \\
          & = & M_{\frac{L_t}{2}-1} + 2 (X_{L_t-1} - M_{\frac{L_t}{2}-1}) 
                 + X_{L_t-1} \\
          & = & 3 X_{L_t-1} - M_{\frac{L_t}{2}-1} \quad {\rm for \ even} \
                 L_t \ .
\label{xevenly}
\eeqs
Setting $L_t = 2\ell$ in eq. (\ref{xevenly}), we have
\beq
X_{2\ell+1} = 3^\ell - 3^{\ell-1}M_0 - 3^{\ell-2}M_1 - ... - M_{\ell-1} =
X_{2\ell+2} \ .
\label{x2l+1}
\eeq
We observe from eq. (\ref{xevenly}) that $2X_{2\ell+1}$ has the same
recursion relation and initial values as $N_{P,L_t+1,\lambda}$ in eq.
(5.1) in \cite{cf}, which proves eq. (\ref{2npsqff-nptff}). \ $\Box$  

\bigskip

\noindent We remark that this proof evidently links together results for two
different sets of longitudinal boundary conditions.

\bigskip

{\bf Theorem 2} \quad
\beq
N_{P,sq,FF,L_t} = \frac{1}{2}M_{L_t-1} + \frac{1}{2}
(L_t^\prime-1)!\sum_{j=0}^{[ \frac{L_t^\prime}{2} ]}
\frac{(L_t^\prime-j)}{(j!)^2(L_t^\prime-2j)!}
\label{npsqff}
\eeq
where
\beq
L_t^\prime = \left [ \frac{L_t+1}{2} \right ] \ .
\label{ltprime}
\eeq

{\sl Proof} \quad This follows from Theorem 1 and the expressions for 
$N_{P,tri,FF,L_t}$ and $N_{P,sqtri,FP,[(L_t+1)/2]}$. 
\ $\Box$.

\bigskip

{\bf Corollary 1}. \quad $N_{P,sq,FF,L_t}$ has the asymptotic behavior
\beq
N_{P,sq,FF,L_t} \sim 3^{1/2}2^{-2}\pi^{-1/2}L_t^{-3/2} 3^{L_t}\bigg [ 1 +
O(L_t^{-1}) \bigg ] \quad {\rm as} \ \ L_t \to \infty \ .
\label{npsqffa}
\eeq
{\sl Proof}. \quad 
This follows directly from the asymptotic behaviors given above for the Motzkin
number and for $N_{P,sqtri,FP,L_t}$, which show that as $L_t \to \infty$, the
third term, $N_{P,sqtri,FP,[(L_t+1)/2]}$, is negligibly small compared with the
first two in eq. (\ref{2npsqff-nptff}).  The latter, in turn, is a consequence
of the fact that the leading exponential asymptotic behavior, as $L_t \to
\infty$, of the dominant terms is $\sim 3^{L_t}$ while that of the third term 
is $\sim 3^{L_t/2}$. 

\medskip

{\bf Corollary 2}. \quad  
\beq
\lim_{N_t \to \infty} \frac{N_{P,sq,FF,N_t}}{N_{P,tri,FF,N_t}} =
\frac{1}{2} \ .
\label{npsqtri}
\eeq
{\sl Proof}.  This follows from (\ref{2npsqff-nptff}) and the asymptotic
behaviors given above. 

\medskip

In \cite{ss00} it was conjectured that $\lim_{L_t \to \infty}
\hbox{\rm SqFree}(L_t)/\hbox{\rm TriFree}(L_t) = 1/2$.  
Our theorem and corollary 2 prove this conjecture. 

\medskip

We next consider the strips with cylindrical boundary conditions and the
quantity $2N_{P,sq,PF,L_t}-N_{P,tri,PF,L_t}$.  We find that the following 
relation holds for $1 \le L_t \le 14$ and conjecture that it holds for all
widths $L_t$: 

\bigskip
  
{\bf Conjecture 1} \quad For arbitrary $L_t$,
\beq
2N_{P,sq,PF,L_t}-N_{P,tri,PF,L_t} = \cases{ \frac{1}{2}N_{P,sqtri,FP,
\frac{L_t}{2}} &
for even $L_t$ \cr
\frac{1}{4}N_{P,sqtri,FP,\frac{L_t+1}{2}} - \frac{1}{2} 
R_{\frac{L_t-1}{2}} & for odd $L_t \ge 3$ }
\label{2npsqpf-nptpf}
\eeq
where $R_n$ is the Riordan number, which may be defined via the generating
function \cite{stanley},\cite{bernhart}
\beq
\frac{1+x-(1-2x-3x^2)^{1/2}}{2x(1+x)} = \sum_{n=0}^\infty R_n x^n
\label{rn}
\eeq
or, for $n \ge 2$, by \cite{rmis} 
\beq
R_n = (-1)^n\sum_{j=1}^{n-1} (-1)^{j+1}M_j \quad {\rm for} \ n \ge 2 \ .
\label{rnsum}
\eeq
Note that $R_{L_t}$ is equal to the quantity $d_{L_t}$ given in Table
\ref{nptable} for $L_t \ge 2$.

\bigskip

{\bf Theorem 3}. \quad 
\beq
N_{P,tri,PF,L_t} = \frac{d_{L_t}+L_t-1}{L_t} \quad {\rm for \ prime} \ L_t \
 .
\label{nptpf}
\eeq
{\sl Proof} \quad  As shown in \cite{ss00}, $d_{L_t}$ is the number of
non-crossing non-nearest-neighbor partitions of the vertex set $\{1, ..., L_t
\}$ with periodic boundary conditions on the set. Now, when $L_t$ is a prime
number, all partitions except the partition $1$ (i.e., all blocks being 
singletons) have an $L_t$-fold degeneracy. Therefore,
\beq
L_tN_{P,tri,PF,L_t}-d_{L_t} = L_t-1 \quad {\rm for \ prime} \ L_t 
\label{lynptpf-d}
\eeq
and the theorem follows. \ $\Box$

In \cite{ss00} it was conjectured that the number 
$\hbox{\rm TriCyl}(L_t)$, i.e., 
$N_{P,tri,PF,L_t}$, behaves asymptotically like $d_{L_t}/L_t$ as $L_t
\to \infty$.  Our Theorem 3 (\ref{nptpf}) proves this conjecture for prime
$L_t$. 

\bigskip

 From eqs. (\ref{2npsqpf-nptpf}) and (\ref{nptpf}), we infer

\bigskip
 
{\bf Conjecture 2} \quad For prime $L_t \ge 3$,
\beq
N_{P,sq,PF,L_t} = \frac{1}{2} \bigg (\frac{d_{L_t}+L_t-1}{L_t} +
\frac{1}{4}N_{P,sqtri,FP,\frac{L_t+1}{2}} 
- \frac{1}{2} R_{\frac{L_t-1}{2}} \bigg ) \ .
\label{npsqpf}
\eeq
Note that Conjecture 2 is consistent with the conjecture made in \cite{ss00}
that in the limit as $L_t \to \infty$ the ratio $SqCyl(L_t)/TriCyl(L_t) \equiv
N_{P,sq,PF,L_t}/N_{P,tri,PF,L_t}$ is equal to 1/2.  For non-prime $L_t$, the
right-hand sides of eqs. (\ref{nptpf}) and (\ref{npsqpf}) are the lower bounds
for $N_{P,tri,PF,L_t}$ and $N_{P,sq,PF,L_t}$, respectively.  However, we have
not yet obtained formulas for $N_{P,tri,PF,L_t}$ or $N_{P,sq,PF,L_t}$ for
arbitrary $L_t$.

\begin{table}
\caption{\footnotesize{Numbers of $\lambda$'s for the Potts model partition
function for the strips of the triangular and square lattices having free and
cylindrical boundary conditions and various widths $L_t$.}}
\begin{center}
\begin{tabular}{|c|c|c|c|c|c|c|c|}
\hline\hline $L_t$ & $N_{Z,tri,FF,L_t}$ & $N_{Z,sq,FF,L_t}$ &
$2N_{Z,sq,FF,L_t}$ & $N_{Z,tri,PF,L_t}$ & $N_{Z,sq,PF,L_t}$ &
$2N_{Z,sq,PF,L_t}$ & $L_tN_{Z,tri,PF,L_t}$ \\ 
 & & & $-N_{Z,tri,FF,L_t}$ & & & $-N_{Z,tri,PF,L_t}$ & $-N_{Z,tri,FF,L_t}$ \\
\hline\hline  
 1  &      1 &      1 &    1 &    1 &    1 &   1 &    0  \\ \hline
 2  &      2 &      2 &    2 &    2 &    2 &   2 &    2  \\ \hline
 3  &      5 &      4 &    3 &    3 &    3 &   3 &    4  \\ \hline
 4  &     14 &     10 &    6 &    6 &    6 &   6 &   10  \\ \hline
 5  &     42 &     26 &   10 &   10 &   10 &  10 &    8  \\ \hline
 6  &    132 &     76 &   20 &   28 &   24 &  20 &   36  \\ \hline
 7  &    429 &    232 &   35 &   63 &   49 &  35 &   12  \\ \hline
 8  &   1430 &    750 &   70 &  190 &  130 &  70 &   90  \\ \hline
 9  &   4862 &   2494 &  126 &  546 &  336 & 126 &   52  \\ \hline
10  &  16796 &   8524 &  252 & 1708 &  980 & 252 &  284  \\ \hline
11  &  58786 &  29624 &  462 & 5346 & 2904 & 462 &   20  \\ \hline 
12  & 208012 & 104468 &  924 &17428 & 9176 & 924 & 1124  \\ \hline
13  & 742900 & 372308 & 1716 &57148 &29432 &1716 &   24  \\ \hline
\end{tabular}
\end{center}
\label{nztable}
\end{table}

\subsection{$N_{Z,G,BC,L_t}$}

Here we carry out a similar analysis for the total number of $\lambda$'s in the
Potts model partition function or equivalently, the Tutte polynomial for
various lattice strips.  First, we list as the left-most column in Table
\ref{nztable} the number of $\lambda$'s that enter in this partition function
for the open strip of the triangular lattice of width $L_t$, i.e.,
$N_{Z,tri,FF,L_t}$.  This number is equal to the number of non-crossing
partitions of the set $\{1,2,...,L_t\}$, which is precisely the Catalan number.
From this observations, it follows that \cite{ss00}
\beq
N_{Z,tri,FF,L_t} = C_{L_t} \ .
\label{nztottriff}
\eeq
In the zero-temperature limit for the antiferromagnetic case, one can work 
in a subspace that corresponds, after a suitable change of basis, to 
only non-nearest-neighbor partitions \cite{ss00}. In this limit, the number of
$\lambda$'s is thus reduced to the non-crossing, non-nearest-neighbor
partitions given by $M_{L_t-1}$ in (\ref{nptottriff}). 

Just as was the case for the chromatic polynomial, the number of $\lambda$'s
for the open square-lattice strip with a given width $L_t$ is less than for the
open triangular strip of this width because of the symmetry under
reflection about the longitudinal axis, which renders some of the partitions
equivalent to each other.  We proceed to derive this number.  Because
$2N_{Z,sq,FF,L_t}-N_{Z,tri,FF,L_t}$ gives the number of non-crossing partitions
for a slice of the transverse vertices which is symmetric under the reflection
symmetry, we have

\medskip   
 
{\bf Theorem 4} \quad
\beq
2N_{Z,sq,FF,L_t}-N_{Z,tri,FF,L_t} = \cases{ N_{Z,sqtri,FP,\frac{L_t}{2}} & 
for even $L_t$ \cr  
\frac{1}{2}N_{Z,sqtri,FP,\frac{L_t+1}{2}} & for odd $L_t$ }                
\label{2nzsqff-nztff}   
\eeq
where $N_{Z,sqtri,FP,L_t}$ was listed in eq. (\ref{nztotcyc}) from \cite{cf}. 

\medskip

{\sl Proof} \quad Let us denote $2N_{Z,sq,FF,L_t}-N_{Z,tri,FF,L_t}$ by
$Y_{L_t}$ for simplicity, and list,
for small $L_t$, the sets ${\bf P}_{Y_{L_t}}$ of partitions having 
reflection symmetry:
${\bf P}_{Y_1} = \{ 1\}$, 
${\bf P}_{Y_2} = \{1, \delta_{1,2}\}$, 
${\bf P}_{Y_3} = \{ 1, \delta_{1,3}, \delta_{1,2,3} \}$, 
${\bf P}_{Y_4} = \{ 1, \delta_{2,3}, \delta_{1,4},\delta_{2,3}\delta_{1,4}, 
                       \delta_{1,2}\delta_{3,4}, \delta_{1,2,3,4,5} \}$, 
${\bf P}_{Y_5} = \{ 1, \delta_{2,4}, \delta_{1,5}, 
    \delta_{2,4}\delta_{1,5}, \delta_{1,2}\delta_{4,5}$, $\delta_{2,3,4},
    \delta_{1,2,4,5}, \delta_{2,3,4}\delta_{1,5}, \delta_{1,3,5}, 
    \delta_{1,2,3,4,5} \}$. 
Here it can be seen that going from an odd $L_t$
to $L_t+1$, the number of partitions is doubled because one can split the
central vertex to two vertices with or without a delta function for these
two vertices. Therefore,
\beq
Y_{L_t+1} = 2 Y_{L_t} \quad {\rm for \ odd} \ L_t \ .
\label{yoddly}
\eeq
Going from the partition for an even $L_t$ to the partition for $L_t+1$, there
are two possibilities. The first occurs if there is no delta function
connecting any vertex in the vertex set $\{1, 2, .., L_t/2\}$ and the
corresponding vertex in vertex set $\{L_t/2+1, L_t/2+2, .., L_t\}$.  Let us
denote the number of these kinds of partitions by $Y_{L_t, a}$, which is equal
to $N_{Z,tri,FF,L_t/2}$.  In this case, one can add a vertex in the middle. The
second possibility occurs if there is more than one delta function connecting
the two vertex sets mentioned above.  Let us denote the number of this kind of
partitions by $Y_{L_t, b}$. Clearly $Y_{L_t, a} + Y_{L_t, b} = Y_{L_t}$ for the
case of even $L_t$ considered here.  For this case one can either add a vertex
in the middle without any further delta function or have one more delta
function connecting this middle vertex and the nearest vertex pair which are
already connected by a delta function.  Using eqs. (\ref{nztottriff}) and
(\ref{yoddly}), we obtain
\beqs 
Y_{L_t+1} & = & Y_{L_t, a} + 2 Y_{L_t, b} \\
          & = & C_{\frac{L_t}{2}} + 2 (2Y_{L_t-1} - C_{\frac{L_t}{2}}) \\
          & = & 4 Y_{L_t-1} - C_{\frac{L_t}{2}} \quad {\rm for \ even}
                \ L_t \ .
\label{yevenly}
\eeqs
Set $L_t = 2\ell$ in eq. (\ref{yevenly}), we have
\beqs
Y_{2\ell+1} & = & 4^\ell - 4^{\ell-1}C_1 - 4^{\ell-2}C_2 - ... - C_{\ell}
 \\ 
            & = & \frac{1}{2}{2\ell+2 \choose \ell+1} \\
\label{y2l+1}
\eeqs
and 
\beq
Y_{2\ell} = {2\ell \choose \ell} \ .
\label{yly}
\eeq
Using (\ref{nztotcyc}), this completes the theorem. $\Box$

\medskip

For even $L_t$, $N_{Z,sqtri,FP,\frac{L_t}{2}}$ is equal to the entry in the
central column of Pascal's triangle (with rows $r\ge 0$ and elements on the
$r$'th row given by ${r \choose s}$ for $0 \le s \le r$).  For odd $L_t$,
$(1/2)N_{Z,sqtri,FP,\frac{L_t+1}{2}}$ is equal to the entry in the
next-to-central column of this triangle \cite{sl}.
                 
\bigskip

{\bf Theorem 5} \quad
\beq
N_{Z,sq,FF,L_t} = \cases{ \frac{1}{2} \left[ C_{L_t} + {L_t \choose
\frac{L_t}{2}} \right] & for even $L_t$  \cr
\frac{1}{2} \left[ C_{L_t} + \frac{1}{2} {L_t+1 \choose \frac{L_t+1}{2}}
\right] & for odd $L_t$ } \ .
\label{nztotsqff}
\eeq

{\sl Proof} \quad The theorem follows from Theorem 4 and
eq. (\ref{nztottriff}).  \ $\Box$.

\bigskip

The sequence of numbers $N_{Z,sq,FF,L_t}$ is equal to the number $N_{br,L_t+1}$
of distinct bracelets (i.e. distinct modulo rotations and reflections) with
$L_t+1$ black beads and $L_t$ white beads (given as sequence A007123 in
\cite{sl} and previously as sequence M1218 in \cite{sp}). This can be
understood as follows.  The number of bracelets enumerated without dividing out
by equivalence classes of reflections, i.e. the number that are distinct up to
rotations only, with $L_t+1$ black beads and $L_t$ white beads, is given by the
Catalan number $C_{L_t}$. Of these, the number that are symmetric
under reflection is ${L_t \choose \frac{L_t}{2}}$ for even $L_t$ and ${L_t
\choose \frac{L_t+1}{2}}$ for odd $L_t$. Therefore, the number of distinct
bracelets is the same as the sequence of numbers $N_{Z,sq,FF,L_t}$.

\medskip

{\bf Corollary 3} \quad 

The number $N_{Z,sq,FF,L_t}$ has the leading asymptotic behavior
\beq
N_{Z,sq,FF,L_t} \sim \frac{1}{2}\pi^{-1/2} L_t^{-3/2} \biggl [ 4^{L_t} +
2^{1/2}L_t 2^{L_t} + O(\frac{1}{L_t}) \biggr ] \quad {\rm as} \ \ L_t \to
\infty \ .
\label{nztotffasymp}
\eeq
{\sl Proof} \quad  This follows from theorem 5 and eq. (\ref{cnasymp}). 
Hence, 

\medskip

{\bf Corollary 4} \quad 
\beq
\lim_{L_t \to \infty} \frac{N_{Z,sq,FF,L_t}}{N_{Z,tri,FF,L_t}} = 
\frac{1}{2} \ .
\label{nztotsqtriratio}
\eeq
{\sl Proof}  \quad This follows from (\ref{nztottriff}) and our new result
(\ref{nztotsqff}). 

\medskip

Similarly, consider the quantity $2N_{Z,sq,PF,L_t}-N_{Z,tri,PF,L_t}$ for the
strips with cylindrical boundary conditions.  We find that this is the same as
$2N_{Z,sq,FF,L_t}-N_{Z,tri,FF,L_t}$ for $1 \le L_t \le 13$ and infer the 
following conjecture:

\medskip

{\bf Conjecture 3} \quad For arbitrary $L_t$,
\beq
2N_{Z,sq,PF,L_t}-N_{Z,tri,PF,L_t} = \cases{ N_{Z,sqtri,FP,\frac{L_t}{2}} & 
for even $L_t$ \cr
\frac{1}{2}N_{Z,sqtri,FP,\frac{L_t+1}{2}} & for odd $L_t$ } \ .
\label{2nzsqpf-nztpf}
\eeq

\bigskip

{\bf Theorem 6} 
\beq
N_{Z,tri,PF,L_t} = \frac{C_{L_t}+2(L_t-1)}{L_t} \quad {\rm for \ prime} \ 
L_t \ .
\label{nztottripf}
\eeq
{\sl Proof} \quad  As shown in \cite{ss00}, $N_{Z,tri,FF,L_t}$ is the number
of non-crossing partitions of vertex set $\{1, ..., L_t \}$.  Now if 
$L_t$ is a prime number, except for the partitions $1$ (i.e., all blocks being
singletons) and $\delta_{1,2,...,L_t}$ (i.e., a unique block), there is an 
$L_t$-fold degeneracy for the rest of the partitions. Therefore,
\beq
L_tN_{Z,tri,PF,L_t}-N_{Z,tri,FF,L_t} = 2(L_t-1) \quad {\rm for \ prime} \ L_t
\label{lynztpf-nztff}
\eeq
and the theorem follows. \ $\Box$
 
\bigskip

 From reference to \cite{sl}, we observe that the numbers $N_{Z,tri,PF,L_t}$
that we have calculated are equal to the sequence listed as A054357 in
\cite{sl} which enumerates the number of tree graphs with $L_t$ edges modulo
equivalence under 2-coloring (so that, for example, this enumeration includes
only one path graph if $L_t$ is odd and includes two path graphs if $L_t$ is
even). We also observe that our result (\ref{nztottripf}) is the special case,
for prime $L_t$, of this sequence.  A general analytic formula for the above
sequence is given in \cite{binc}.  Based on these connections, we conjecture
that

\bigskip

{\bf Conjecture 4} \quad For arbitrary $L_t$ 
\beq
N_{Z,tri,PF,L_t}=\frac{1}{L_t}\biggl [ C_{L_t} + \sum_{d|L_t; \ 1 \le d < L_t}
\phi(L_t/d){2d \choose d} \biggr ]
\label{nztripfgen}
\eeq
where $d|L_t$ means that $d$ divides $L_t$ and $\phi(n)$ is the Euler function,
equal to the number of positive integers not exceeding the positive integer 
$n$ and relatively prime to $n$. 

\bigskip

Substituting eq. (\ref{nztripfgen}) into eq. (\ref{2nzsqpf-nztpf}) then
yields a conjecture for $N_{Z,sq,PF,L_t}$ (again for arbitrary $L_t$), namely 
\beq
N_{Z,sq,PF,L_t} = \cases{ \frac{1}{2}\biggl [ {L_t \choose L_t/2} + 
N_{Z,tri,PF,L_t} \biggr ] & for even $L_t$ \cr
\frac{1}{2}\biggl [ \frac{1}{2}{L_t+1 \choose (L_t+1)/2} + N_{Z,tri,PF,L_t} 
\biggr ] & for odd $L_t$  \ . }
\label{nzsqpfgen}
\eeq
where $N_{Z,tri,PF,L_t}$ is given by eq. (\ref{nztripfgen}).

\vspace{6mm}

Note that for each strip graphs of a given lattice $\Lambda$ with some
specified boundary conditions $BC$ for which one has results from
\cite{cf}, \cite{ss00} and the present work, the ratio of the number of
$\lambda$'s in the chromatic polynomial divided by the number of $\lambda$'s in
the full Potts model partition function decreases exponentially fast for $L_t$
and vanishes in the limit of large $L_t$:
\beq
\lim_{L_t \to \infty} \frac{N_{P,\Lambda,BC,L_t}}{N_{Z,\Lambda,BC,L_t}} = 0 \ .
\label{nptotnztotratio}
\eeq

\section{Potts Model Partition Functions for Square-lattice Strips with Free 
Boundary Conditions}

The Potts model partition function $Z(G,q,v)$, equivalent to the Tutte
polynomial, for a square-lattice strip of width $L_t$ and length $L_\ell=m$
vertices with free boundary conditions is given by
\beqs 
Z_{(L_\ell)_{\rm F} \times (L_t)_{\rm F}}(q,v) &=& 
        \vec{v}^{\rm T} \cdot H \cdot T^{m-1} \cdot \vec{u} \\
    & = & \vec{w}^{\rm T} \cdot T^{m-1} \cdot \vec{u} 
\eeqs  
where $\vec{w}^{\rm T} = \vec{v}^{\rm T} \cdot H$.  (No confusion should result
between the variable $v$ and the vector $\vec{v}$.) Here $T$ is the 
transfer matrix, and $H$ is the part of the transfer matrix corresponding to
the edges (bonds) lying within a single layer. This partition function was
calculated for arbitrary $q$, $v$, and $L_\ell$ with $L_t=2$ in \cite{a} and
with $L_t=3$ in \cite{s3a}.  It was also calculated for arbitrary $q$ and $v$
for large finite strips in \cite{ks}.  We review the calculations for arbitrary
$L_\ell$ with $L_t=2,3$ here in our present notation.  We have also calculated
the Potts model partition function for arbitrary $L_\ell$ for the strip with
widths $L_t=4$ and $L_t=5$ having free boundary conditions and have used these
for the analyses of partition function zeros given later in this paper, but the
results are too lengthy to list here.  They are available from the authors on
request and in the {\tt mathematica} file {\tt transfer\_tutte\_sq.m} that is 
available with the electronic version of 
this paper in the {\tt cond-mat} archive at {\tt http://www.lanl.gov}.

%
%
\subsection{$L_t = 2$} \label{sec2F} 

The number of elements in the basis is trivially equal to two: 
${\bf P} = \{ 1, \delta_{1,2} \}$. In this basis, the transfer matrices and 
the other relevant quantities are given by  
\beqs 
T &=& \left( \begin{array}{cc} 
             q^2 + 3 q v + 3 v^2 & (1 + v) (q + 2 v) \\ 
             v^3                 & v^2 (1+v) 
             \end{array} \right) \\
H &=& \left( \begin{array}{cc}
              1 & 0 \\
              v & 1+v 
             \end{array} \right) \\
\vec{v} &=& \left( \begin{array}{c} 
                   q^2 \\
                    q 
             \end{array} \right) \\
\vec{w} &=&  \left( \begin{array}{c}
                    q (q + v) \\
                    q (1 + v) 
             \end{array} \right) \\
\vec{u} &=&  \left( \begin{array}{c} 
                     1 \\ 
                     0  
             \end{array} \right) 
\eeqs 

In \cite{strip,strip2} a generating function method was presented for chromatic
polynomials and was generalized to the full Potts model partition function in
\cite{a}.  For a strip graph of type $G$, including specification of width and
boundary conditions, with arbitrary length $L_\ell=m$, and denoting the 
particular member of this class of graphs with length $m$ as $G_m$, one has
\beq
\Gamma(G,q,v,z) = \sum_{m=0}^\infty Z(G_m,q,v)z^m \ .
\label{gammazfbc}
\eeq
The generating function is a rational function of the form 
\beq
\Gamma(G,q,v,z) = \frac{ {\cal N}(G,q,v,z)}{{\cal D}(G,q,v,z)}
\label{gammazcalc}
\eeq
with 
\beq
{\cal N}(G,q,v,z) = \sum_{j=0}^{d_{\cal N}} A_{G,j}(q,v) z^j
\label{n}
\eeq
and
\beq
{\cal D}(G,q,v,z) = 1 + \sum_{j=1}^{d_{\cal D}} b_{G,j}(q,v) z^j
\label{d}
\eeq
where the $A_{G,j}$ and $b_{G,j}$ are polynomials in $q$ and $v$ that depend 
on the type of strip (but not its length) of respective degrees $d_{\cal N}$
and $d_{\cal D}$ in the auxiliary expansion variable $z$.  The factorization of
the denominator yields the $\lambda$'s for a given type of strip:
\beq
{\cal D}(G,q,v,z) = \prod_{j=1}^{d_{\cal D}}(1-\lambda_{G,j}(q,v)z) \ .
\label{lambdaform}
\eeq
In \cite{strip}-\cite{a} a convention was used that allowed for a more general
(inhomogeneous) form of strip graph $J(\prod H)I$ in which the end subgraphs
$I$ and $J$ could be different from the subgraph $H$ that is repeated to form
the strip.  Since this generality is not needed for homogeneous lattice strip 
graphs, a convention was introduced in \cite{hca} according to which
the $m=0$ term in the expansion (\ref{gammazfbc}) is taken to be the line graph
$T_{L_t}$ rather than a column of squares of height $L_t$.  This simplifies the
$A_{G,j}$ polynomials in ${\cal N}(G,q,v,z)$ (the denominator 
${\cal D}(G,q,v,z)$ is independent of this convention).  Here we shall use the
convention of \cite{hca}.  Further, to shorten the notation, we shall suppress
the $G$-dependence on ${\cal N}$, ${\cal D}$, the $b_{G,j}$, and $A_{G,j}$
where this is obvious. 
Then for the present strip graph of width $L_t=2$ \cite{strip}
\beqs
{\cal D} &=& 1 + b_1 z + b_2 z^2 \\
{\cal N} &=& A_0 + A_1 z 
\eeqs
where \cite{a} 
\beqs 
b_1 &=& -(v^3+4v^2+3qv+q^2) \\
b_2 &=& v^2(1+v)(q+v)^2 \\[2mm]
A_0 &=& v(q + v) \\
A_1 &=& -q^2 v^2 (1+v)
\eeqs
%
 
%
%
\subsection{$L_t = 3$} 
\label{sec3F}

In this case we have a four-dimensional basis given by ${\bf P} = 
\{ 1, \delta_{1,2} + \delta_{2,3}, \delta_{1,3}, \delta_{1,2,3} \}$. 
In this basis the transfer matrix is given by  
\beq
T = \left( \begin{array}{cccc}
        T_{11} & T_{12} & T_{13} & T_{14} \\ 
        v^3 (q + 2 v) & v^2 (1 + v) (q + 3 v) & v^3 (2 + v) & v^2 (1 + v)^2 \\ 
        v^4 &  2 v^3 (1 + v) &  v^2 (q + 3 v + v^2) & v^2 (1 + v)^2 \\ 
        v^5 &  2 v^4 (1 + v) & v^4 (2 + v) & v^3 (1 + v)^2 
    \end{array} \right) 
\eeq 
where 
\beqs
T_{11}  &=& (q + 2 v) (q^2 + 3 q v + 4 v^2) \\ 
T_{12}  &=& 2 (1 + v) (q^2 + 4 q v + 5 v^2) \\ 
T_{13}  &=& q^2 + 5 q v + 8 v^2 + q v^2 + 3 v^3 \\ 
T_{14}  &=& (1 + v)^2 (q + 3 v) 
\eeqs 

The rest of the matrices and vectors are given by 
\beqs
H &=& \left( \begin{array}{cccc}
                1 &            0 &          0 &          0  \\
                v &        1 + v &          0 &          0  \\
                0 &            0 &          1 &          0  \\
              v^2 &   2 v(1 + v) &  v (2 + v) &  (1 + v)^2  
             \end{array} \right) \\
\vec{v} &=& \left( \begin{array}{c}
                   q^3 \\
                 2 q^2 \\
                   q^2 \\
                   q  
             \end{array} \right) \\
\vec{w} &=& \left( \begin{array}{c}
                q (q + v)^2         \\
                2 q (1 + v) (q + v) \\
                q (q + 2 v + v^2)   \\
                q (1 + v)^2 
             \end{array} \right) \\
\vec{u} &=& \left( \begin{array}{c}
                   1   \\
                   0   \\
                   0   \\
                   0
             \end{array} \right) 
\eeqs

In the generating-function formalism 
\beqs
{\cal D} &=& 1+b_1 z+b_2 z^2+b_3 z^3+b_4 z^4\\
{\cal N} &=& A_0 +A_1 z + A_2 z^2 + A_3 z^3  
\eeqs
where
\beqs 
b_1 &=& -q^3 - 5 q^2 v - 12 q v^2 - 15 v^3 - q v^3 - 6 v^4 - v^5 \\
b_2 &=& v^2 (q + v) (2 q^3 + 13 q^2 v + q^3 v + 30 q v^2 + 8 q^2 v^2 
      + 32 v^3 +21 q v^3 + q^2 v^3 
        \nonumber \\
    & & \qquad 
      + 30 v^4 + 4 q v^4 + 10 v^5 + v^6) \\
b_3 &=& -v^4 (1 + v) (q + v)^2 (q^3 + 9 q^2 v + 19 q v^2 + 4 q^2 v^2 + 
       15 v^3 +10 q v^3 + q^2 v^3 
        \nonumber \\
    & & \qquad 
       + 10 v^4 + 2 q v^4 + 2 v^5) \\ 
b_4 &=& v^7 (1 + v)^3 (q + v)^5 \\[2mm]        
A_0 &=& q (q + v)^2 \\
A_1 &=& -q v^2 (- v^4 + 6 q^2 v^2 + q^2 v^3 + q^3 v + 2 q v^4 +
        7 q v^3 + 9 q^2 v + 9 q v^2 + 2 q^3) \\
A_2 &=& q v^4 (1 + v) (q + v) (2 q v^3 - v^3 + q^2 v^3 + 4 q v^2 
        + 4 q^2 v^2 + 7 q^2 v + q^3) \\
A_3 &=& -q^3 v^7 (1 + v)^3 (q + v)^2
\eeqs
%

%
%
\section{Potts Model Partition Functions for Strips with Cylindrical Boundary 
         Conditions}

The Potts model partition function for a square-lattice strip of fixed width 
$L_t$ vertices and arbitrarily great length $L_\ell=m$ vertices, with 
cylindrical boundary conditions, is given by
\beqs
Z_{(L_t)_{\rm P} \times (L_\ell)_{\rm F}}(q;v) &=&
        \vec{v}^{\rm T} \cdot H \cdot T^{m-1} \cdot \vec{u} \\
    & = & \vec{w}^{\rm T} \cdot T^{m-1} \cdot \vec{u}
\eeqs
where $\vec{w}^{\rm T} = \vec{v}^{\rm T} \cdot H$.  Here we give results for
the cases $L_t=2$ and $L_t=3$. We have also calculated this partition
function for $L_t=4,5$, but the results are too lengthy to list here.  
They are available from the authors on
request and in the {\tt mathematica} file {\tt transfer\_tutte\_sq.m} that is
available with the electronic version of
this paper in the {\tt cond-mat} archive at {\tt http://www.lanl.gov}.
 
%
%
\subsection{$L_t = 2$} \label{sec2P}

The square-lattice strip of width $2_{\rm P}$ is equivalent to the
square-lattice strip of width $2_{\rm F}$ when $v=-1$ (chromatic polynomial),
but this equivalence does not hold for nonzero temperature.  Furthermore,
the square-lattice strip of width $2_{\rm P}$ with coupling $v$ is easily seen
to be equivalent to an inhomogeneous square-lattice strip of width $2_{\rm F}$
with coupling $v$ in the longitudinal direction and $v' = v(2+v)$ in the
transverse direction.  

The number of elements in the basis is two:
${\bf P} = \{ 1, \delta_{1,2} \}$. In this basis, the matrices and vectors 
are given by   
\beqs
T &=& \left( \begin{array}{cc}
            q^2 + 4 q v + 5 v^2 + q v^2 + 2 v^3 & 
            (1 + v)^2 (q + 2 v) \\ 
            v^3 (2 + v) &  v^2 (1 + v)^2  
             \end{array} \right) \\
H &=& \left( \begin{array}{cc}
              1 & 0 \\ 
              v(2 + v) &  (1+v)^2  
             \end{array} \right) 
   = \left( \begin{array}{cc}
              1 & 0 \\
              v &  (1+v)
             \end{array} \right) ^2 \\
\vec{v} &=& \left( \begin{array}{c}
                   q^2 \\
                    q
             \end{array} \right) \\
\vec{w} &=&  \left( \begin{array}{c}
                   q (q + 2 v + v^2) \\ 
                   q (1 + v)^2  
             \end{array} \right) \\
\vec{u} &=&  \left( \begin{array}{c}
                     1 \\
                     0
             \end{array} \right)
\eeqs

In the generating-function language we have 
\beqs
{\cal D} &=& 1 + b_1 z + b_2 z^2 \\
{\cal N} &=& A_0 + A_1 z 
\eeqs
where
\beqs 
b_1 &=& -(q^2 + 4 q v + 6 v^2 + q v^2 + 4 v^3 + v^4) \\
b_2 &=& v^2(1+v)^2(q+v)^2 \\[2mm]
A_0 &=& q(q + v^2 + 2 v) \\
A_1 &=& -q^2 v^2 (1+v)^2
\eeqs
%

%
%
\subsection{$L_t = 3$} \label{sec3P}

For the strip with width $L_t=3$ and cylindrical boundary conditions (PF3)
there are three elements in our basis: ${\bf P} = \{ 1, \delta_{1,2} +
\delta_{2,3} + \delta_{1,3}, \delta_{1,2,3} \}$.  In this basis the transfer
matrix is given by
\beq
T = \left( \begin{array}{ccc}
        T_{11} & T_{12} & T_{13} \\
        v^3 (q + 4 v + v^2) &  v^2 (1 + v) (q + 7 v + 3 v^2) &  v^2 (1 + v)^3 \\ 
        v^5 (3 + v)         &  3 v^4 (1 + v) (2 + v)         &  v^3 (1 + v)^3 
    \end{array} \right)
\eeq
where
\beqs
T_{11}  &=& q^3 + 6 q^2 v + 15 q v^2 + 16 v^3 + q v^3 + 3 v^4 \\ 
T_{12}  &=& 3 (1 + v) (q^2 + 5 q v + 8 v^2 + q v^2 + 3 v^3) \\ 
T_{13}  &=& (1 + v)^3 (q + 3 v)  
\eeqs

The rest of the matrices and vectors are given by
\beqs
H &=& \left( \begin{array}{ccc}
                1 &                    0 &          0  \\
                v &                1 + v &          0  \\
      v^2 (3 + v) &  3 v (1 + v) (2 + v) &  (1 + v)^3 
             \end{array} \right) \\
\vec{v} &=& \left( \begin{array}{c}
                   q^3 \\
                 3 q^2 \\
                   q
             \end{array} \right) \\
\vec{w} &=& \left( \begin{array}{c}
                    q (q^2 + 3 q v + 3 v^2 + v^3) \\
                    3 q (1 + v) (q + 2 v + v^2)   \\
                    q (1 + v)^3
             \end{array} \right) \\
\vec{u} &=& \left( \begin{array}{c}
                   1   \\
                   0   \\
                   0
             \end{array} \right) 
\eeqs

In the generating-function language we have
\beqs
{\cal D} &=& 1 + b_1z + b_2z^2 + b_3z^3\\
{\cal N} &=& A_0 + A_1z + A_2z^2 
\eeqs
where
\beqs
b_1 &=& 
-(q^3 + 6 q^2 v + 16 q v^2 + 24 v^3 + 2 q v^3 + 16 v^4 + 6 v^5 + v^6) 
    \\
b_2 &=& 
v^2 (1 + v) (q + v) (q^3 + 10 q^2 v + 26 q v^2 + 5 q^2 v^2 + 24 v^3 +
        20 q v^3 + q^2 v^3 + 26 v^4 
        \nonumber \\
    & & \qquad
        + 5 q v^4 + 10 v^5 + v^6) \\ 
b_3 &=& -v^5 (1 + v)^4 (q + v)^4 \\[2mm] 
A_0 &=& q (3v^2 + 3 q v + v^3 + q^2) \\
A_1 &=& -q v^2 (1 + v) (3 q v^4 - 2 v^4 + q^2 v^3 + 10 q v^3 - 3 v^3 +
       5 q^2 v^2 + 9 q v^2 \cr
    & & + 8 q^2 v + q^3) \\ 
A_2 &=& q^3 v^5 (1 + v)^4 (q + v)
\eeqs

\section{Partition Function Zeros in the $\lowercase{q}$ Plane}

In this section we shall present results for zeros and the continuous
accumulation set (singular locus for the free energy) ${\cal B}$ in the $q$
plane for the partition function of the Potts antiferromagnet on square-lattice
strips with free longitudinal and free or periodic transverse boundary
conditions.

\subsection{Free Transverse Boundary Conditions}

\begin{figure}[hbtp]
  \centering
  \epsfxsize=380pt
  \epsffile{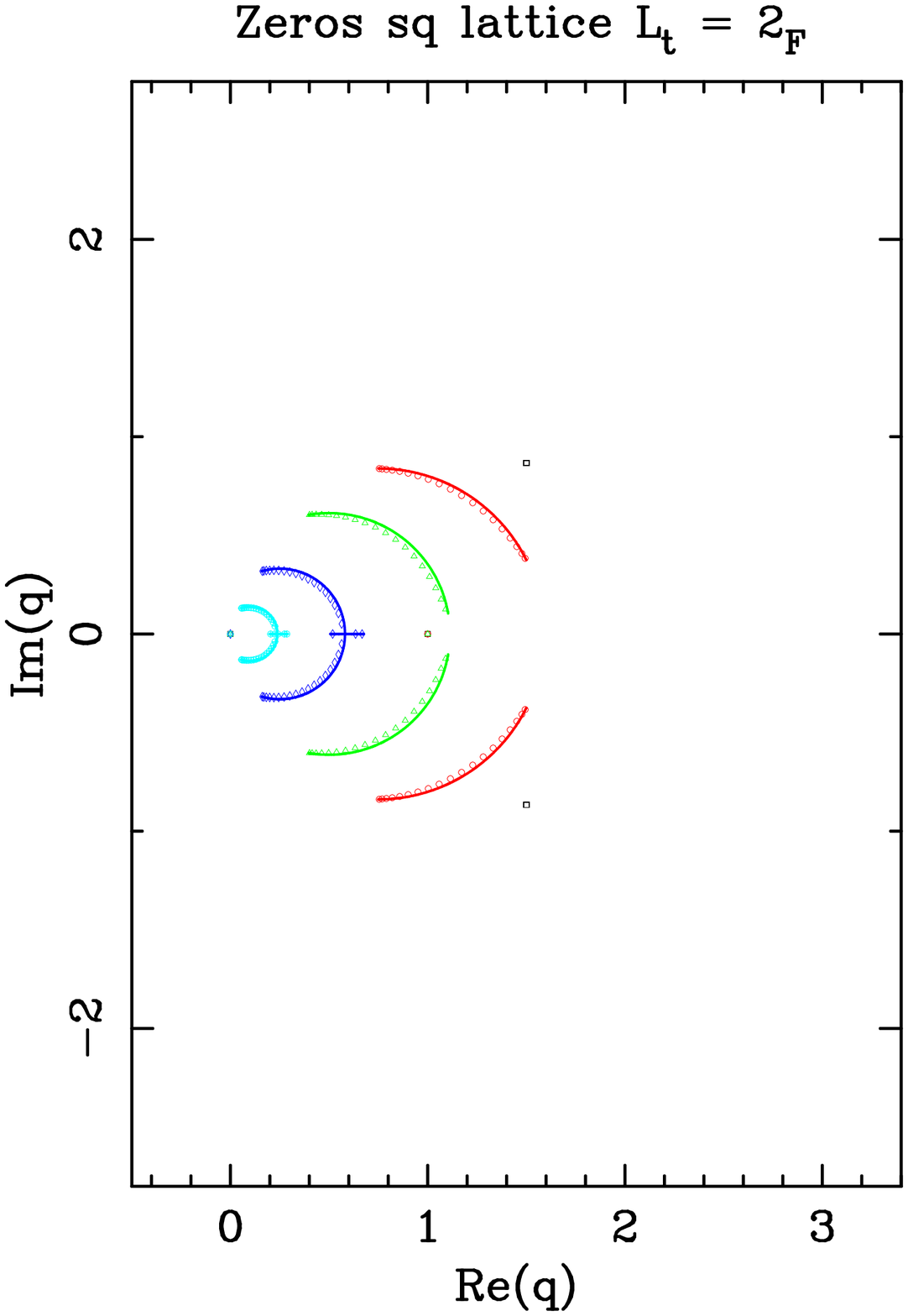}
  \caption[a]{
  \protect\label{zeros_sq_2F_q} 
  Partition-function zeros in the complex $q$ plane for the
  $2_{\rm F} \times 20_{\rm F}$ square-lattice strip and resultant accumulation
  set ${\cal B}$ for the $2_{\rm F} \times \infty_{\rm F}$ strip (singular 
  locus for the free energy) for several values of 
  the temperature-like parameter $v$: $-1$ ($\Box$), $-0.75$ ($\circ$), 
  $-0.50$ ($\triangle$), $-0.25$ ($\Diamond$), and $-0.10$ ($\oplus$), 
  where the symbols correspond to the zeros calculated for the finite strip. 
  The zero at $q=0$ is present for all $v$. 
  } 
\end{figure}

The $L_t=2$ strip with free boundary conditions was previously
studied in \cite{a}.  It will be useful to review some of these results.  In
Fig. \ref{zeros_sq_2F_q} we show the accumulation set ${\cal B}$ formed by the
partition function zeros in the $q$ plane for the Potts antiferromagnet on the
free strip of width $L_t=2$ in the limit of infinite length.  For comparison we
show the partition function zeros calculated for a finite strip with length
$L_\ell=20$.  Evidently, except for obvious discrete zeros such as the zero
always present at $q=0$, these zeros lie close to the asymptotic curves.  For
$v=-1$, the continuous locus ${\cal B}= \emptyset$, with the zeros accumulating
at the discrete complex-conjugate points $q=e^{i \pi/3}$.  As $v$ increases
above $-1$, i.e. the temperature for the Potts antiferromagnet increases above
zero, ${\cal B}$ forms complex-conjugate arcs whose endpoints nearest to the
real axis approach this axis.  As was noted in \cite{a}, when $a=e^K=v+1$ 
increases through the value $(3/4)^2$, i.e. $v=-7/16$, these endpoints touch
the real axis, and in the interval $-7/16 \le v < 0$, ${\cal B}$ consists of
the above-mentioned complex-conjugate arcs together with a real line segment.
As $v \to 0^-$, the locus ${\cal B}$ shrinks and finally degenerates to the
point at $q=0$ for $v=0$ (see Figs. 3,4 in \cite{a}).  The ferromagnetic case
$1 \le v \le \infty$ has been studied in \cite{a} and will not be discussed
here.

\begin{figure}[hbtp]
  \centering
  \epsfxsize=380pt
  \epsffile{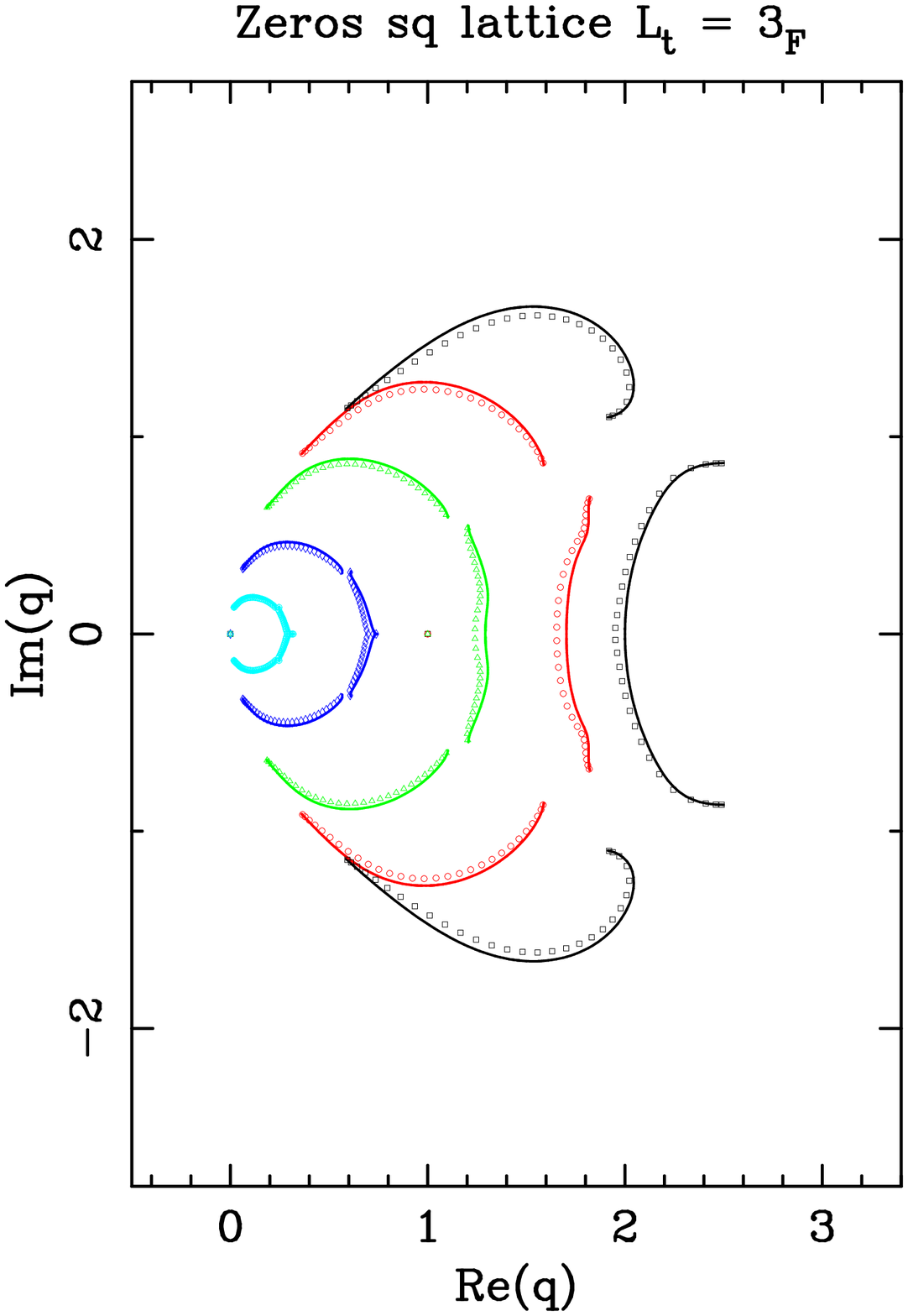}
  \caption[a]{
  \protect\label{zeros_sq_3F_q}
  Partition-function zeros in the complex $q$ plane for the
  $3_{\rm F} \times 30_{\rm F}$ square-lattice strip for several values of
  the temperature-like parameter $v$: $-1$ ($\Box$), $-0.75$ ($\circ$),
  $-0.50$ ($\triangle$), $-0.25$ ($\Diamond$), and $-0.10$ ($\oplus$).  The
  loci ${\cal B}$ for the $3_{\rm F} \times \infty_{\rm F}$ strip for these 
  values of $v$ are also shown.
  } 
\end{figure}

\begin{figure}[hbtp]
  \centering
  \epsfxsize=380pt
  \epsffile{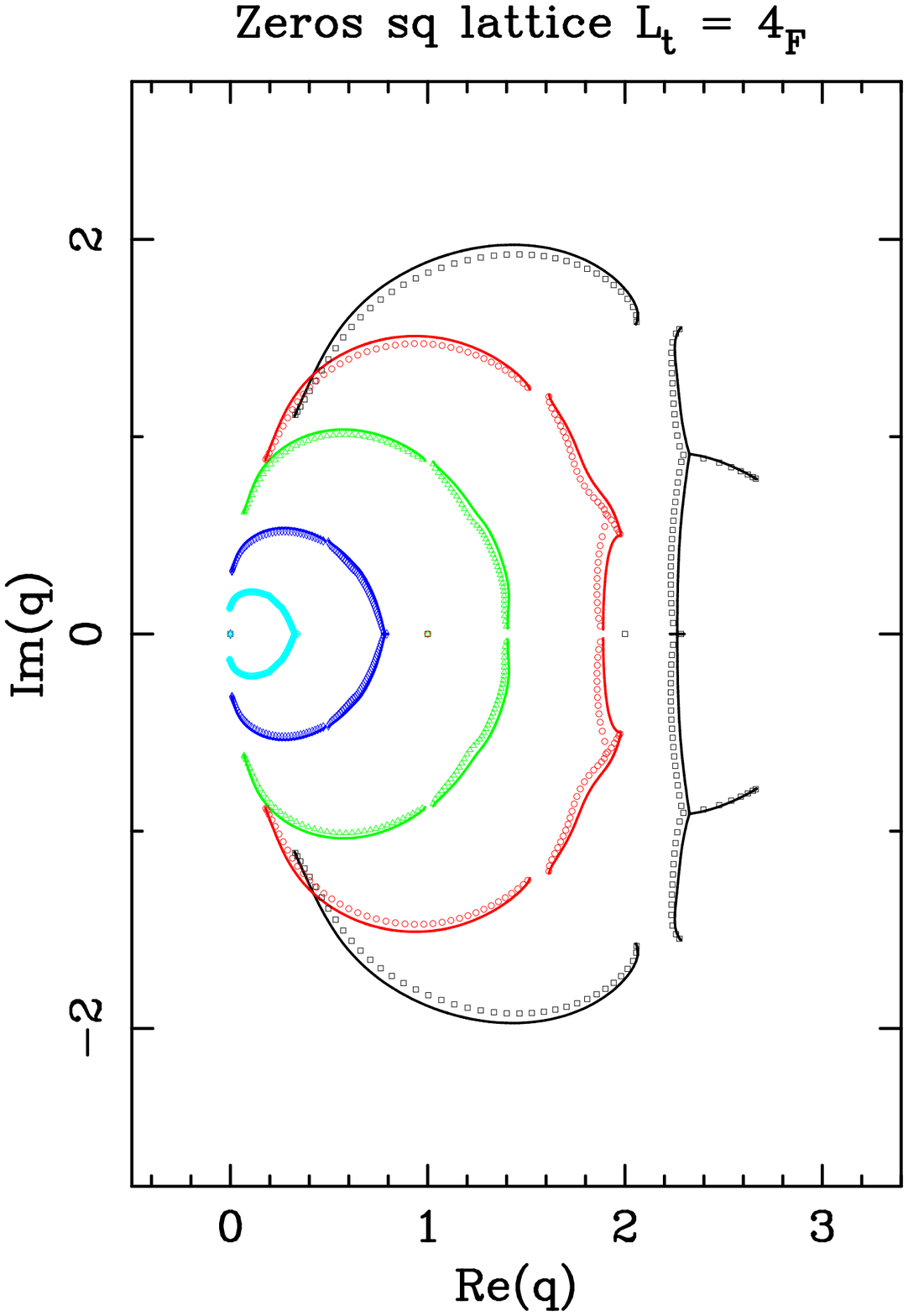}
  \caption[a]{
  \protect\label{zeros_sq_4F_q}
  Partition-function zeros in the complex $q$ plane for the
  for the $4_{\rm F} \times 40_{\rm F}$ strip for several values of
  the temperature-like parameter $v$: $-1$ ($\Box$), $-0.75$ ($\circ$),
  $-0.50$ ($\triangle$), $-0.25$ ($\Diamond$), and $-0.10$ ($\oplus$). The
  loci ${\cal B}$ for the $4_{\rm F} \times \infty_{\rm F}$ strip for these 
  values of $v$ are also shown.
 }
\end{figure}

Partition functions and zeros in the $q$ plane for finite temperature were
previously studied in \cite{ks} for the free $L_t=3$ and $L_t=4$ strips. In
Fig. \ref{zeros_sq_3F_q} we show ${\cal B}$ in the $q$ plane for the
infinite-length limit of the $L_t=3$ free strip and, for comparison, partition
function zeros calculated for a finite strip with length $L_\ell=30$, for the
Potts antiferromagnet.  The zeros in this figure are similar to those in
Fig. 3.3 in \cite{ks}.  For zero temperature, i.e., $v=-1$, ${\cal B}$ is
comprised of the union of a self-conjugate arc passing through $q=2$ and a
complex-conjugate pair of arcs (see Fig. 3(a) in \cite{strip}).  As the
temperature increases above zero, the arcs tend to move together and contract
toward the origin. In Fig. \ref{zeros_sq_4F_q} we show ${\cal B}$ for the
infinite-length limit of the free $L_t=4$ strip and, for comparison, partition
function zeros calculated for a finite strip with length $L_\ell=40$, for the
Potts antiferromagnet.  The zeros in this figure are similar to those in
Fig. 3.9 in \cite{ks}.  For zero temperature, ${\cal B}$ consists of several
arcs together with a small real line segment (see Fig. 3(b) in \cite{strip}).
There are complex-conjugate triple points evident on this locus.  Such triple
points were studied in \cite{z6} in the context of complex-temperature phase
diagrams, and it was shown how they arise when curves on the singular locus
${\cal B}$ meet in such a manner that, as one travels along given curve(s) past
the intersection point the eigenvalues whose degeneracy in magnitude defines
the curve cease to be dominant eigenvalues (see Figs. 1 and 2 of
\cite{z6}). The nature of these triple points was further analyzed in
\cite{ss00}.  If one considers ${\cal B}$ as the union of the various curves
and line segments that comprise it, then a triple point is a multiple point
(intersection point) on ${\cal B}$, since it lies on multiple branches of
${\cal B}$.  In a different nomenclature, if one considers each of the
algebraic curves comprising ${\cal B}$ individually, including the portions
where the pairs of degenerate-magnitude $\lambda$'s are not dominant so that
these portions are not on ${\cal B}$, then such triple points are not multiple
points on each individual algebraic curve, since these individual curves pass
through the triple point as shown in Fig. 2 in \cite{z6}.  As was found for the
strips with $L_t=2,3$, as the temperature increases, the curves forming ${\cal
B}$ contract toward the origin.  One can observe, in particular, that for
sufficiently high temperature the triple points disappear.

\begin{figure}[hbtp]
  \centering
  \epsfxsize=380pt
  \epsffile{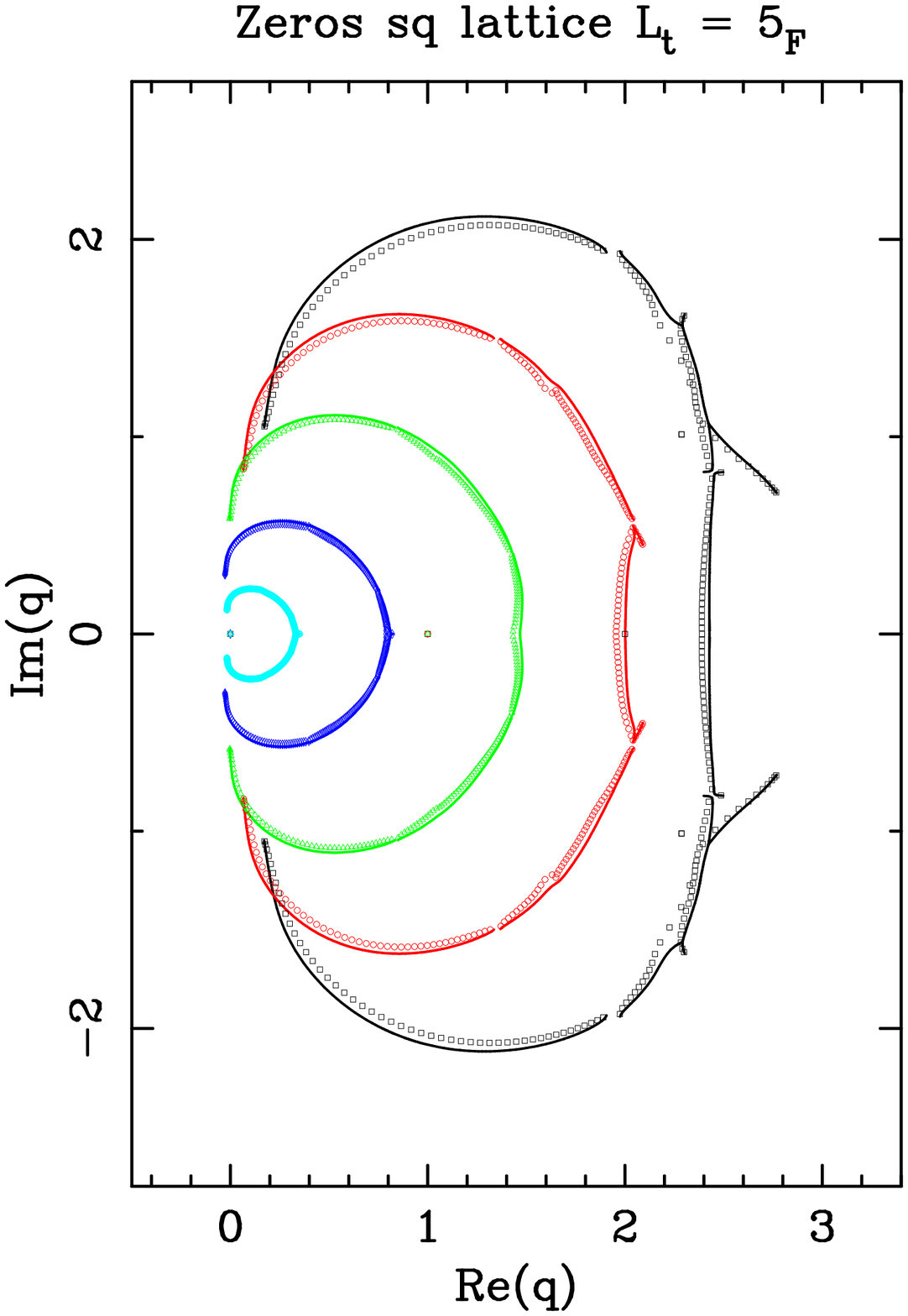}
  \caption[a]{
  \protect\label{zeros_sq_5F_q}
  Partition-function zeros in the complex $q$ plane for the
  $5_{\rm F} \times 40_{\rm F}$ square-lattice strip for several values of
  the temperature-like parameter $v$: $-1$ ($\Box$), $-0.75$ ($\circ$),
  $-0.50$ ($\triangle$), $-0.25$ ($\Diamond$), and $-0.10$ ($\oplus$). The
  loci ${\cal B}$ for the $5_{\rm F} \times \infty_{\rm F}$ strip for these 
  values of $v$ are also shown. 
} 
\end{figure}

We proceed to the free strip of width $L_t=5$.  In Fig. \ref{zeros_sq_5F_q} we
show ${\cal B}$ for the infinite-length limit of the free $L_t=5$ strip and,
for comparison, partition function zeros calculated for a finite strip with
length $L_\ell=40$, for the Potts antiferromagnet. The special case $v=-1$
(chromatic polynomial) was studied in \cite{s4,ss00}.  The results exhibit the
general trend that for $v=-1$ as $L_t$ increases, the arcs on ${\cal B}$ move
together, and the endpoints nearest to the origin approach this point.  Just as
was seen for narrower widths $L_t$, as the temperature increases, the arcs tend
to come together further, the prongs protruding to the right contract and
eventually disappear, and the entire locus contracts toward the origin.  While
${\cal B}$ has an oval-like shape for $v=-1$, it becomes more round as $v \to
0^-$.

\subsection{Periodic Transverse Boundary Conditions}

\begin{figure}[hbtp]
  \centering
  \epsfxsize=380pt
  \epsffile{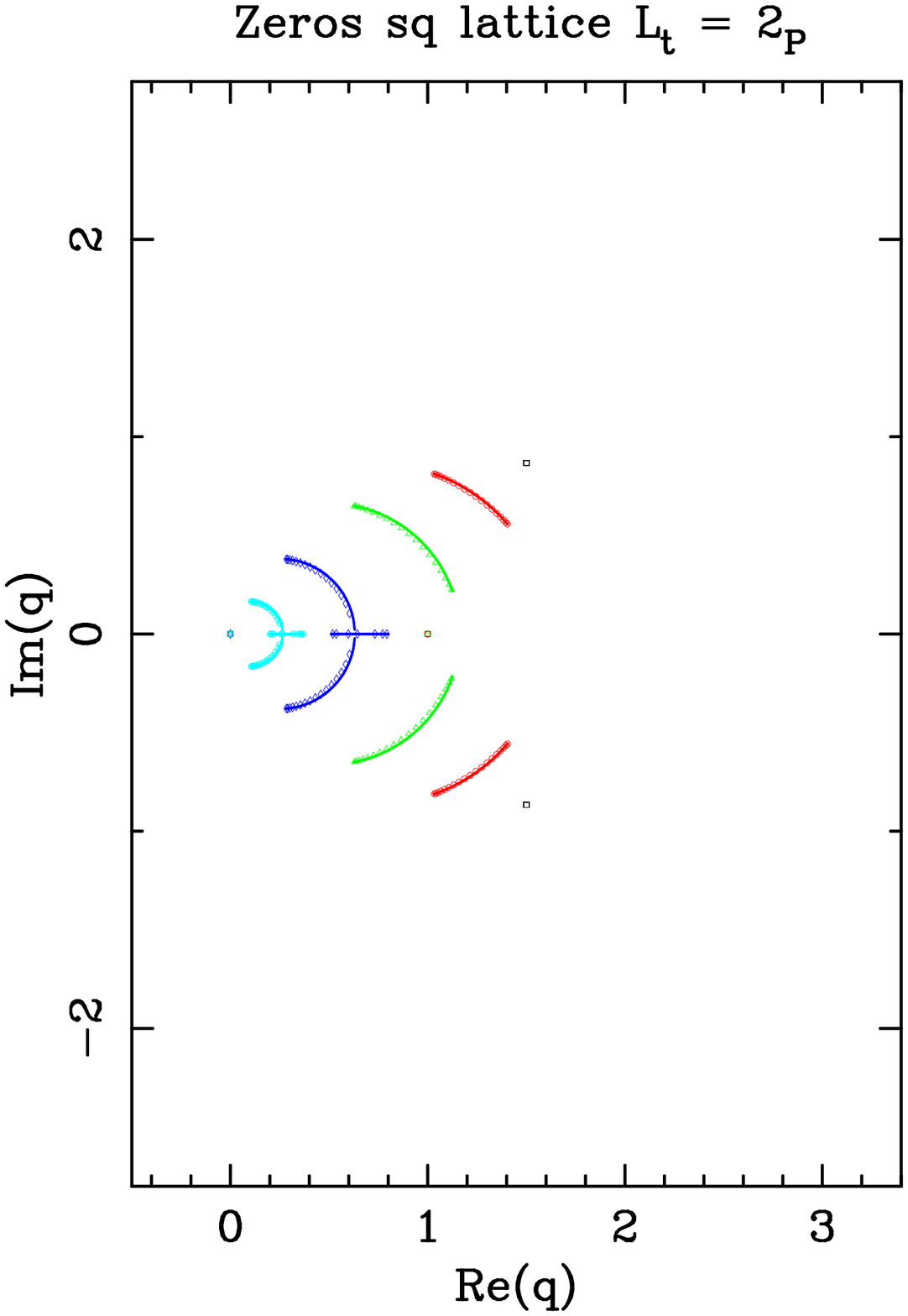}
  \caption[a]{
  \protect\label{zeros_sq_2P_q}
  Partition-function zeros in the complex $q$ plane for the
  $2_{\rm P} \times 20_{\rm F}$ square-lattice strip for several values of
  the temperature-like parameter $v$: $-1$ ($\Box$), $-0.75$ ($\circ$),
  $-0.50$ ($\triangle$), $-0.25$ ($\Diamond$), and $-0.10$ ($\oplus$). The
  loci ${\cal B}$ for the $2_{\rm P} \times \infty_{\rm F}$ strip for these 
  values of $v$ are also shown.
} 
\end{figure}

\begin{figure}[hbtp]
  \centering
  \epsfxsize=380pt
  \epsffile{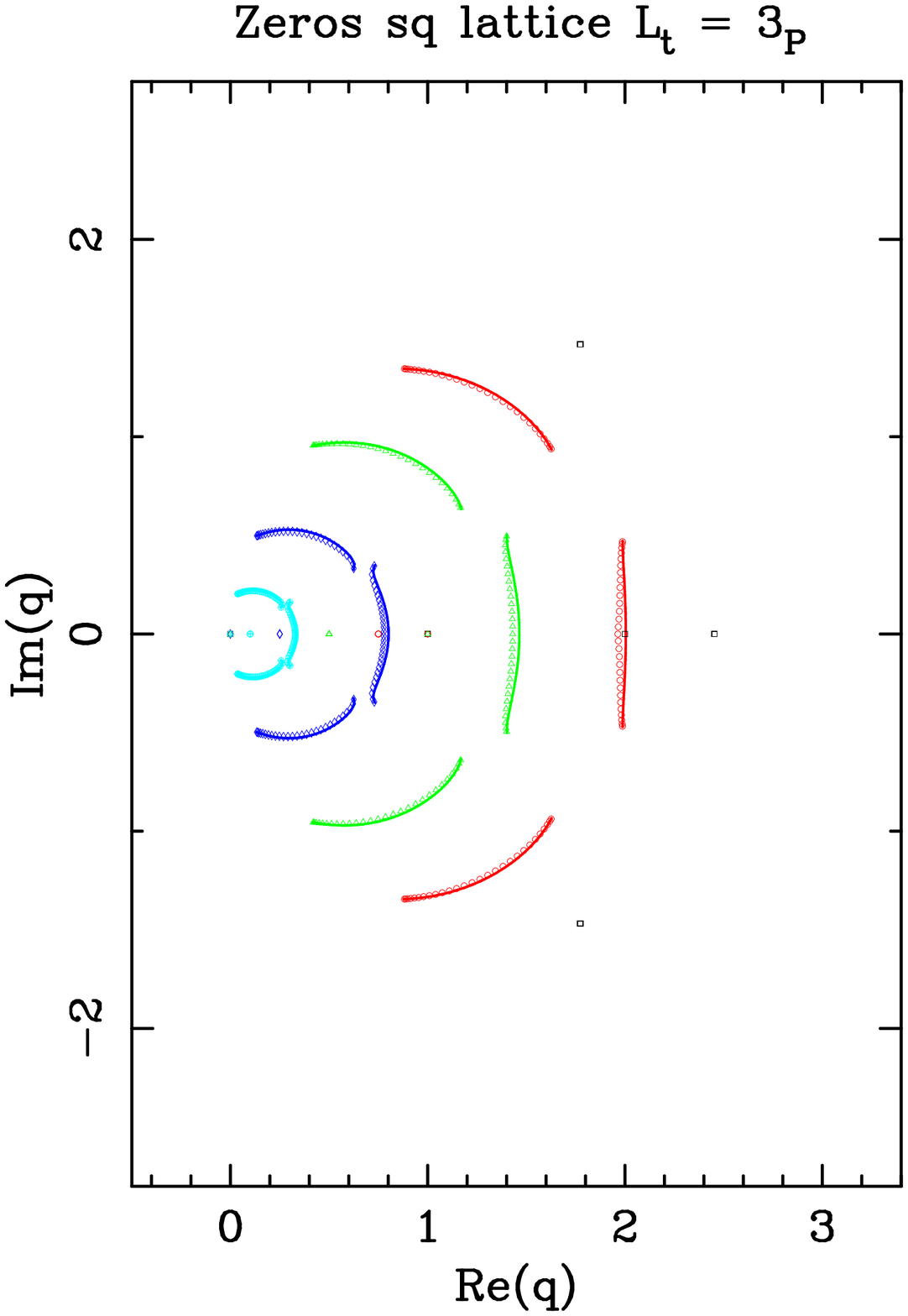}
  \caption[a]{
  \protect\label{zeros_sq_3P_q}
  Partition-function zeros in the complex $q$ plane for the
  $3_{\rm P} \times 30_{\rm F}$ square-lattice strip for several values of
  the temperature-like parameter $v$: $-1$ ($\Box$), $-0.75$ ($\circ$),
  $-0.50$ ($\triangle$), $-0.25$ ($\Diamond$), and $-0.10$ ($\oplus$). The
  loci ${\cal B}$ for the $3_{\rm P} \times \infty_{\rm F}$ strip for these 
  values of $v$ are also shown. } 
\end{figure}

\begin{figure}[hbtp]
  \centering
  \epsfxsize=380pt
  \epsffile{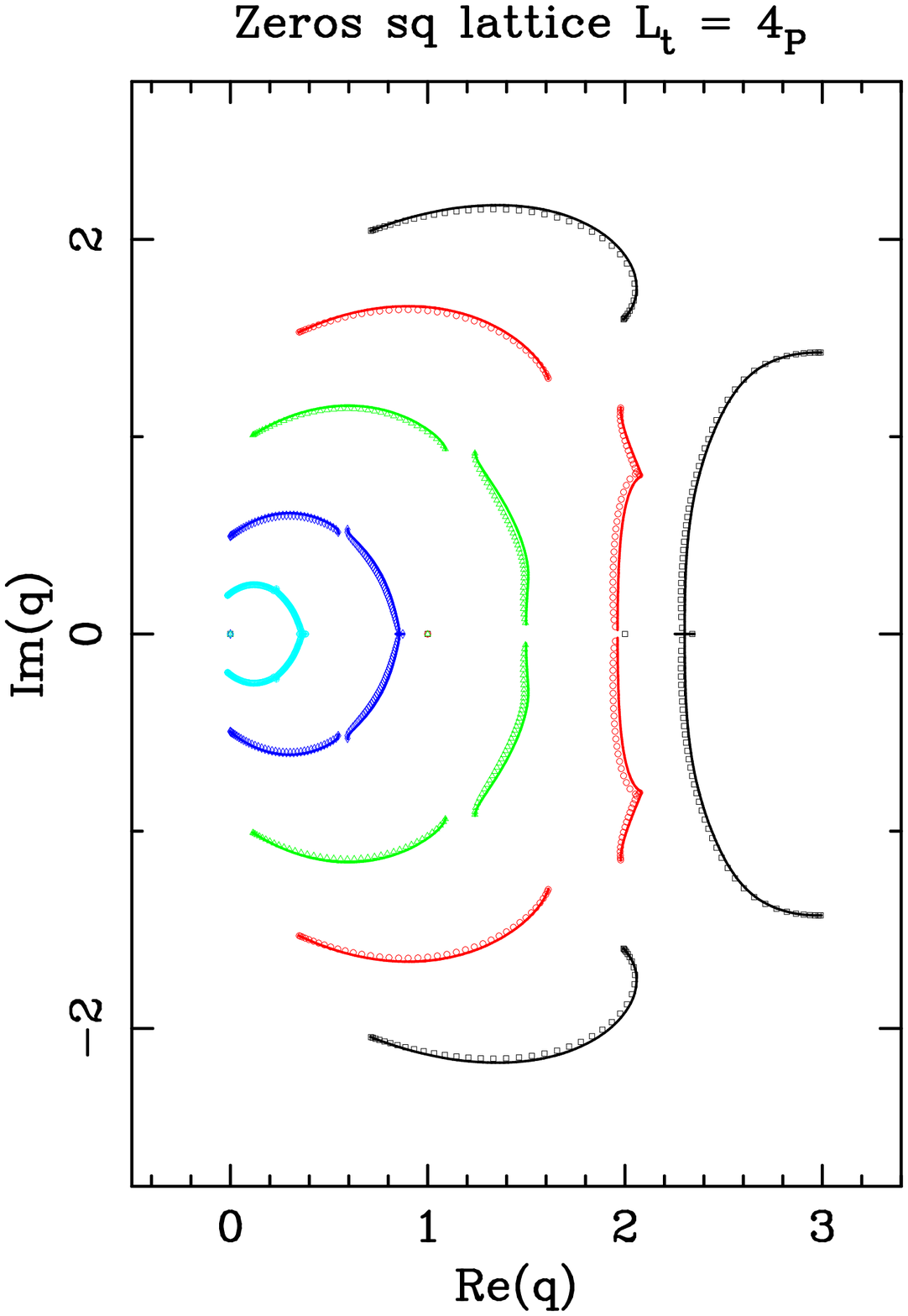}
  \caption[a]{
  \protect\label{zeros_sq_4P_q}
  Partition-function zeros in the complex $q$ plane for the
  $4_{\rm P} \times 40_{\rm F}$ square-lattice strip for several values of
  the temperature-like parameter $v$: $-1$ ($\Box$), $-0.75$ ($\circ$),
  $-0.50$ ($\triangle$), $-0.25$ ($\Diamond$), and $-0.10$ ($\oplus$). The
  loci ${\cal B}$ for the $4_{\rm P} \times \infty_{\rm F}$ strip for these 
  values of $v$ are also shown.
  } 
\end{figure}

\begin{figure}[hbtp]
  \centering
  \epsfxsize=380pt
  \epsffile{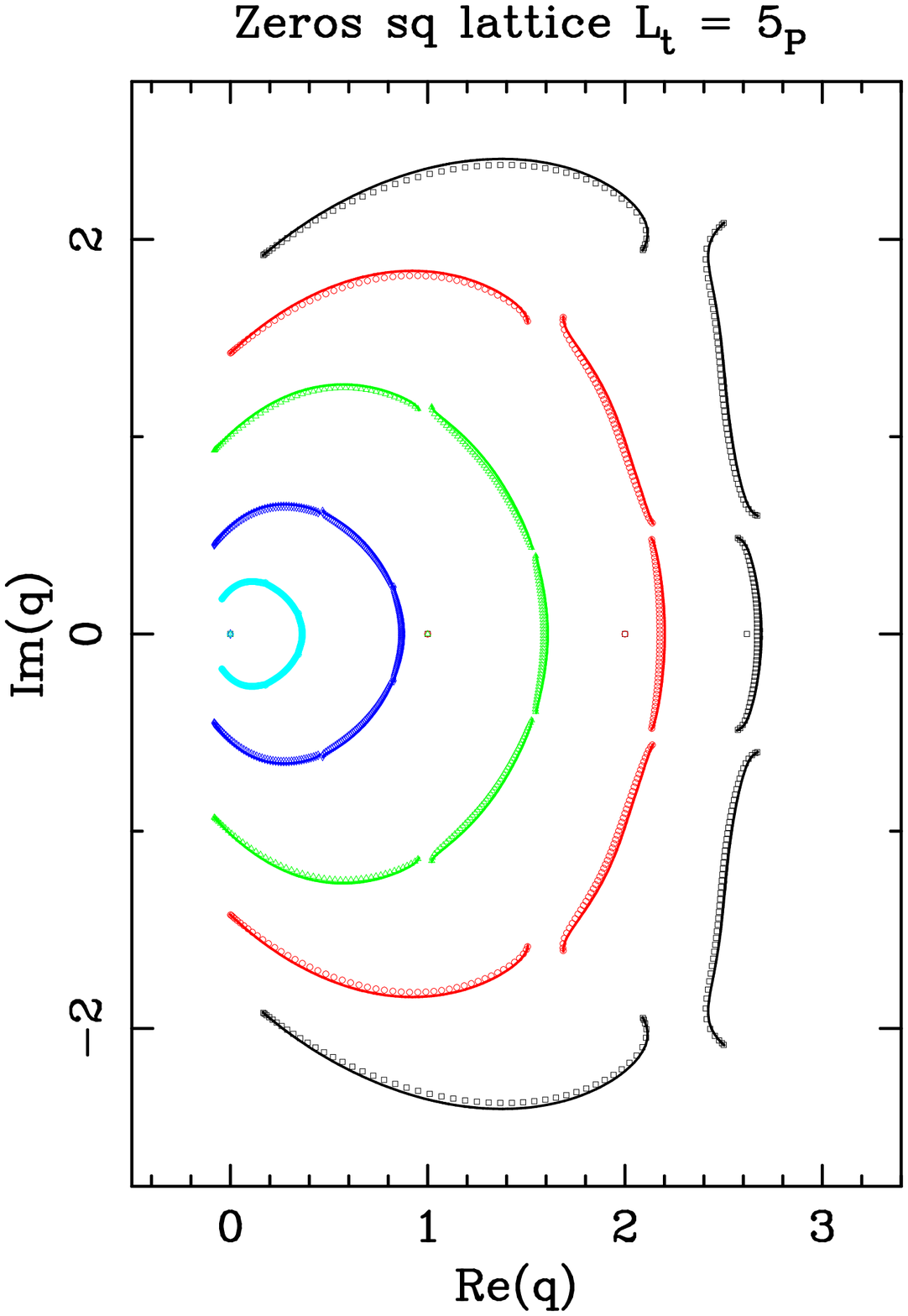}
  \caption[a]{
  \protect\label{zeros_sq_5P_q}
  Partition-function zeros in the complex $q$ plane for the
  $5_{\rm P} \times 40_{\rm F}$ square-lattice strip for several values of
  the temperature-like parameter $v$: $-1$ ($\Box$), $-0.75$ ($\circ$),
  $-0.50$ ($\triangle$), $-0.25$ ($\Diamond$), and $-0.10$ ($\oplus$). The
  loci ${\cal B}$ for the $5_{\rm P} \times \infty_{\rm F}$ strip for these 
  values of $v$ are also shown.
  } 
\end{figure}

We next consider the strips with cylindrical boundary conditions.  In
Figs. \ref{zeros_sq_2P_q}-\ref{zeros_sq_5P_q} we show ${\cal B}$ in the $q$
plane for the infinite-length limits of the cylindrical strips with widths
$L_t=2$ through $L_t=5$ and partition function zeros on long finite strips of
the respective widths, for the Potts antiferromagnet.  For the $v=-1$ special
case with $L_t=2,3$, ${\cal B}$ degenerates to discrete points.  The $v=-1$
special case for $L_t=4$ was previously studied in \cite{strip2} and for
$L_t=5$ in \cite{s4,ss00} (as well as higher widths in these papers, going up
to $L_t=7$ in \cite{ss00} and $L_t=13$ in \cite{sqcyl}).  The cases $L_t=3$ and
$L_t=4$ for arbitrary temperature were studied in \cite{ks} and our zeros for
$L_t=4$ are similar to those in Fig. 3.12 of \cite{ks}.  Our present results
show that the locus ${\cal B}$ tends to become somewhat more complicated as
$L_t$ increases.

\section{Partition Function Zeros in the $\lowercase{v}$ Plane}

\begin{figure}[hbtp]
  \centering
  \begin{tabular}{cc}
     \epsfxsize=200pt
     \epsffile{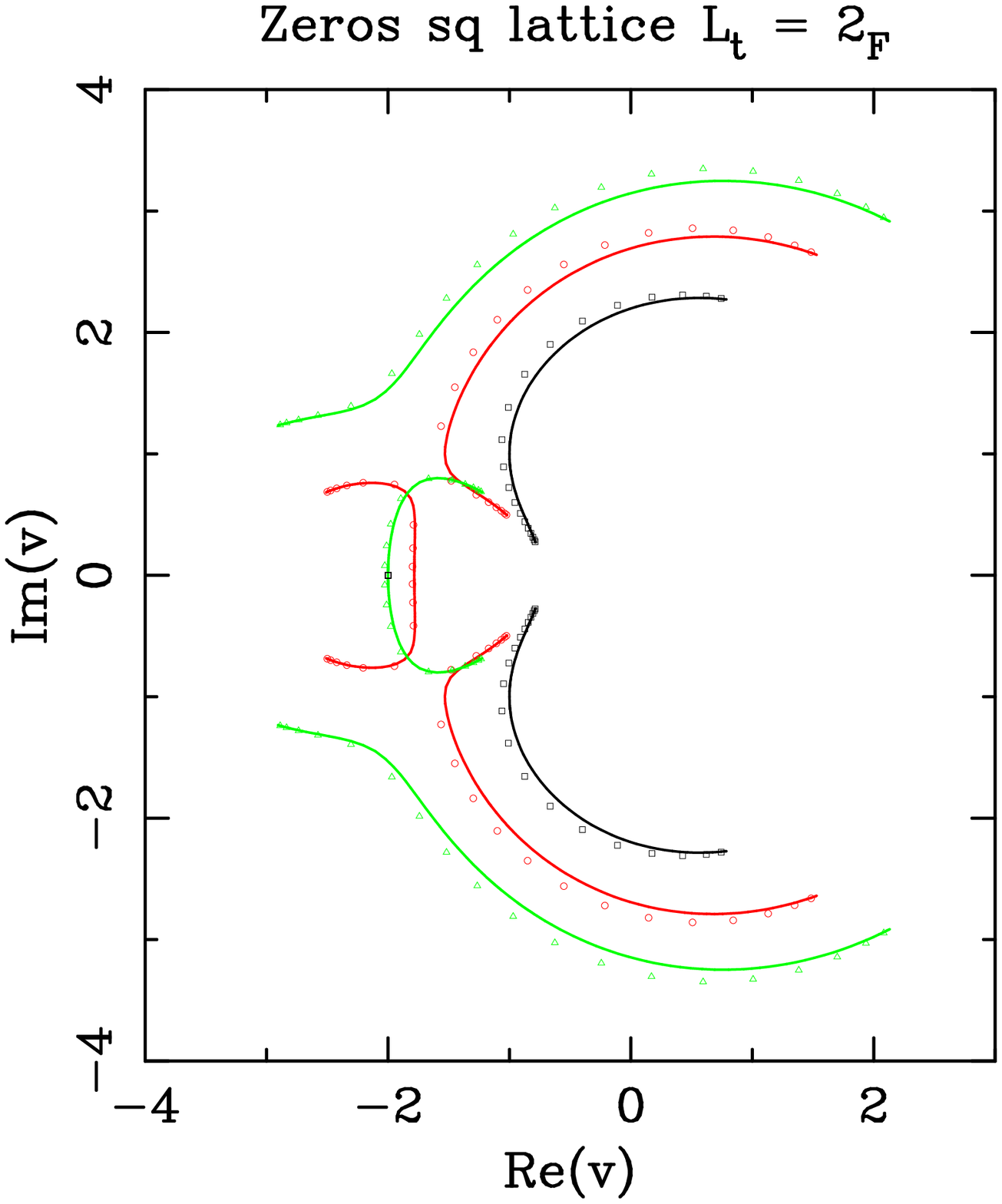}  & 
     \epsfxsize=200pt
     \epsffile{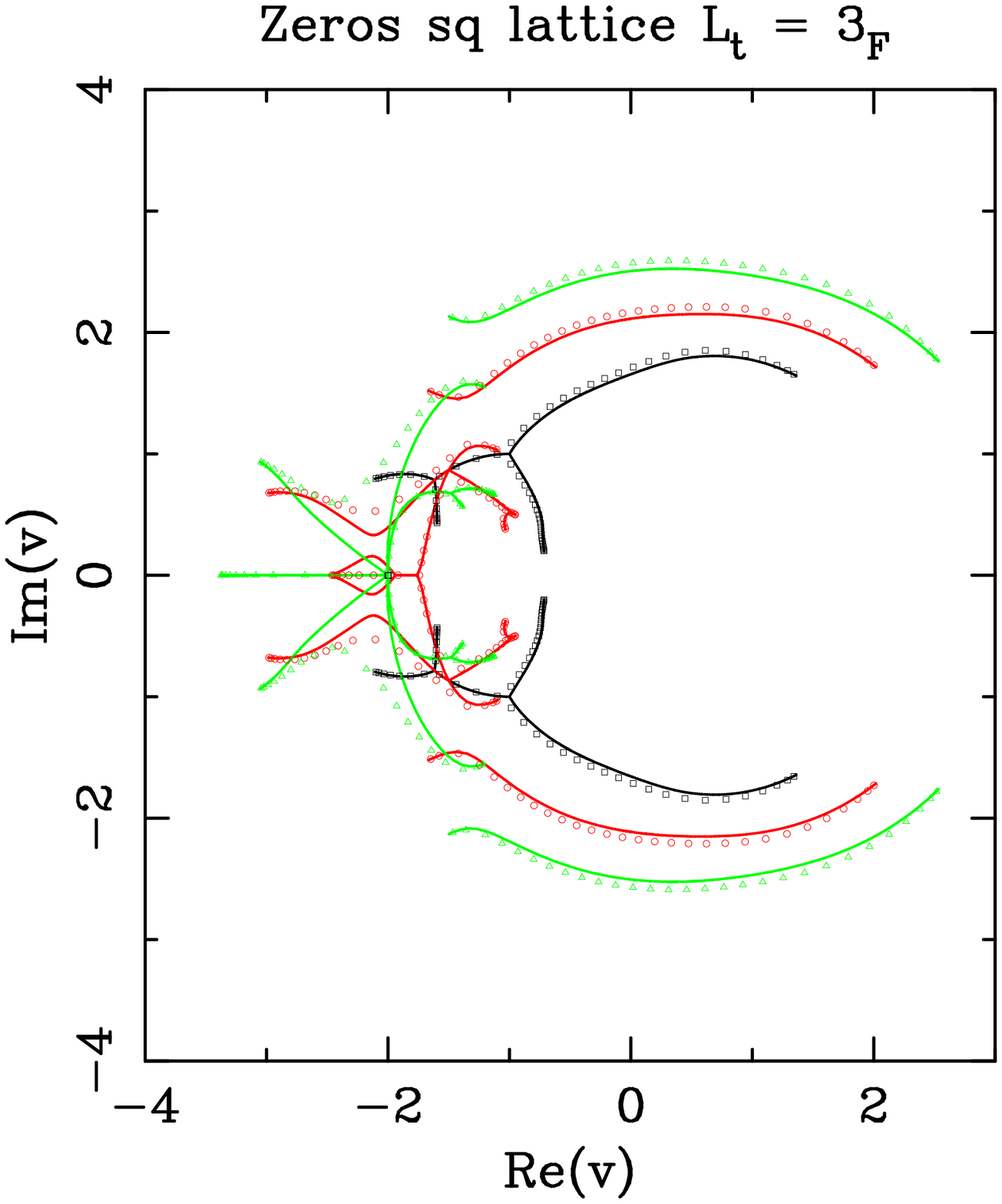}  \\[1mm]
     \phantom{(((a)}(a)    & \phantom{(((a)}(b) \\[5mm]
     \epsfxsize=200pt
     \epsffile{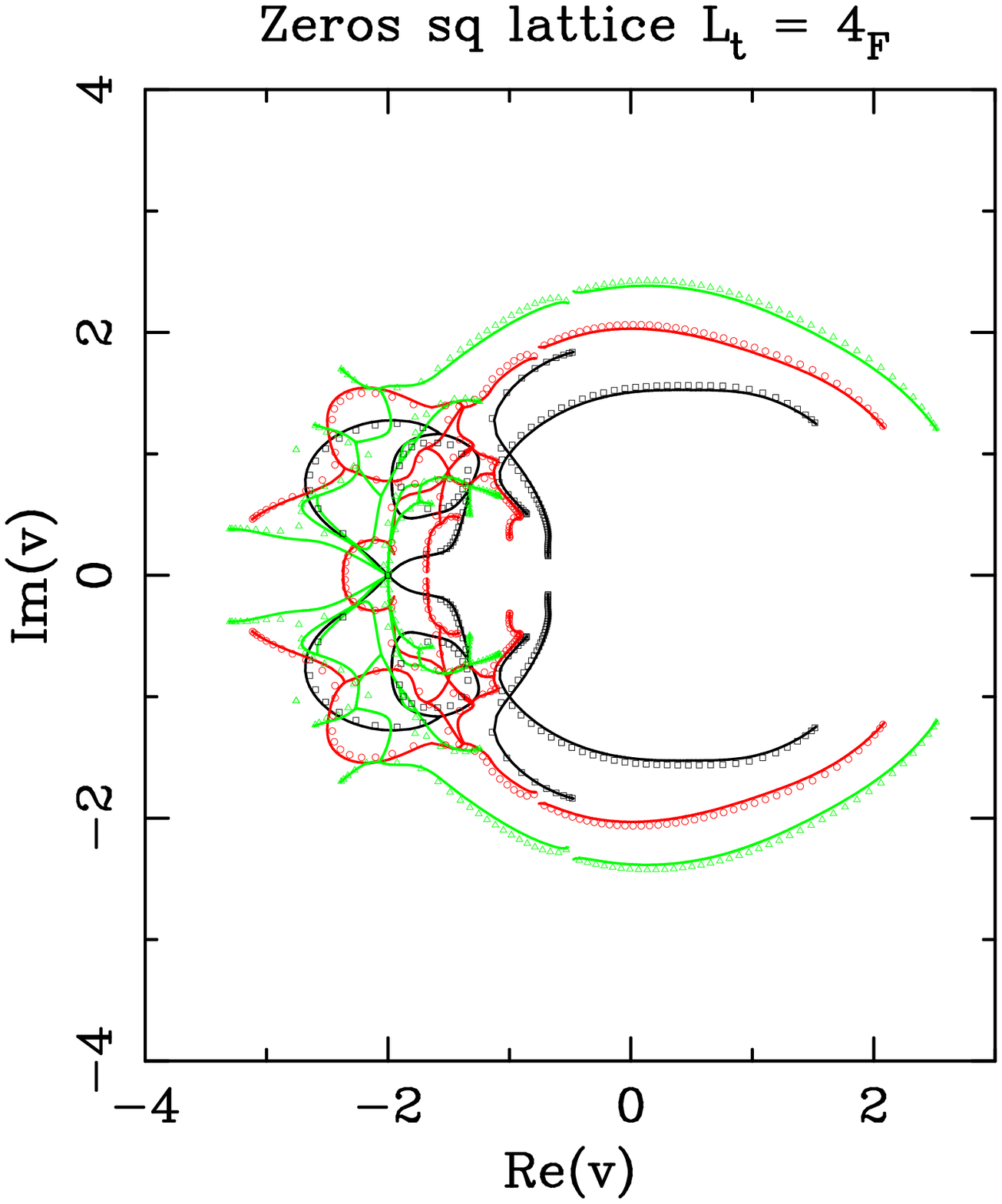}  & 
     \epsfxsize=200pt
     \epsffile{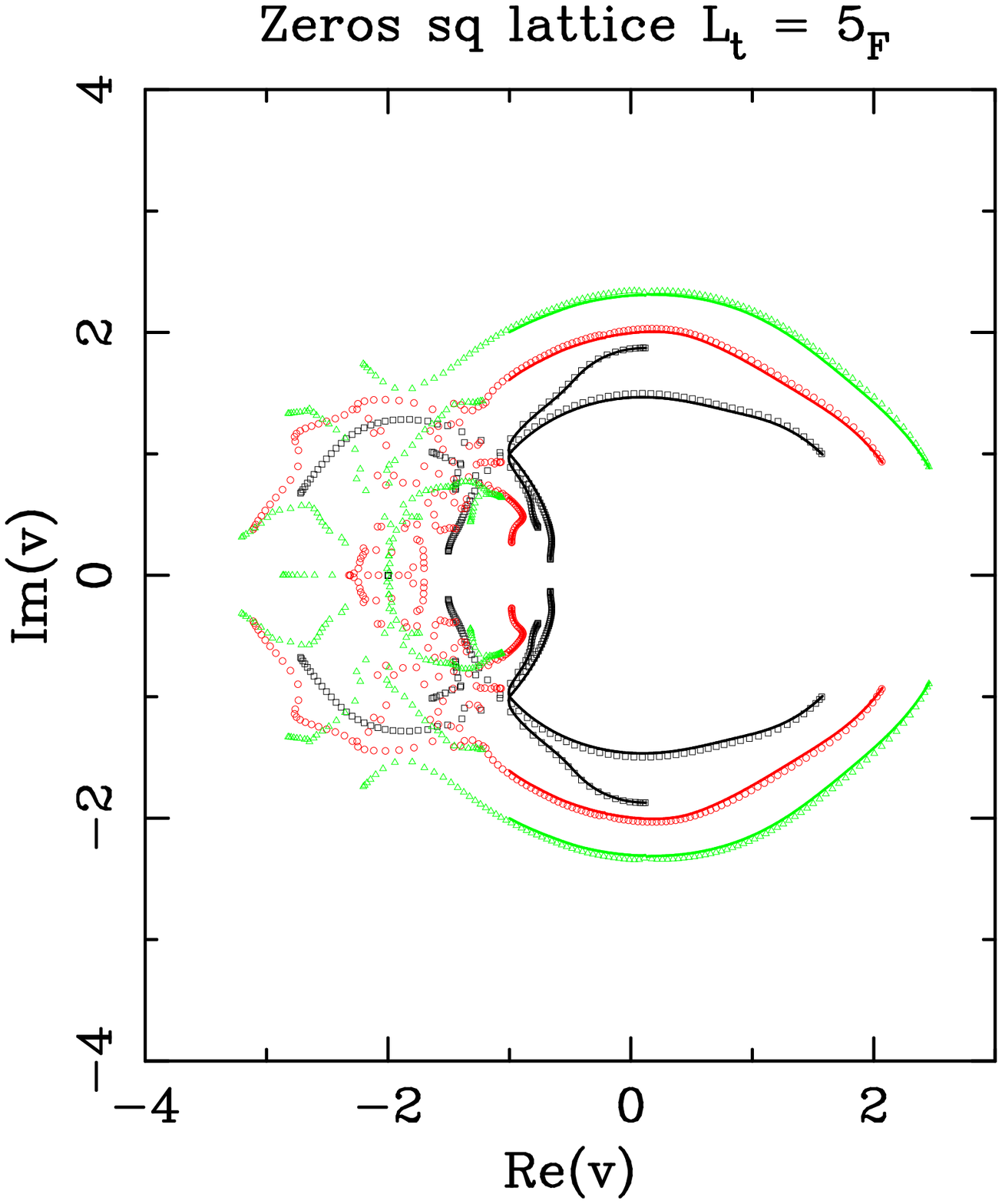} \\
     \phantom{(((a)}(c)    & \phantom{(((a)}(d) \\
  \end{tabular}
  \caption[a]{
  \protect\label{zeros_sqF_v}
  Partition function zeros in the complex $v$ plane for several 
  square-lattice strips: (a) $2_{\rm F} \times 20_{\rm F}$, 
  (b) $3_{\rm F} \times 30_{\rm F}$, (c) $4_{\rm F} \times 40_{\rm F}$, and
  (d) $5_{\rm F} \times 50_{\rm F}$. In each plot we show the zeros for 
  several values of the parameter $q$: $2$ ($\Box$, black), $3$ ($\circ$, red),
  and $4$ ($\triangle$, green) and the corresponding loci ${\cal B}$ for these
  values of $q$.  
  } 
\end{figure}

\begin{figure}[hbtp]
  \centering
  \begin{tabular}{cc}
     \epsfxsize=200pt
     \epsffile{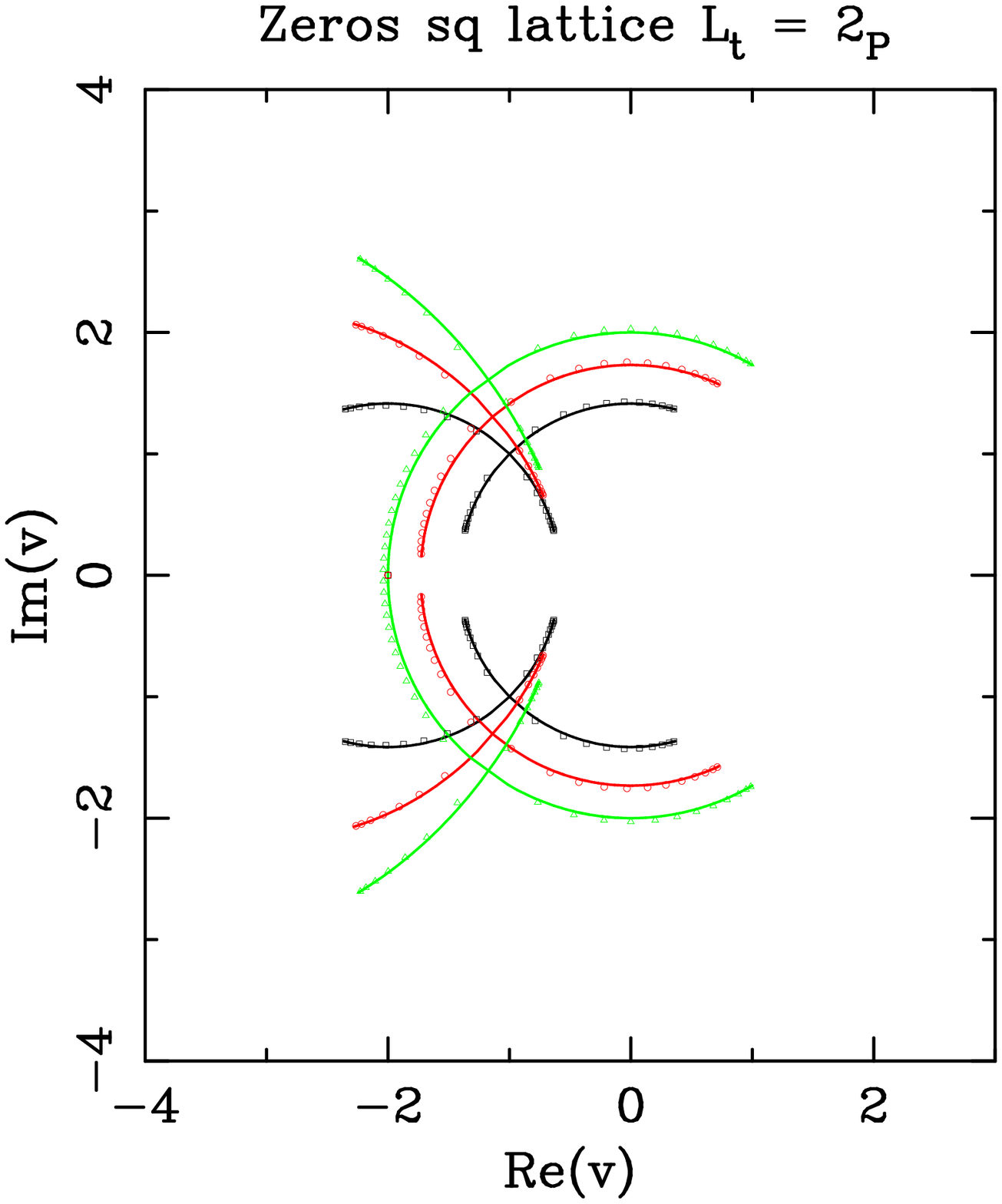}  &
     \epsfxsize=200pt
     \epsffile{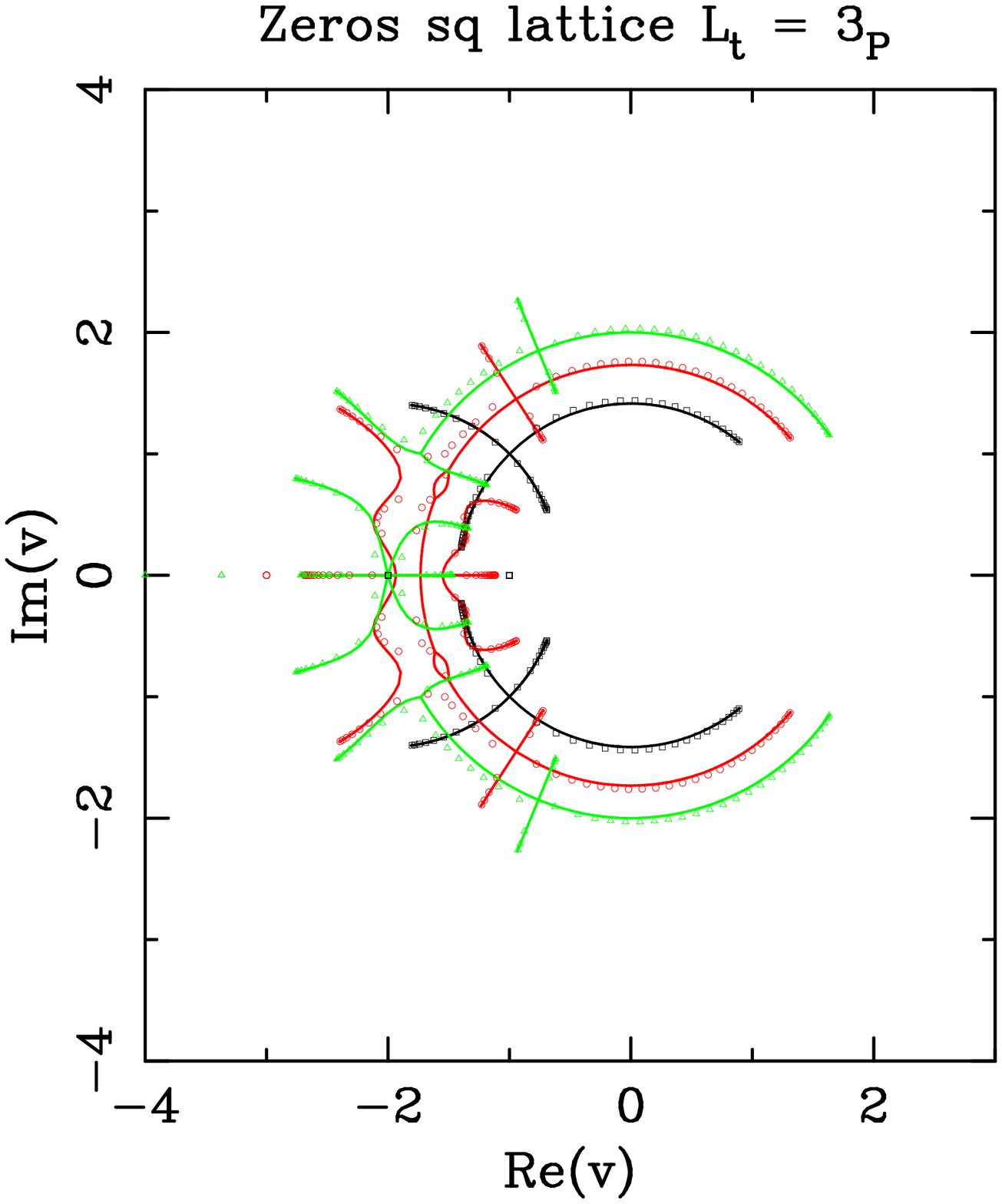}  \\[1mm]
     \phantom{(((a)}(a)    & \phantom{(((a)}(b) \\[5mm]
     \epsfxsize=200pt
     \epsffile{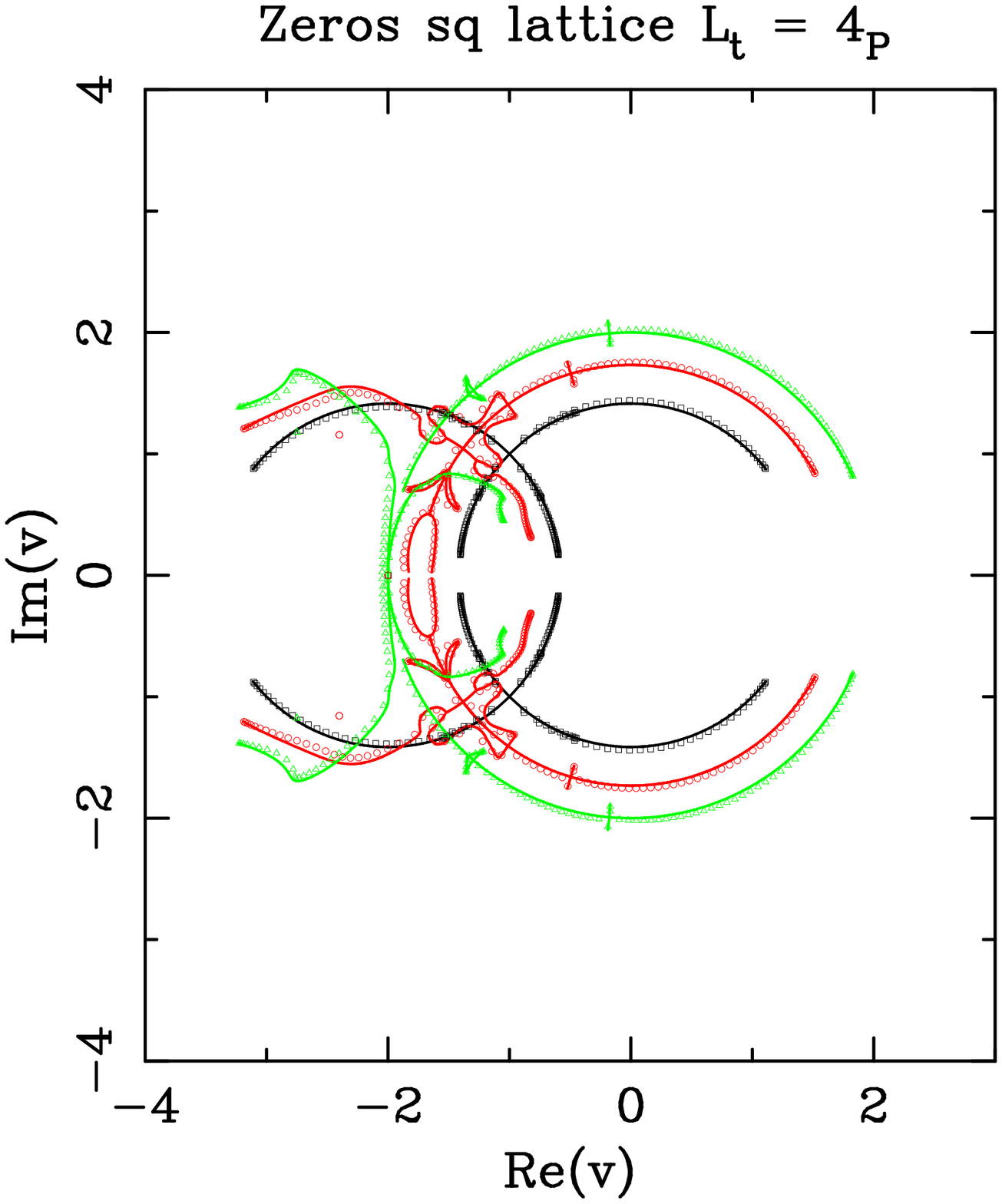}  &
     \epsfxsize=200pt
     \epsffile{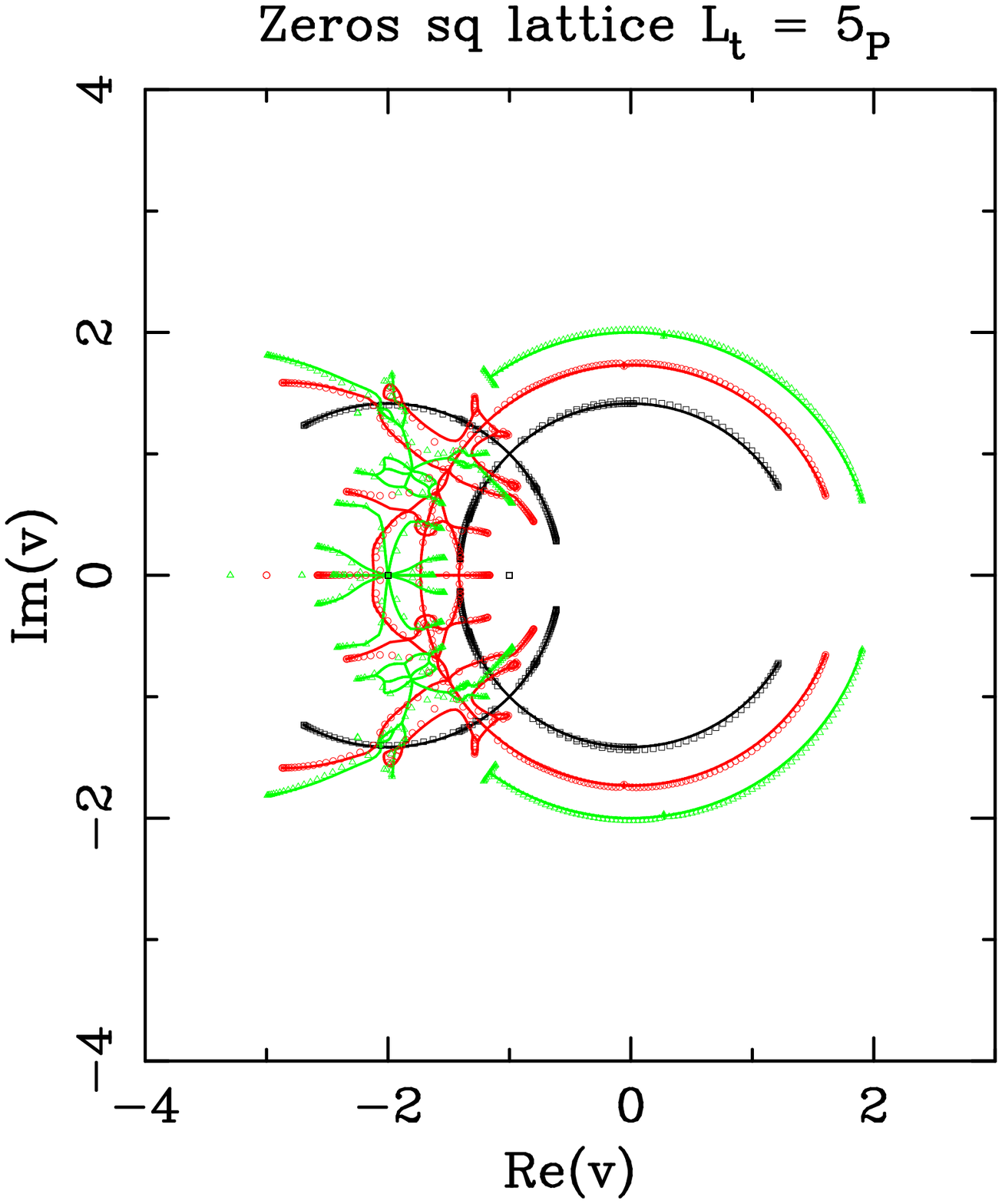} \\
     \phantom{(((a)}(c)    & \phantom{(((a)}(d) \\
  \end{tabular}
  \caption[a]{
  \protect\label{zeros_sqP_v}
  Partition function zeros in the complex $v$ plane for several
  square-lattice strips: (a) $2_{\rm P} \times 20_{\rm F}$,
  (b) $3_{\rm P} \times 30_{\rm F}$, (c) $4_{\rm P} \times 40_{\rm F}$, and
  (d) $5_{\rm P} \times 50_{\rm F}$. In each plot we show the zeros for
  several values of the parameter $q$: $2$ ($\Box$, black), $3$ ($\circ$, red),
  and $4$ ($\triangle$, green) and the corresponding loci ${\cal B}$ for these
  values of $q$.
  }
\end{figure}

\begin{figure}[hbtp]
  \centering
  \begin{tabular}{cc}
     \epsfxsize=200pt
     \epsffile{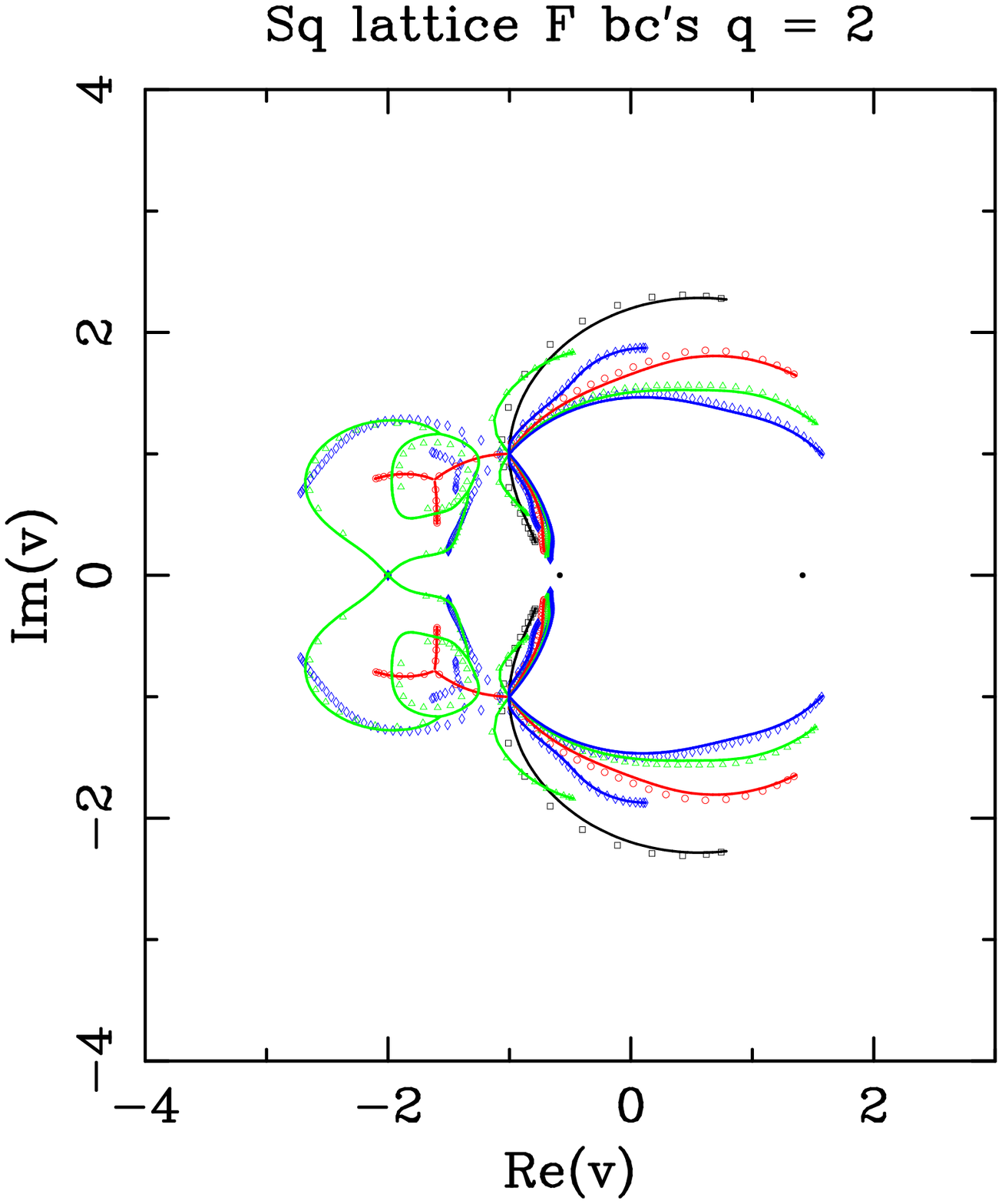}  &
     \epsfxsize=200pt
     \epsffile{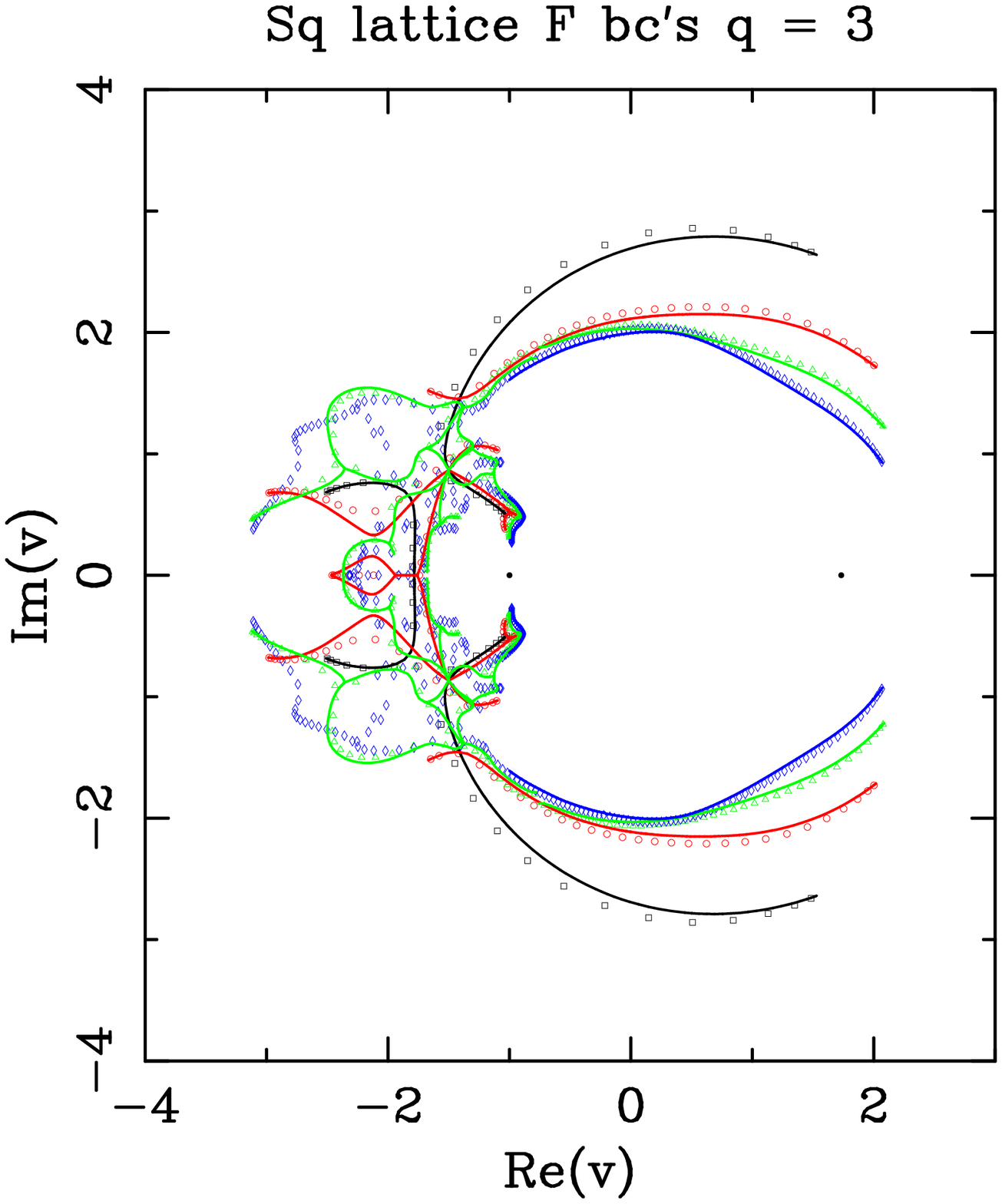}  \\[1mm]
     \phantom{(((a)}(a)    & \phantom{(((a)}(b) \\[5mm]
     \multicolumn{2}{c}{ 
        \epsfxsize=200pt
        \epsffile{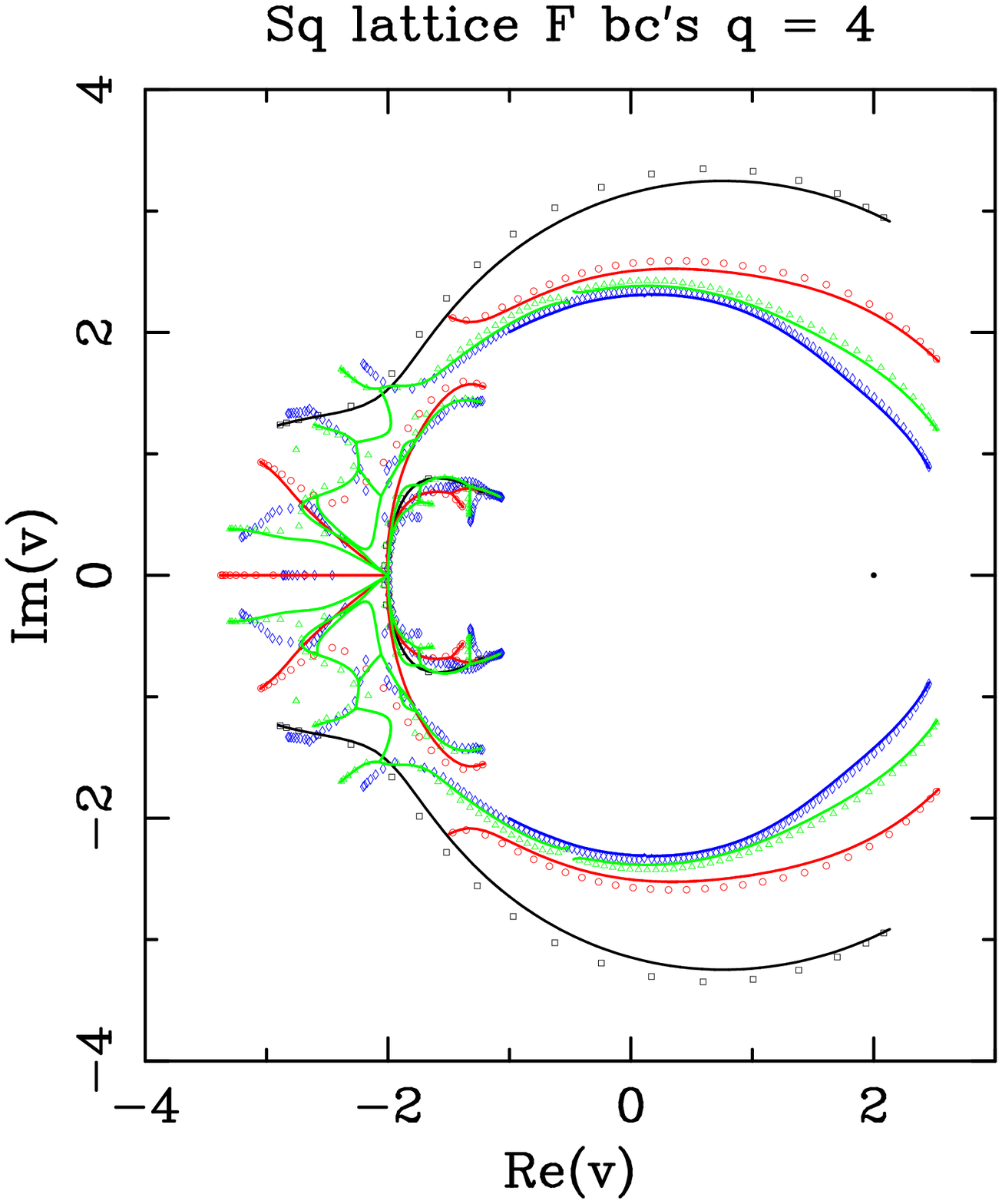} 
     } \\[1mm]
     \multicolumn{2}{c}{
        \phantom{(((a)}(c)  
     } \\
  \end{tabular}
  \caption[a]{
  \protect\label{zeros_sq_allF_v}
  Limiting curves forming the singular locus ${\cal B}$, in the $v$ plane, for
  the free energy of the $q=2$ (a) $q=3$ (b), and $q=4$ (c) Potts model on
  the $(L_t)_{\rm F} \times \infty_{\rm F}$
  square-lattice strips with the following widths $L_t$: 2 (black),
  3 (red), 4 (green), and 5 (blue). We also show the partition-function zeros
  for the strips $(L_t)_{\rm F} \times (10 L_t)_{\rm F}$ for the same values 
  of $L_t$: 2 ($\Box$, black), 3 ($\circ$, red), 4 ($\triangle$, green), and
  5 ($\diamond$, blue). 
  }
\end{figure}

\begin{figure}[hbtp]
  \centering
  \begin{tabular}{cc}
     \epsfxsize=200pt
     \epsffile{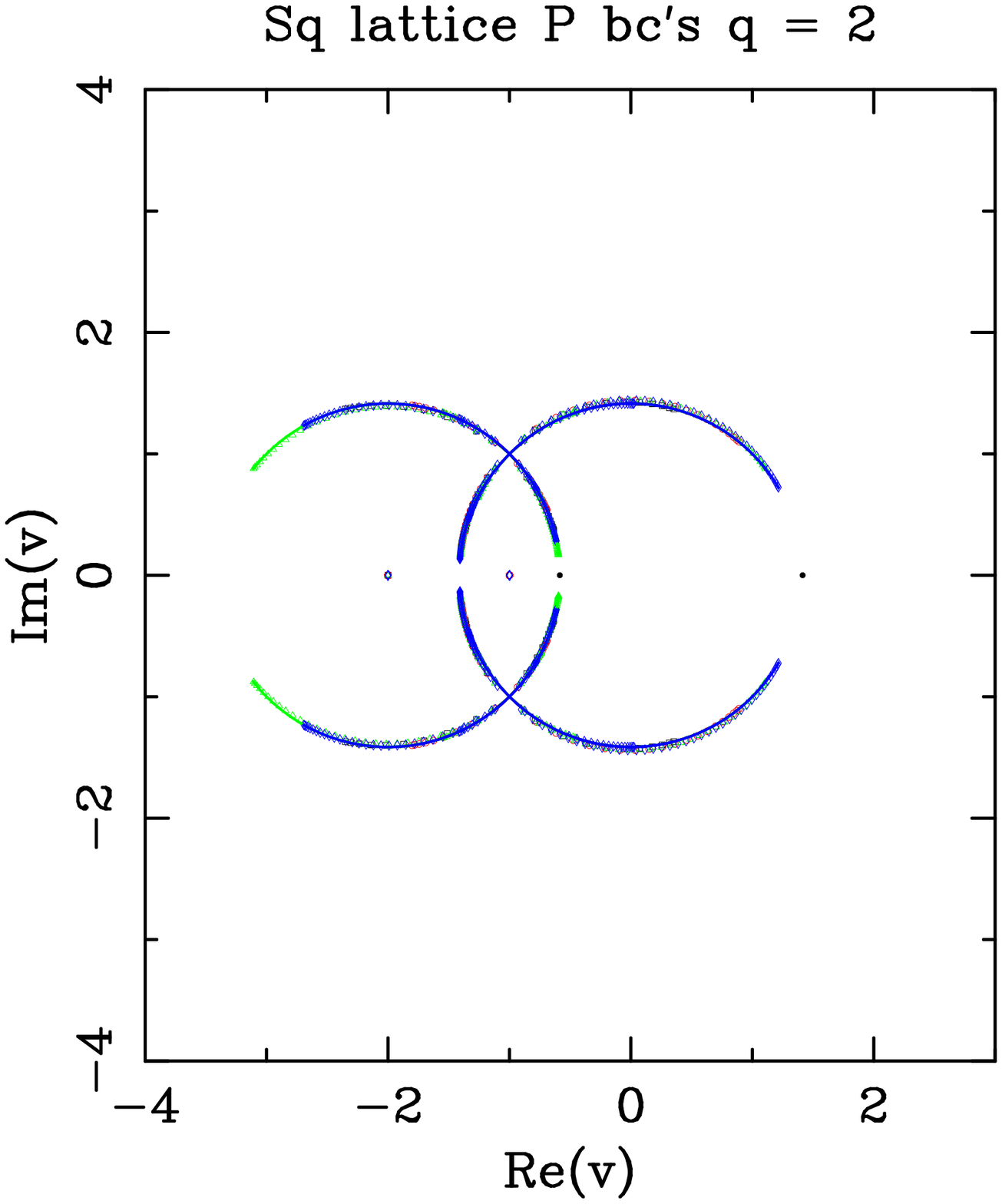}  &
     \epsfxsize=200pt
     \epsffile{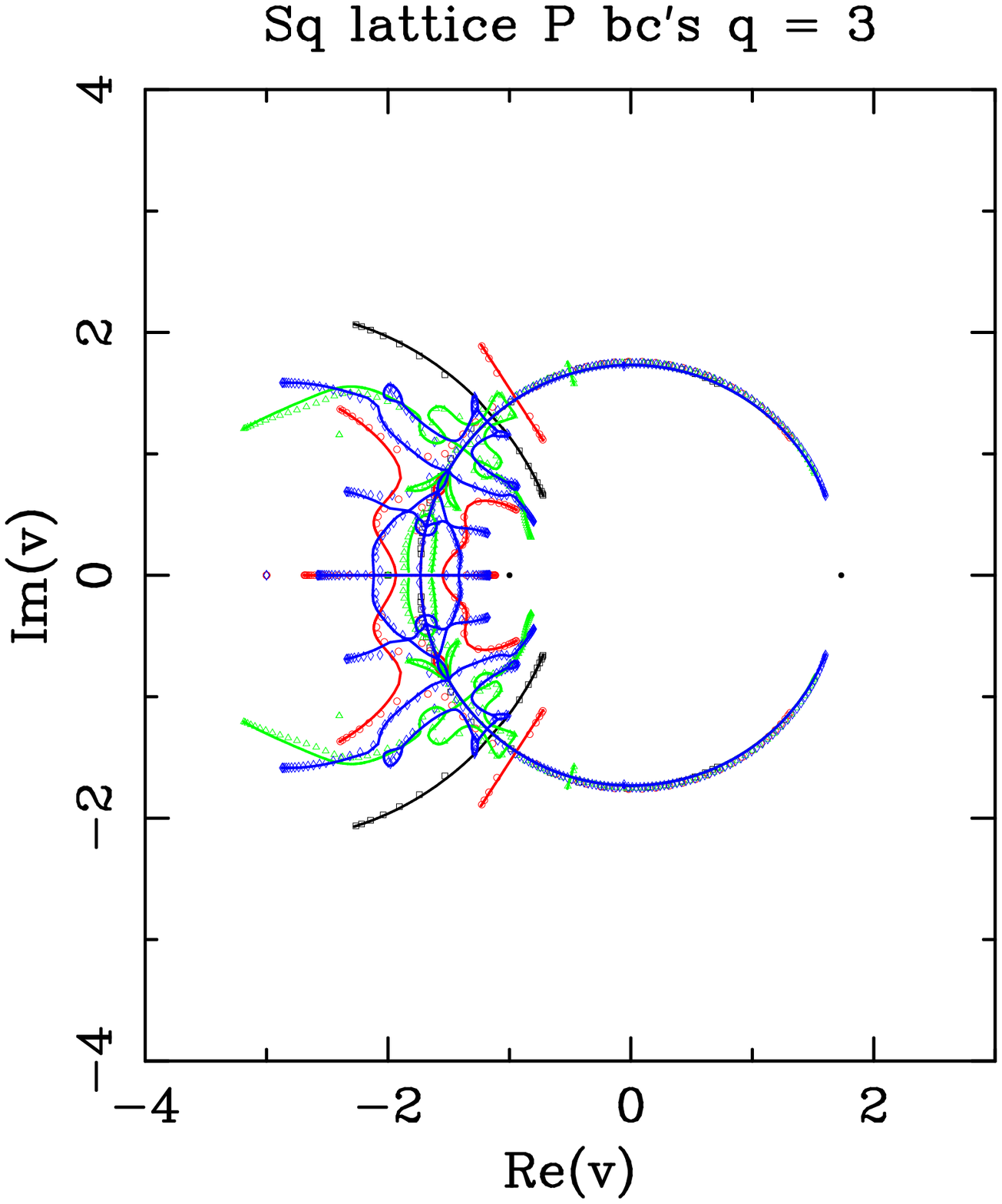}  \\[1mm]
     \phantom{(((a)}(a)    & \phantom{(((a)}(b) \\[5mm]
     \multicolumn{2}{c}{
        \epsfxsize=200pt
        \epsffile{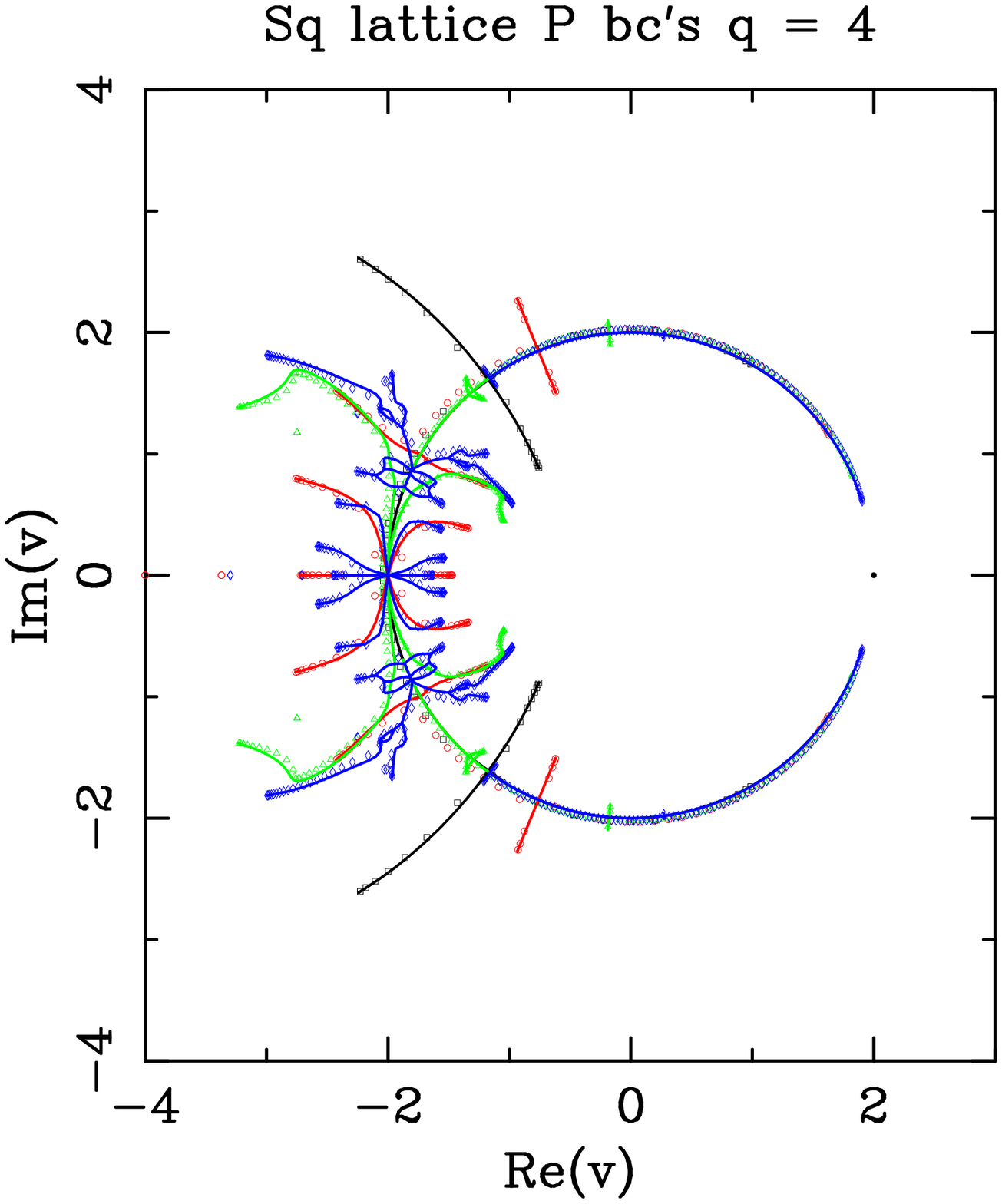}
     } \\[1mm]
     \multicolumn{2}{c}{
        \phantom{(((a)}(c)
     } \\
  \end{tabular}
  \caption[a]{
  \protect\label{zeros_sq_allP_v}
  Limiting curves forming the singular locus ${\cal B}$, in the $v$ plane, for
  the free energy of the $q=2$ (a) $q=3$ (b), and $q=4$ (c) Potts model on
  the $(L_t)_{\rm P} \times \infty_{\rm F}$
  square-lattice strips with the following widths $L_t$: 2 (black),
  3 (red), 4 (green), and 5 (blue). We also show the partition-function zeros
  for the strips $(L_t)_{\rm P} \times (10 L_t)_{\rm F}$ for the same values
  of $L_t$. The color code is as in Figure~\protect\ref{zeros_sq_allF_v}.
  }
\end{figure}

We next present results for complex-temperature (Fisher) zeros and the
continuous accumulation set (singular locus for the free energy) ${\cal B}$ in
the $v$ plane for the partition function of the Potts model on square-lattice
strips with free longitudinal boundary conditions and free or periodic
transverse boundary conditions.  Since the infinite-length limits of the strips
considered here are effectively one-dimensional systems and since a
one-dimensional spin system with short-range interactions does not have any
finite-temperature phase transition, it follows that the locus ${\cal B}$ where
the free energy is singular does not intersect the real $v$--axis in the
interval $-1 < v < \infty$ corresponding to nonzero temperature.  Hence there
cannot be any physical finite-temperature phase with ferromagnetic or
antiferromagnet ordering, and the complex-temperature phase diagram involves
only the (complex-temperature extension of the) paramagnetic phase, together
with possible O phases, in the nomenclature of \cite{chisq}.  In
Fig.~\ref{zeros_sqF_v} we show the Fisher zeros for long finite strips with
free boundary conditions, and the approximate respective loci ${\cal B}$ for
the infinite-length limits of these strips with $L_t$ ranging from 2 through 5.
(In Fig. \ref{zeros_sqF_v}(d) the locus ${\cal B}$ is shown only for the
physical range ${\rm Re}(v) \ge -1$.)  Corresponding results are presented in
Fig.~\ref{zeros_sqP_v} for strips with cylindrical boundary conditions.  For
each specific width and type of boundary condition we show curves and zeros for
$q=2$, 3, and 4.  Although some of these plots are rather complicated, the
reader can distinguish the curves for different values of $q$ by their close
association with the zeros, which are labelled with different symbols, namely
squares, circles, and triangles for $q=2$, $q=3$, and $q=4$, respectively.
Furthermore, in the postscript files for the figures in the cond-mat archive,
the curves for different $q$ values have different colors: black for $q=2$, red
for $q=3$, and green for $q=4$, thereby rendering them easily distinguishable.
To complement the types of comparisons that can be made with these plots, we
show in Fig.~\ref{zeros_sq_allF_v} and \ref{zeros_sq_allP_v} the loci ${\cal
B}$ and zeros for various widths plotted together for each value of $q$ from 2
to 4.

There are a number of interesting features that are evident in these plots.  In
cases where one uses boundary conditions that are self-dual, a subset of the
Fisher zeros lie exactly on the circle $|v|=\sqrt{q}$
\cite{mbook},\cite{chw,pfef,kc,dg}.  It has also been proved \cite{wuetal}
that in the limit $q \to \infty$, ${\cal B}$ consists of the unit circle in the
complex $\zeta$ plane, where 
\beq
\zeta = \frac{v}{\sqrt{q}} \ . 
\label{zeta}
\eeq
Furthermore, the approach to this limit has been studied using exact solutions
for $Z(G,q,v)$ on infinite-length, finite-width self-dual strips in \cite{dg}.
Although the open and cylindrical strip graphs with finite $L_\ell$ used here
are not self-dual, in the case of cylindrical boundary conditions, the effect
of this non-self-duality is significantly reduced in the limit $L_\ell \to
\infty$.  Indeed, we find that a subset of the Fisher zeros lie on the
respective circles $|v|=\sqrt{q}$, as is evident in the figures.  In
\cite{mbook,chw,pfef,ks,kc} these zeros were studied for finite slices of the
two-dimensional square lattice, and the resultant Fisher zeros form a circular
pattern in the $v$ plane which one may infer would become, in the thermodynamic
limit, the right-hand boundary of the (complex-temperature extension of the)
paramagnetic phase, separating it on the right from the analogous extension of
the ferromagnetic phase.  Since the Potts model does not have a ferromagnetic
phase on the infinite-length finite-width self-dual strip graphs used in
\cite{dg}, what was found there was that the arcs on ${\cal B}$ in the $v$
plane end at certain angles that depend on $L_t$ and $q$.  Let us introduce
polar coordinates for the variable $\zeta$ defined in (\ref{zeta}): $\zeta=
|\zeta|e^{i\theta}$ and denote the value of $\theta$ at the right-hand endpoint
of the arc of ${\cal B}$ with $|\zeta|=1$ and ${\rm Im}(v) > 0$ as
$\theta_{ae}$.  For fixed $L_t$, the endpoint angle $\theta_{ae}$ decreases
with increasing $q$, and for fixed $q$, $\theta_{ae}$ decreases with increasing
$L_t$. Further, for fixed $L_t$, $\lim_{q \to \infty} \theta_{ae}= 0$, and for
fixed $q$, $\lim_{L_t \to \infty} \theta_{ae}= 0$ \cite{wuetal,dg}.  For our
present strips, we find the following results showing the same quantitative
trends.  For each strip we list three approximate arc endpoint angles
corresponding to $q=2$, $q=3$, and $q=4$: for $2_{\rm P} \times \infty_{\rm
F}$, $\theta_{ae} \simeq 74^\circ$, $65^\circ$, and $60^\circ$; for $3_{\rm P}
\times \infty_{\rm F}$, $\theta_{ae} \simeq 50^\circ$, $40^\circ$, and
$35^\circ$; for $4_{\rm P} \times \infty_{\rm F}$, $\theta_{ae} \simeq
38^\circ$, $29^\circ$, and $24^\circ$; and for $5_{\rm P} \times \infty_{\rm
F}$, $\theta_{ae} \simeq 31^\circ$, $23^\circ$, and $18^\circ$.

For $q=2$ in the figure comparing different widths for cylindrical boundary
conditions, Fig. \ref{zeros_sq_allP_v}(a), one can see that the zeros lie not
just on the circle $|v|=\sqrt{2}$ but also on the other of the Fisher circles,
$|v+2|=\sqrt{2}$, and one can observe how the endpoints on these circles move
closer to the real axis as $L_t$ increases.  As $L_t \to \infty$, these would
then close to produce the full Fisher circles for the model on the infinite
square lattice.  Finally, for the values $q=3,4$, one can observe prongs on the
left-most part of ${\cal B}$ that are reminiscent of analogous prongs whose
endpoints were determined from analyses of low-temperature series expansions
for the infinite 2D square lattice in \cite{pfef}.
%
%
\section{Internal Energy and Specific Heat} 

 From the partition function and the resultant reduced free energy for the
finite and infinite-length strip, (\ref{ffinite}) and (\ref{f}), it is
straightforward to compute the (internal) energy $E$ and specific heat $C$ as
\beq
E = -\frac{\partial f}{\partial \beta} = -J(v+1)\frac{\partial f}{\partial v}
\label{e}
\eeq
and
\beq
C = \frac{\partial E}{\partial T} = k_B K^2(v+1)\left [
\frac{\partial f}{\partial v} + (v+1)\frac{\partial^2 f}{\partial v ^2} 
\right ] \ . 
\label{c}
\eeq
In the limit $L_\ell \to \infty$, since only the dominant eigenvalue 
$\lambda_d$ of the transfer matrix contributes to the free energy, one has 
\beq
f = {1 \over L_t} \ln \lambda_d
\label{flam}
\eeq
\beq
E = -\frac{J(v+1)}{L_t \lambda_d}\frac{\partial \lambda_d}{\partial v} 
\label{elam}
\eeq
and
\beq
C =   \frac{k_BK^2(v+1)}{L_t \lambda_d}
       \left [ (v+1) \frac{\partial^2 \lambda_d}{\partial v^2} 
    - \frac{(v+1)}{\lambda_d}\left ( 
      \frac{\partial \lambda_d}{\partial v} \right )^2 
     +  \frac{\partial \lambda_d}{\partial v} \right ] \ .
\eeq
As noted, since the infinite-length limits of the strips considered here are
quasi-one-dimensional systems with free energies that are analytic for all
finite temperatures, it follows that the dominant eigenvalue $\lambda_d$ is the
same on the whole semi-axis ${\rm Im}(v) = 0$, ${\rm Re}(v) \geq -1$.
Furthermore, as discussed in \cite{a}, the free energy and thermodynamic
functions such as the internal energy and specific heat for infinite-length
finite width lattice strips are independent of the longitudinal boundary
conditions (although they depend on the transverse boundary conditions).

For convenience we define a dimensionless internal energy 
\beq
E_r = -\frac{E}{J} = (v+1)\frac{\partial f}{\partial v} \ .
\label{er}
\eeq
Note that $\sgn(E_r)$ is (i) opposite to $\sgn(E)$ in the ferromagnetic case
where $J > 0$ for which the physical region is $0 \le v \le \infty$ and (ii)
the same as $\sgn(E)$ in the antiferromagnet case $J<0$ for which the physical
region is $-1 \le v \le 0$.  We recall the high-temperature
(equivalently, small--$|K|$) expansion for an infinite lattice of
dimensionality $d \ge 2$ with coordination number $\Delta$: 
\beq
-\frac{E}{J} = E_r = \frac{\Delta}{2}\left [ \frac{1}{q} + \frac{(q-1)K}{q^2} +
O(K^3) \right ] \ .
\label{ehightemp}
\eeq
In passing, it should be noted that in papers on the $q=2$ Ising special case,
the Hamiltonian is usually defined as ${\cal H}_I = -J_I\sum_{\langle i j
\rangle} \sigma_i \sigma_j$ with $\sigma_i = \pm 1$ rather than the Potts model
definition (\ref{ham}); the isomorphism between these conventions involves the
rescaling $2K_I=K$, where $K_I = \beta J_I$.  Furthermore, $E_I =
-J\langle \sigma_i \sigma_j \rangle$ rather than the Potts definition $E =
-J\langle \delta_{\sigma _i\sigma _j} \rangle$, where $\langle i j
\rangle$ are adjacent
vertices.  Hence, for example, for $q=2$, with the usual Ising model
definitions, $E_I(v=0)=0$ rather than $E=-J\Delta/(2q)$ and the
high-temperature expansion is $E_I = -J(\Delta/2)[K + O(K^3)]$
rather than the $q=2$ form of (\ref{ehightemp}).

We also define the reduced function
\beq 
C_H = \frac{C}{k_B K^2} \ .
\label{cr}
\eeq

\begin{figure}[hbtp]
  \centering
  \begin{tabular}{cc}
     \epsfxsize=200pt
     \epsffile{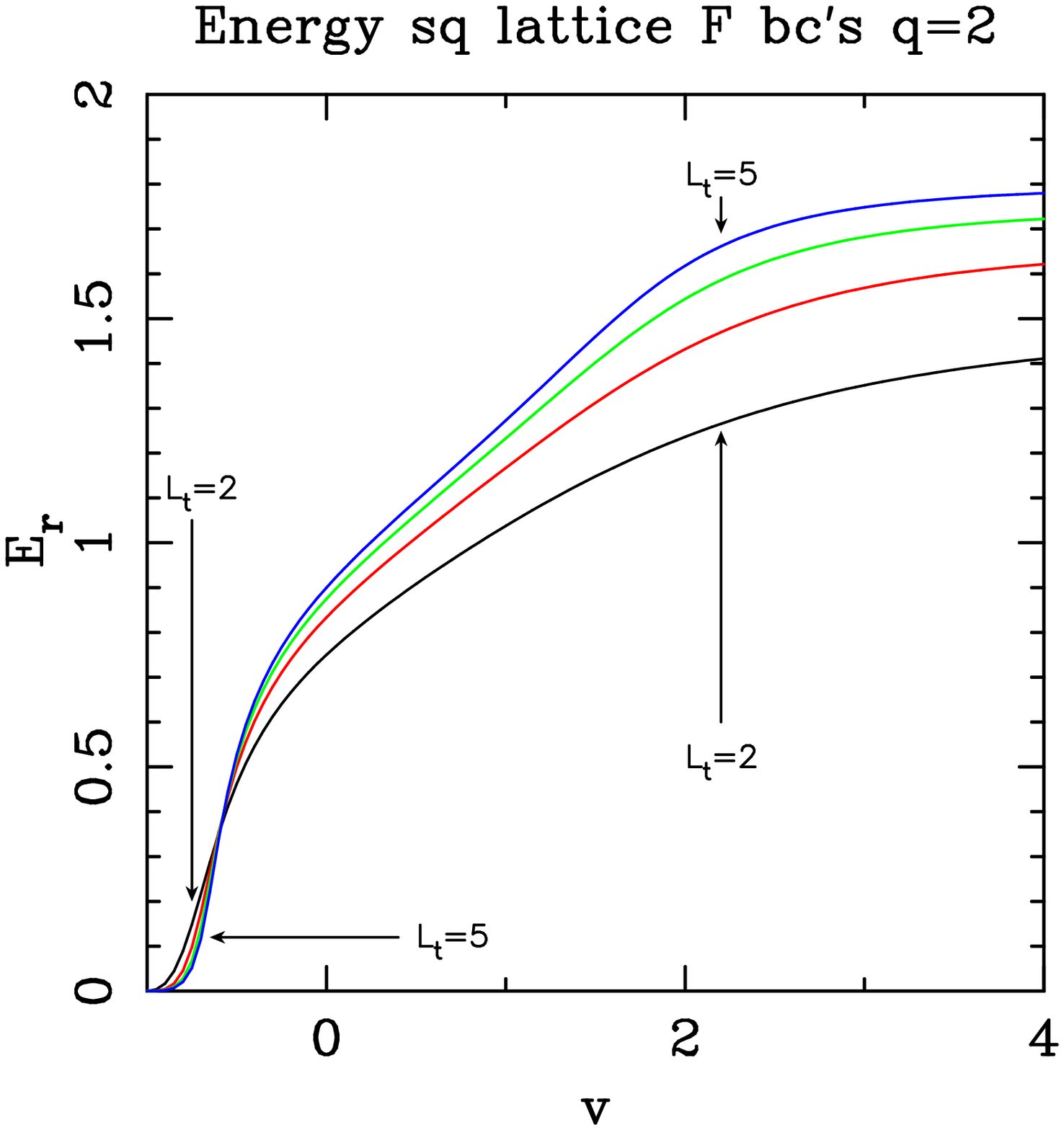}  &
     \epsfxsize=200pt
     \epsffile{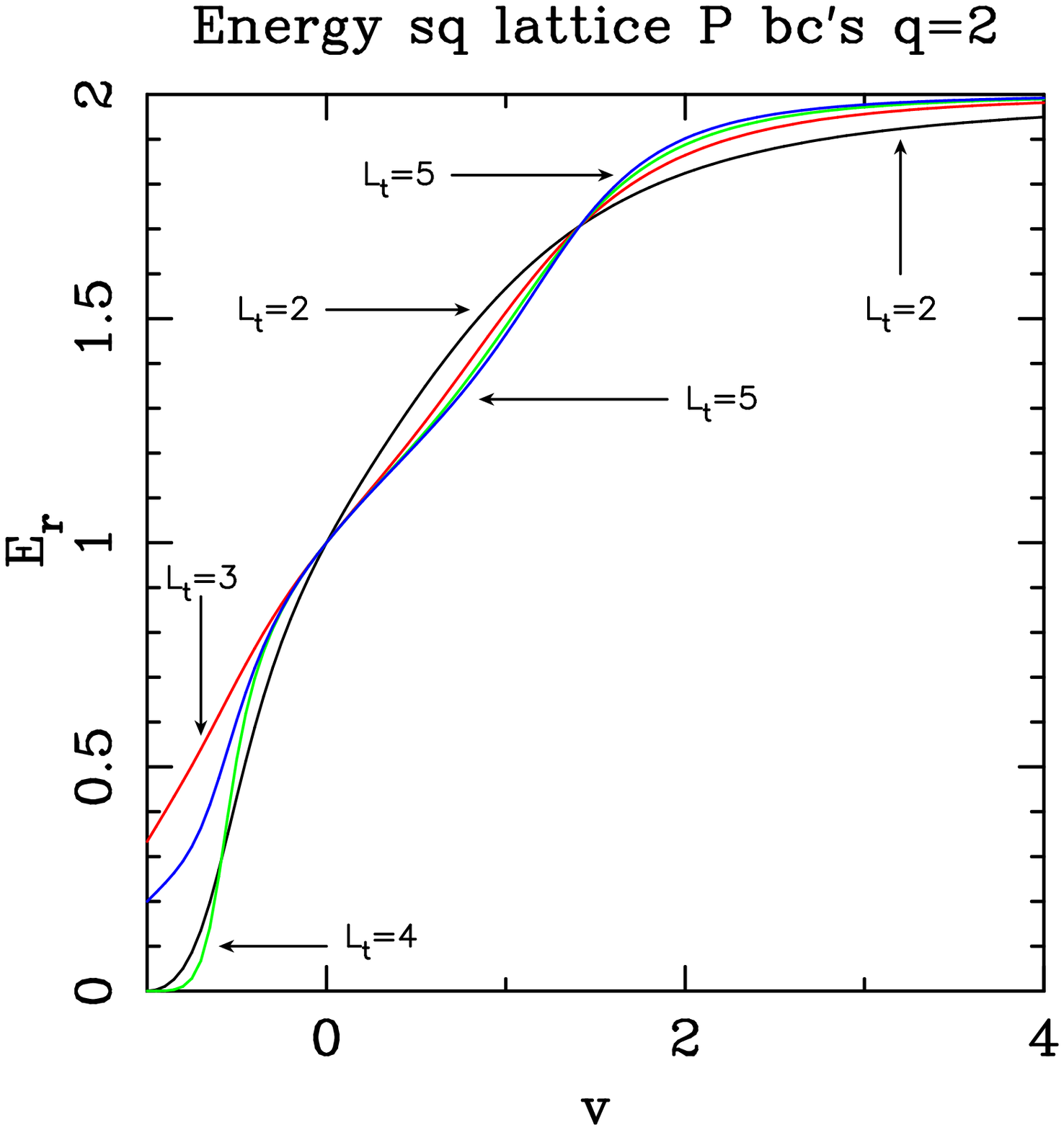}  \\[1mm]
     \phantom{(((a)}(a)    & \phantom{(((a)}(b) \\[5mm]
     \epsfxsize=200pt
     \epsffile{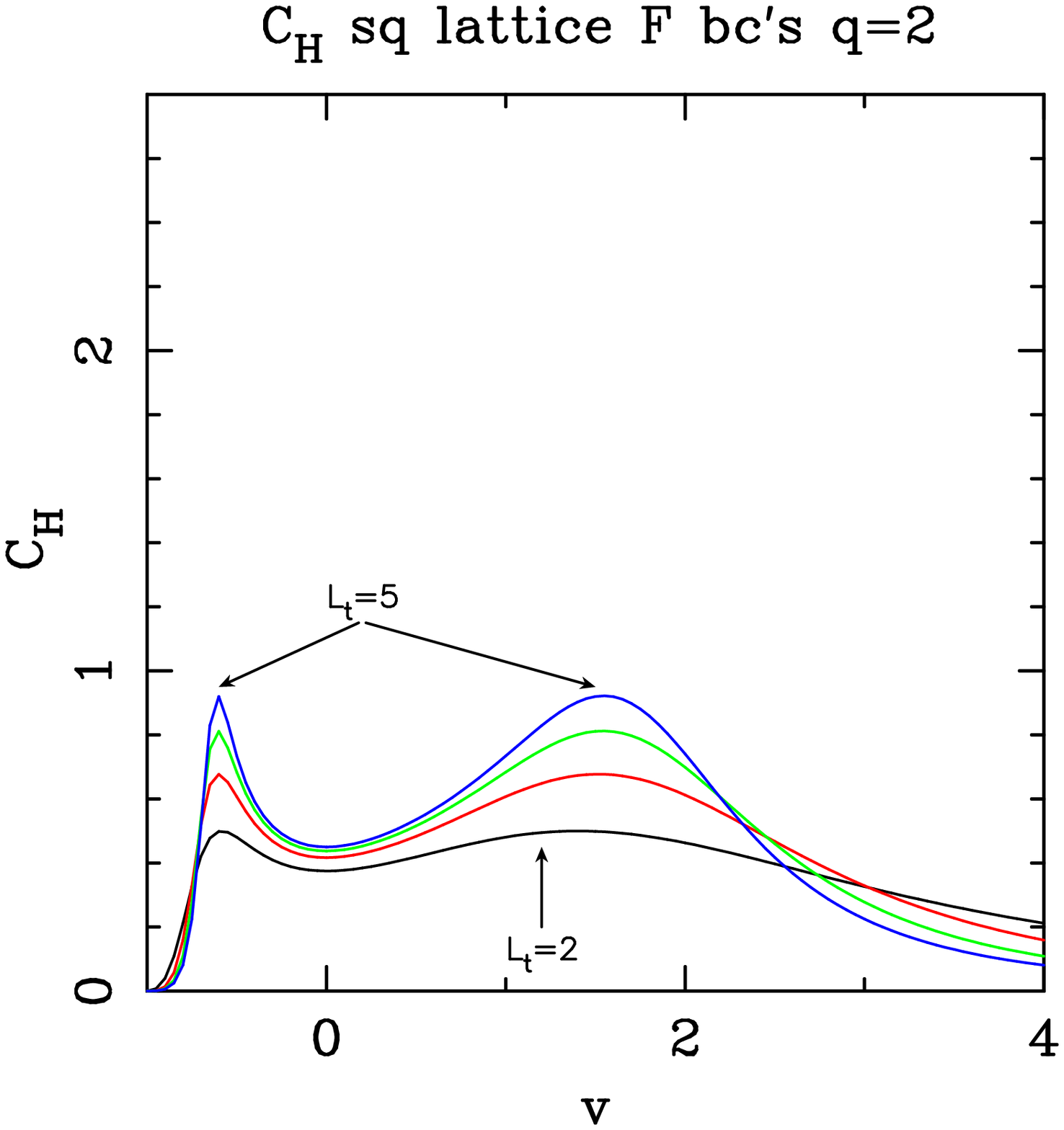}  &
     \epsfxsize=200pt
     \epsffile{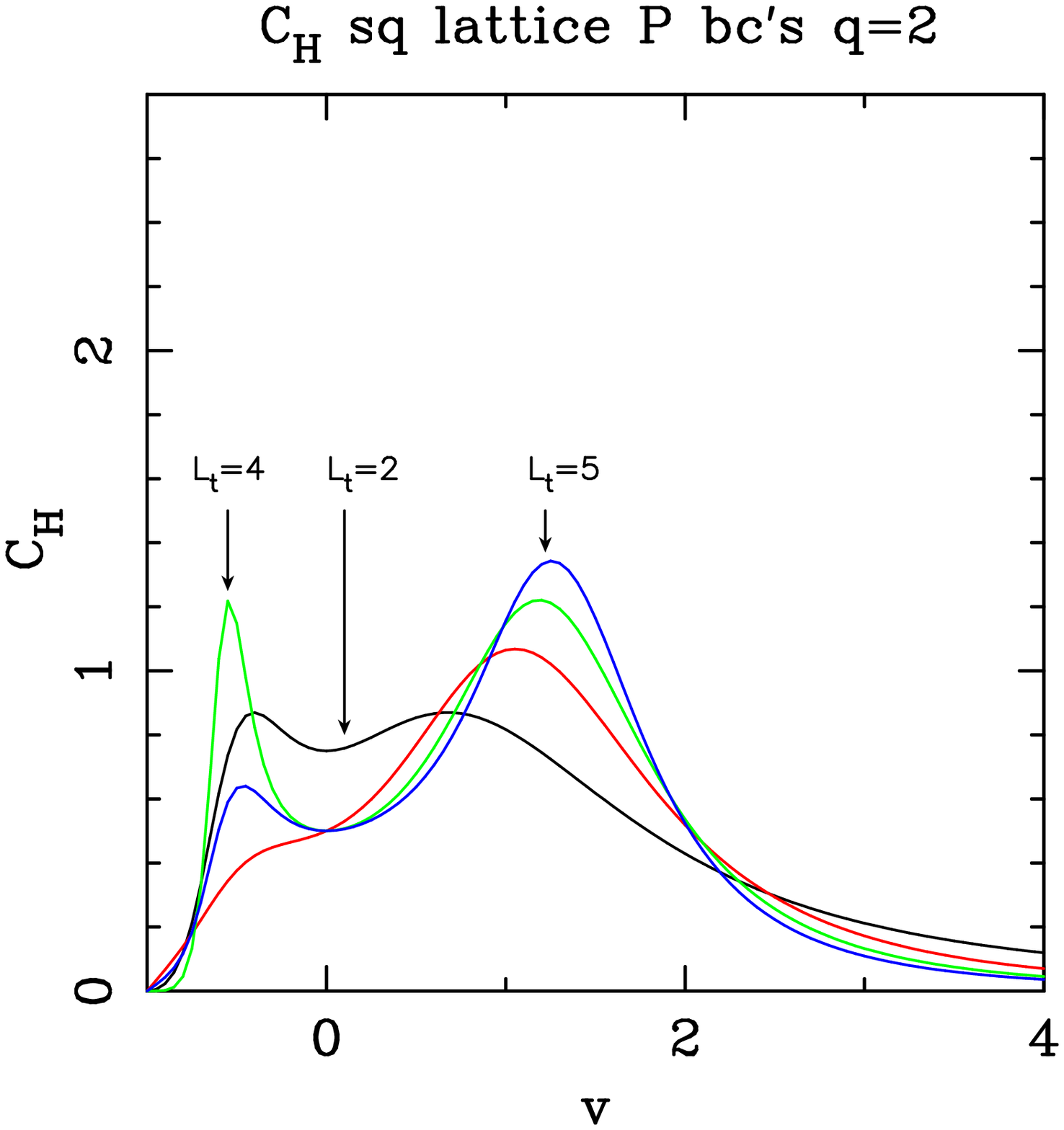} \\
     \phantom{(((a)}(c)    & \phantom{(((a)}(d) \\
  \end{tabular}
  \caption[a]{
  \protect\label{energy_sq_q=2}
Reduced internal energy $E_r=-E/J$ and specific heat $C_H$ as functions of
the temperature-like parameter $v$ for the $q=2$ Potts model on
square-lattice strips of size $(L_t)_{\rm F} \times \infty_{\rm F}$ and
$(L_t)_{\rm P} \times \infty_{\rm F}$ with $2\leq L_t \leq 5$. The plot
includes both the ferromagnetic and antiferromagnetic Potts models, for
which the temperature ranges are $0 \le v \le \infty$ and $-1 \le v \le
0$, respectively. }
\end{figure}

\begin{figure}[hbtp]
  \centering
  \begin{tabular}{cc}
     \epsfxsize=200pt
     \epsffile{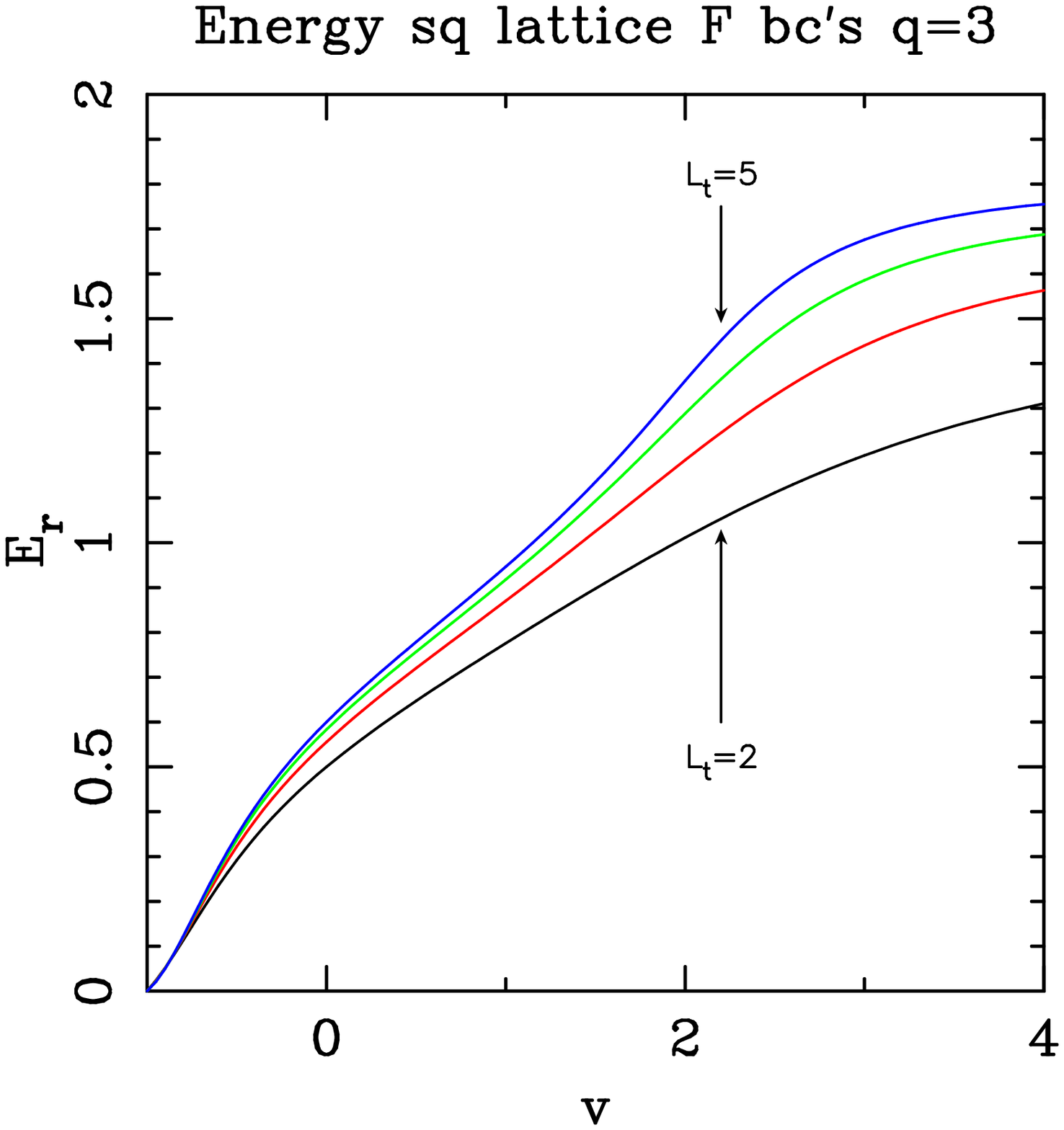}  &
     \epsfxsize=200pt
     \epsffile{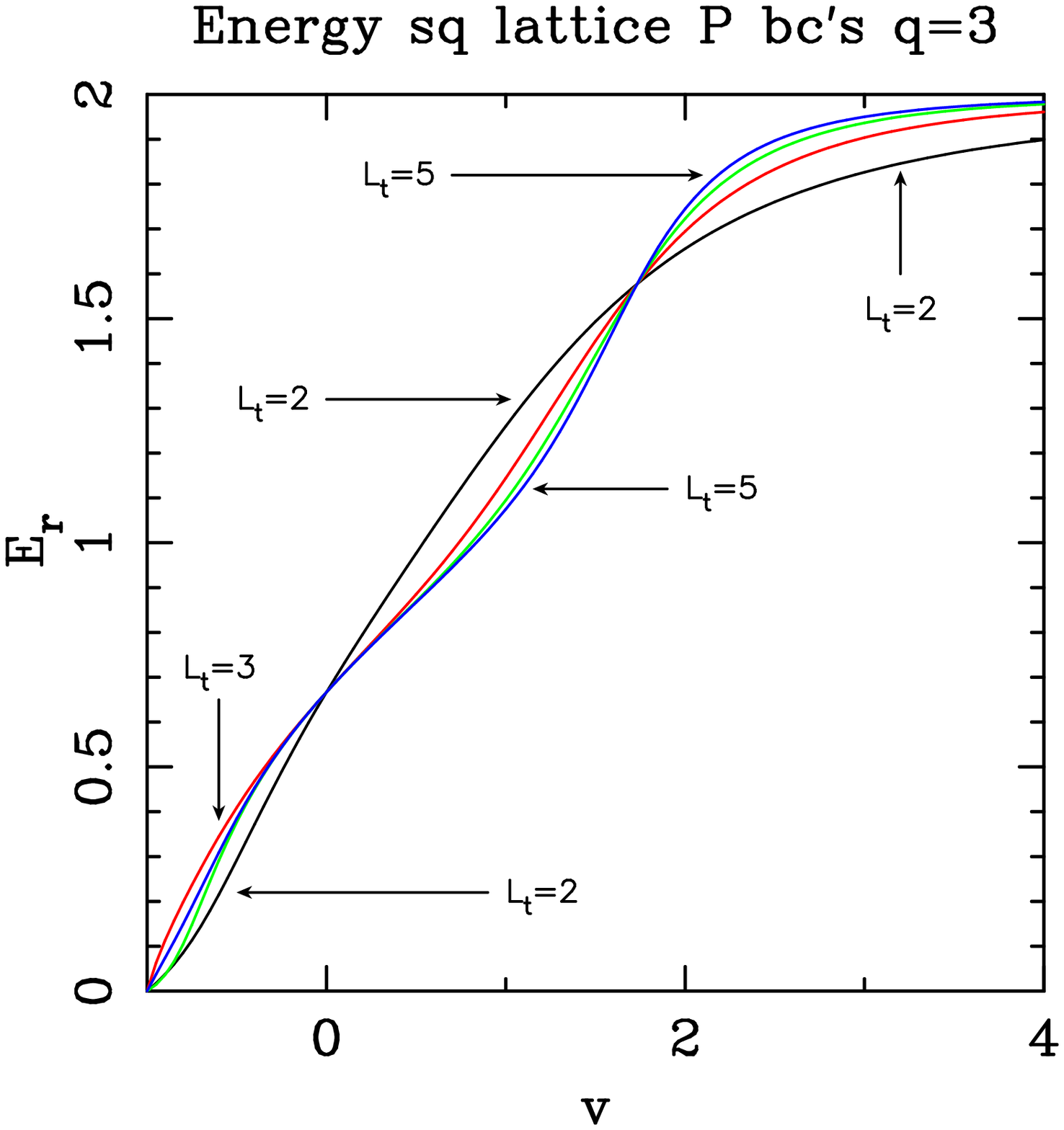}  \\[1mm]
     \phantom{(((a)}(a)    & \phantom{(((a)}(b) \\[5mm]
     \epsfxsize=200pt
     \epsffile{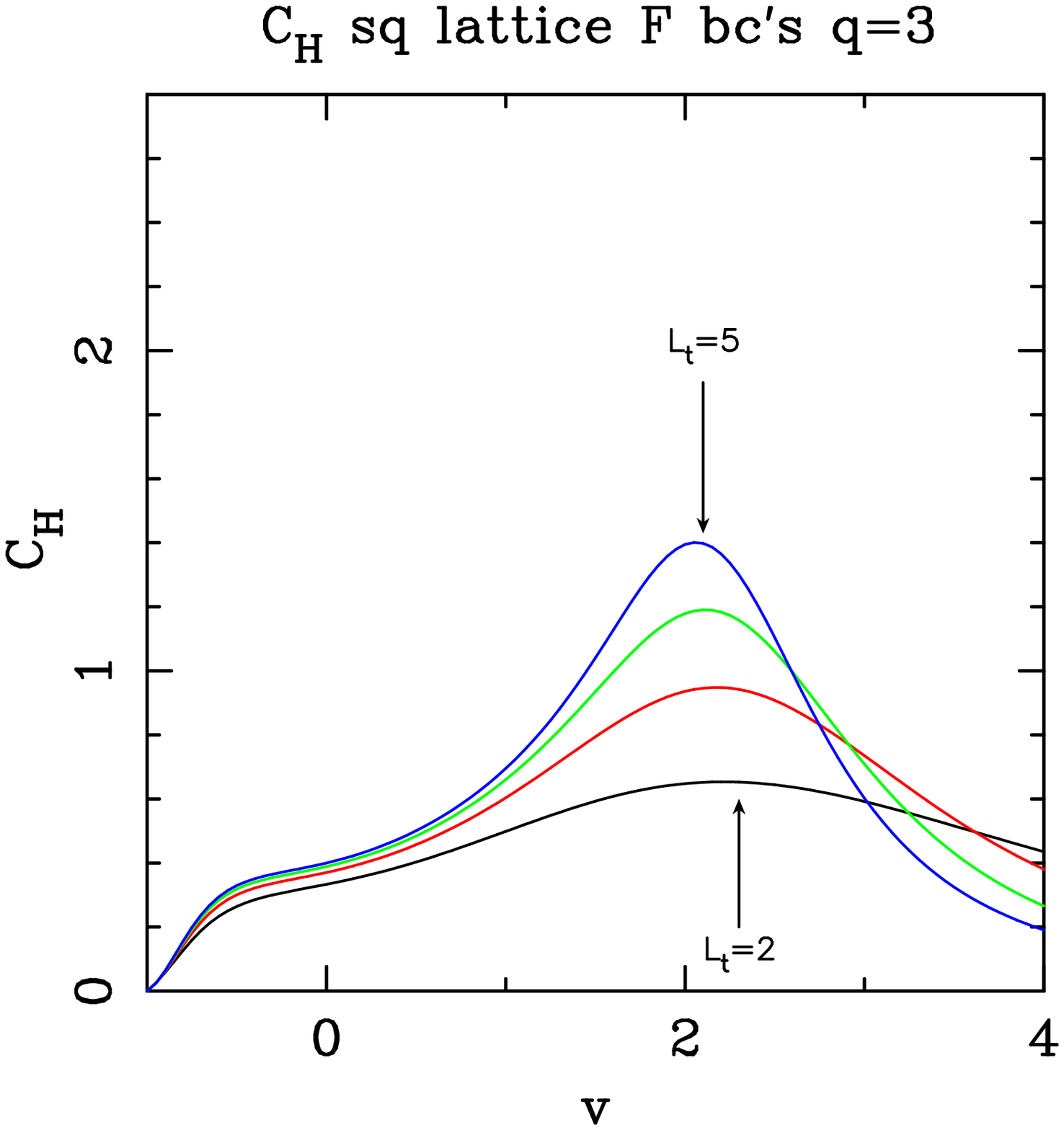}  &
     \epsfxsize=200pt
     \epsffile{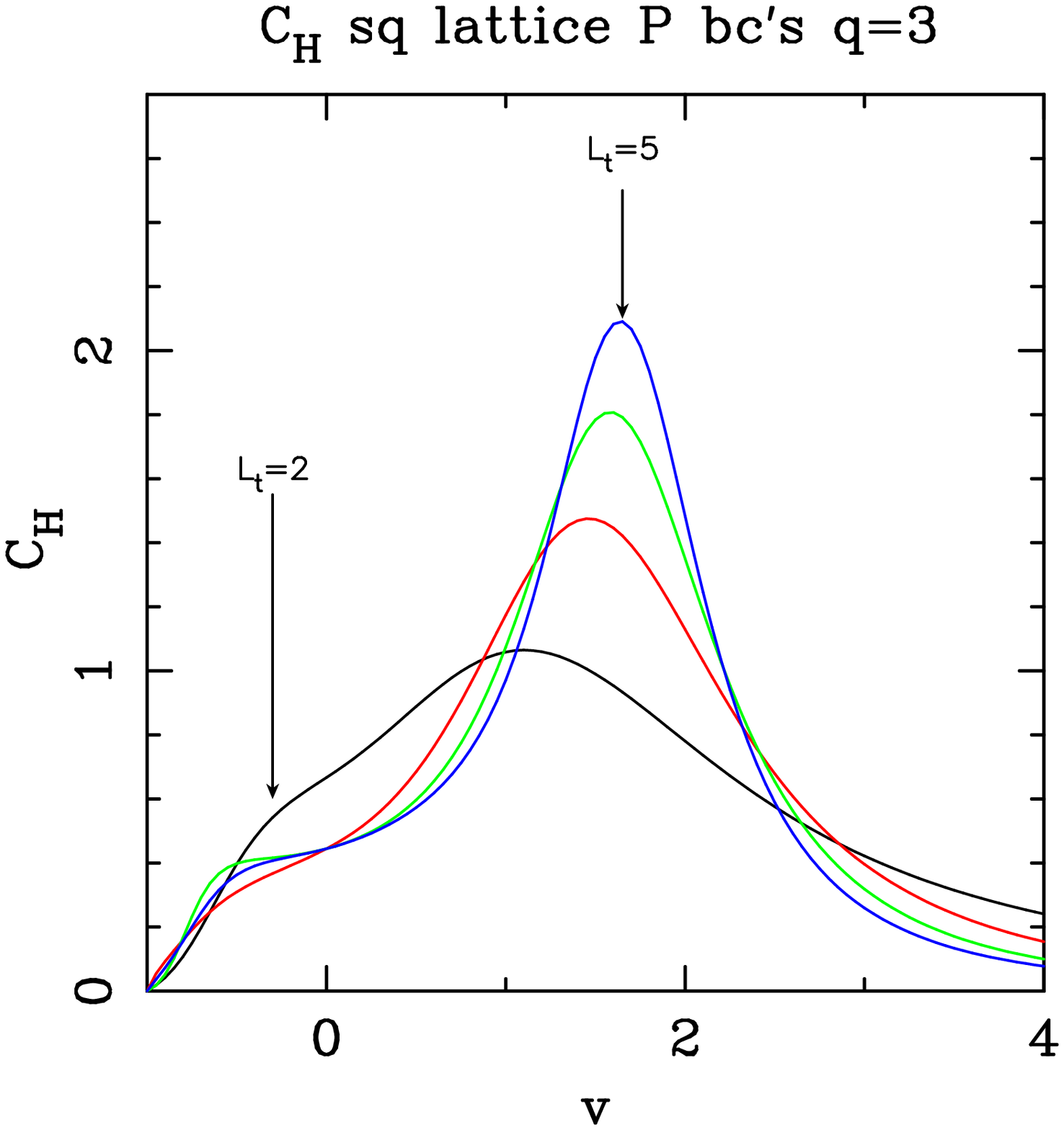} \\
     \phantom{(((a)}(c)    & \phantom{(((a)}(d) \\
  \end{tabular}
  \caption[a]{
  \protect\label{energy_sq_q=3}
Reduced internal energy $E_r=-E/J$ and specific heat $C_H$ as functions of
$v$ for the $q=3$ Potts model on square-lattice strips of size $(L_t)_{\rm
F} \times \infty_{\rm F}$ and $(L_t)_{\rm P} \times \infty_{\rm F}$ with
$2\leq L_t \leq 5$. Notation is as in Fig. \ref{energy_sq_q=2}.
  }
\end{figure}

\begin{figure}[hbtp]
  \centering
  \begin{tabular}{cc}
     \epsfxsize=200pt
     \epsffile{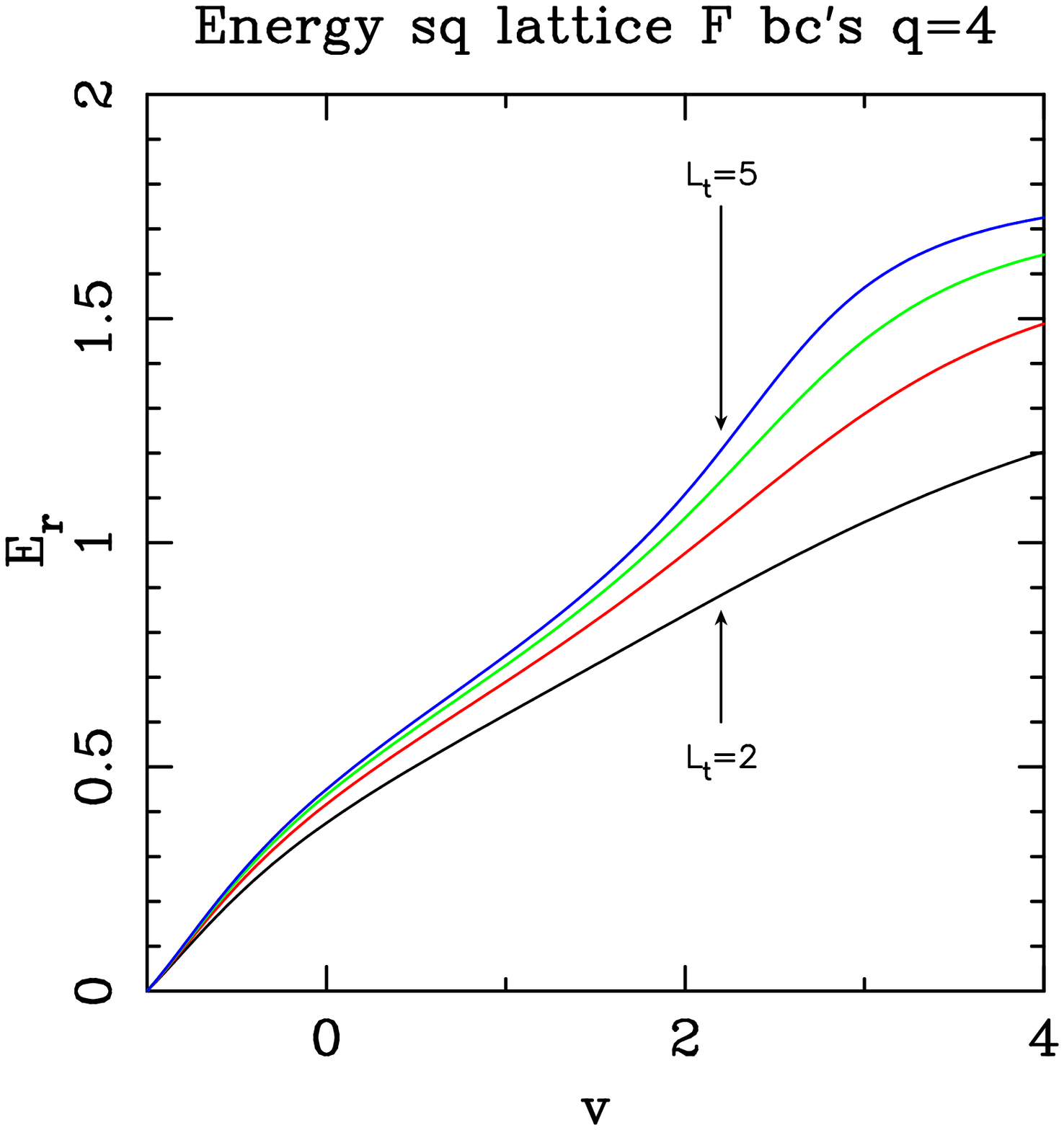}  &
     \epsfxsize=200pt
     \epsffile{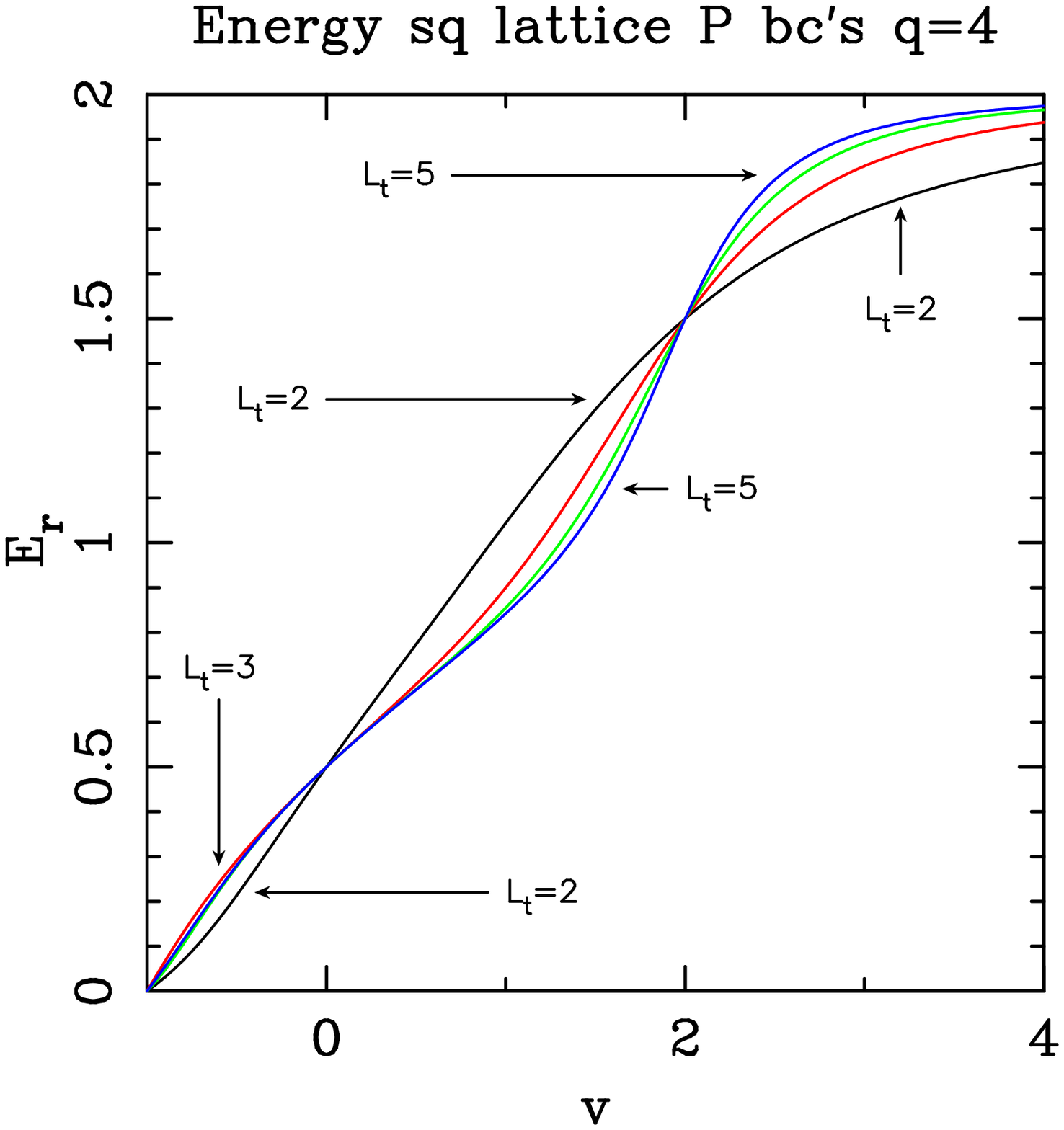}  \\[1mm]
     \phantom{(((a)}(a)    & \phantom{(((a)}(b) \\[5mm]
     \epsfxsize=200pt
     \epsffile{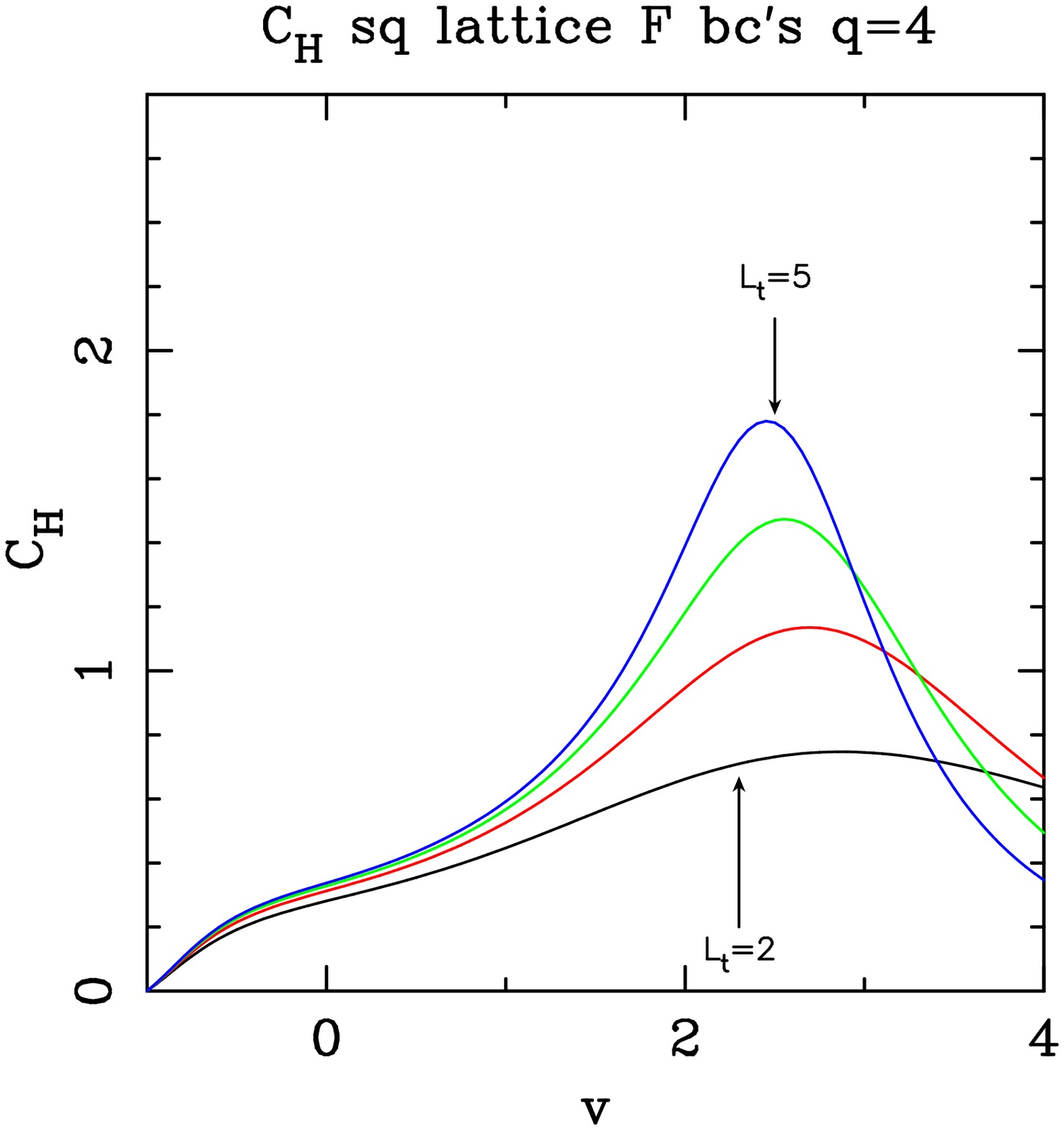}  &
     \epsfxsize=200pt
     \epsffile{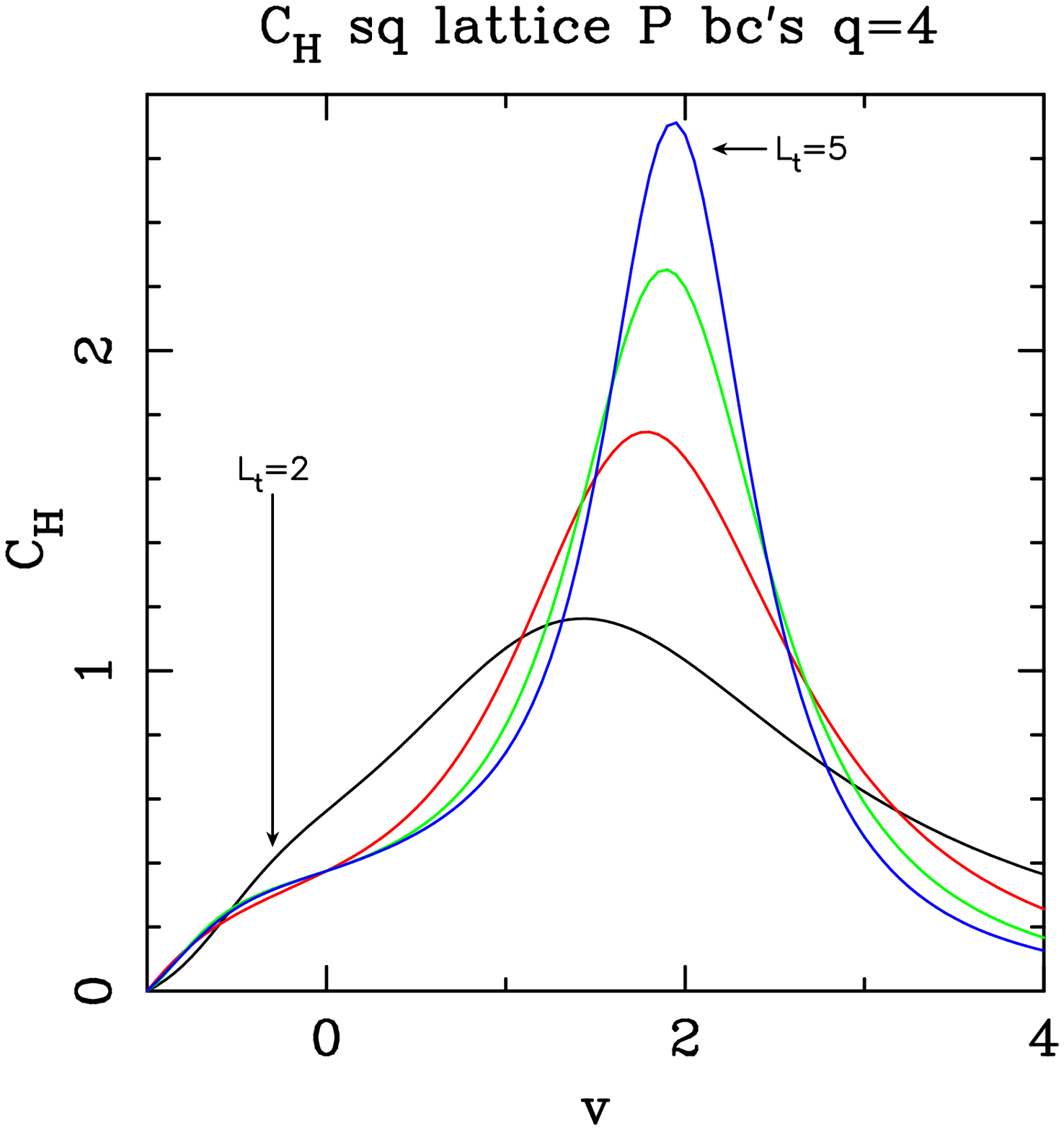} \\
     \phantom{(((a)}(c)    & \phantom{(((a)}(d) \\
  \end{tabular}
  \caption[a]{
  \protect\label{energy_sq_q=4}
Reduced internal energy $E_r=-E/J$ and specific heat $C_H$ as functions of
$v$ for the $q=4$ Potts model on square-lattice strips of size $(L_t)_{\rm
F} \times \infty_{\rm F}$ and $(L_t)_{\rm P} \times \infty_{\rm F}$ with
$2\leq L_t \leq 5$. Notation is as in Fig. \ref{energy_sq_q=2}. 
  }
\end{figure}

We show in Figures \ref{energy_sq_q=2}--\ref{energy_sq_q=4} the reduced energy
$E_r$ and the function $C_H$ that enters in the specific heat for $q=2$ through
$q=4$ for the square-lattice strips of widths $2 \leq L_t \leq 5$ with free and
cylindrical boundary conditions.  We recall that the Potts model on the
infinite square lattice has a phase transition from the paramagnetic
high-temperature phase in the interval $0 \le v \le v_{p,FM}$ to the
low-temperature phase with ferromagnetic long-range order in the interval
$v_{p,FM} \le v \le \infty$, where
\beq
v_{p,FM} = \sqrt{q} \ .
\label{vcfm}
\eeq
This phase transition is continous (second-order) for $q \le 4$ and first-order
for $q > 4$.  In the figures plotting the function $C_H$ entering in the
specific heat one can see how the maxima at the respective values of $v \simeq
v_{p,FM}$ for various $q$ increase as the width of the strip $L_t$ increases.
In the limit $L_t \to \infty$, these maxima diverge, corresponding to the
divergence of the specific heat at $v_{p,FM}$.  Since the free strips and
the cylindrical strips with even $L_t$ are bipartite graphs, there is
a well-known isomorphism between the ferromagnetic and antiferromagnetic Ising
models, and one sees corresponding maxima at the values of $v$ obtained by the
replacement $K \to -K$.  
 
A general property for $E_r$ calculated for the infinite-length limit of the
lattice strips with free transverse boundary conditions is that $E=0$ for
$v=-1$, i.e., the zero-temperature limit of the Potts antiferromagnet.  This is
evident in the figures.  For (the infinite-length limit of) the square lattice
strips with periodic transverse boundary conditions, the value of $E=0$ at
$v=-1$ for $q \ge \chi$, the chromatic number for these strips, which is 2 if
$L_t$ is even and 3 if $L_t$ is odd.  For $q=2$ and $L_t$ odd, there is
frustration, and hence $E$ is larger than zero at $v=-1$.  

In the limit of infinite temperature on the infinite square lattice, $E_r =
2/q$, as is evident from the expansion (\ref{ehightemp}).  The infinite-length
strips with cylindrical boundary conditions have uniform coordination number
$\Delta=4$, and one sees the agreement with the formula $E_r=2/q$ in the
figures.  The infinite-length strips with open boundary conditions do not have
a uniform coordination number (this is equal to 3 for the vertices on the 
upper and lower sides and 4 for the vertices in the interior of the strip). 
However, one can see the approach to the above formula as $L_t$ increases. 

For the $q$-state Potts ferromagnet on the infinite-length finite-width strips
with cylindrical boundary conditions, we find that the curves for the energy
$E$ cross at a unique value of $v$ depending on $q$ but independent of the
width $L_t$; furthermore, this value is equal to the value $v=v_{p,FM}$ at 
which the phase transition occurs on the infinite square lattice.  Let us
denote 
\beq
E_p = E(v=v_{p,FM})
\label{epdef}
\eeq
for the internal energy of the infinite-length finite-width strips evaluated at
the value of $v$ given in (\ref{vcfm}).  A careful evaluation of the value of 
$E_p$ reveals that for the range of $q \le 4$ where the Potts ferromagnet on
the square lattice has a continuous second-order phase transition, this value
is independent of the width $L_t$ and is equal to the value for the infinite
square lattice at the ferromagnetic critical point, given by \cite{qge5} 
\beq
E_c = -J(1 + q^{-1/2}) \ .
\label{ercrit}
\eeq
This behavior could have been anticipated in the Ising case, $q=2$. Finite-size
relations for statistical mechanical models using methods of conformal field
theory have been discussed in \cite{cardy}-\cite{fms}. For the Ising
model, the asymptotic expansion of the internal energy evaluated at
$v=v_{p,FM}$ on a torus of length $L_\ell$ and width $L_t$ takes the
following form (where we shall use the subscript $c$ to denote the fact that
this would be the critical value in two-dimensional thermodynamic limit)  
\cite{Ferdinand_Fisher}, \cite{cardy,Izmailian_00,Salas_00}
\beq
E_c(L_t,\rho) = E_c + \sum_{i=0}^\infty {E_{2 i + 1}(\rho) \over (L_t)^{2i+1} }
\eeq
where
\beq
\rho=\frac{L_\ell}{L_t}
\label{rho}
\eeq
is the aspect ratio of the torus.  The infinite-length torus corresponds to the
limit $\rho \to \infty$. It has been shown 
\cite{Ferdinand_Fisher,Izmailian_00,Salas_00} that in this limit
\beq
E_1(\infty) = E_3(\infty) = E_5(\infty) = 0 \ .
\eeq
Furthermore, one can see that all the correction terms have a factor
proportional to $\theta_2 \theta_3 \theta_4$ where $\theta_i$ is the usual
$\theta$-function $\theta_i(z,\tau)$ (e.g., \cite{abram}) evaluated at $z=0$,
where $\tau=i \rho$, i.e. $q_{nome} \equiv e^{i \tau} = e^{-\pi \rho}$
(see, e.g., Appendix B of \cite{Salas_00}). In the limit $\rho
\rightarrow \infty$, from the basic definitions of these theta functions, it
follows that $\theta_2 \rightarrow 0$ and $\theta_3,\theta_4 \rightarrow
1$. Hence, for the Ising model on an infinitely long cylinder 
\beq
E_c(L_t,\infty) = E_c
\eeq
for all widths $L_t$. This (together with the property that the thermodynamic
functions are independent of the longitudinal boundary conditions in the limit
of infinite length) explains why all of the curves for $E_r$ on the
infinite-length, finite-width strips with cylindrical boundary conditions cross
at the point $v=v_{p,FM}$ where the the energy attains the critical value on
the infinite square lattice, $E_c$.  Our numerical results suggest the
inference that all the finite-size corrections to the internal energy at the
respective ferromagnetic critical points also vanish (for an infinitely long
cylinder) for the other two values of $q$, i.e., $q=3$ and $q=4$, where the
Potts ferromagnet has a second-order phase transition on the infinite square
lattice.  We have also checked and confirmed that the value of the internal
energy at this crossing, $(E_r)_c(v=v_{p,FM})$ is equal to $1+q^{-1/2}$
for $q > 4$, where the model on the infinite square lattice has a
first-order
transition, so that the limits of the internal energy at $v=v_{p,FM}$ are
different when approached from high and low temperature \cite{qge5}.

Given the bipartite property of the strip graphs with cylindrical boundary
conditions and even $L_t$ and the consequent equivalence of the Ising
ferromagnet and antiferromagnet, and taking into account our discussion above,
it follows that for $q=2$ the internal energy curves should exhibit a crossing
in the antiferromagnetic regime for even $L_t$, and our results agree with
this. The point where this occurs is $v = v_{p,AFM}=\sqrt{2}-2$, and at this
point the internal energy is precisely equal to the value on the infinite
square lattice, $E_r = 1 - 1/\sqrt{2}$.

\bigskip

Acknowledgment: The research of R.S. was supported in part by the NSF grant 
PHY-9722101. The research of J. S. was partially supported by CICyT
(Spain) grant AEN99-0990.  One of us (R.S.) wishes to acknowledge H. Kluepfel 
for related collaborative work. 

\bigskip
\bigskip

\section{Appendix}

In this appendix we recall the connection between the Potts model partition
function $Z(G,q,v)$ and the Tutte (also called Tutte/Whitney) polynomial
$T(G,x,y)$ for a graph $G=G(V,E)$, given by \cite{tutte1}-\cite{boll}
\beq
T(G,x,y)=\sum_{G^\prime \subseteq G} (x-1)^{k(G^\prime)-k(G)}
(y-1)^{c(G^\prime)}
\label{tuttepol}
\eeq 
where $G^\prime$ is a spanning subgraph of $G$, and $k(G^\prime)$,
$e(G^\prime)$, and $n(G^\prime)=n(G)$ denote the number of components, edges,
and vertices of $G^\prime$, and
\beq c(G^\prime) =
e(G^\prime)+k(G^\prime)-n(G^\prime)
\label{ceq}
\eeq
is the number of independent circuits in $G^\prime$. 
As stated in the text, $k(G)=1$ for the graphs of interest here.  Now let
\beq
x=1+\frac{q}{v}
\label{xdef}
\eeq
and
\beq
y=a=v+1
\label{ydef}
\eeq
so that
\beq
q=(x-1)(y-1) \ .
\label{qxy}
\eeq
Then
\beq
Z(G,q,v)=(x-1)^{k(G)}(y-1)^{n(G)}T(G,x,y) \ .
\label{ztutte}
\eeq
For a planar graph $G$ the Tutte polynomial satisfies the duality relation
\beq
T(G,x,y) = T(G^*,y,x)
\label{tuttedual}
\eeq
where $G^*$ is the (planar) dual to $G$.

There are several special cases of the Tutte polynomial that are of interest.
One that we have commented on in the text and in previous papers is the
chromatic polynomial $P(G,q)$.  This is obtained by setting $y=0$, i.e.,
$v=-1$, so that $x=1-q$; the correspondence is $P(G,q) =
(-q)^{k(G)}(-1)^{n(G)}T(G,1-q,0)$.  A second special case is the flow
polynomial \cite{bbook,welsh,boll} $F(G,q)$, obtained by setting $x=0$ and
$y=1-q$: $F(G,q) = (-1)^{e(G)-n(G)+k(G)}T(G,0,1-q)$ For planar $G$, given the
relation (\ref{tuttedual}), the flow polynomial is, up to a power of $q$,
proportional to the chromatic polynomial on the dual graph $G$: $F(G,q) \propto
P(G^*,q)$.  A third special case for $x=1$ is the reliability polynomial (e.g.,
\cite{welsh}).  Consider a connected graph $G(V,E)$ and now let each edge of
$G$ be present with probability $p$ (and hence absent with probability $1-p$);
then the probability that exists a path connecting any two vertices in $G$ is
given by the (all-terminal) reliability polynomial 
$R(G,p)=\sum_{H \subseteq G} p^{e(H)}(1-p)^{e(G)-e(H)}$, where $H$ denotes
a connected spanning subgraph of $G$.  In turn, this is
expressed as a special case of the Tutte polynomial according to
\beq
R(G,p) = p^{n(G)-1}(1-p)^{e(G)-n(G)+1}T(G,1,(1-p)^{-1})
\label{reliability}
\eeq
Thus, our results for $Z(G,q,v)$ and hence, via (\ref{ztutte}), for $T(G,x,y)$
for these lattice strips yield, as a special case, the reliability polynomials
for the strips.  A conjecture for the complex zeros of the reliability 
polynomial was given in \cite{bc} (see \cite{sokalzero} for a recent
discussion). 

For a given graph $G=(V,E)$, at certain special values of the arguments $x$ and
$y$, the Tutte polynomial $T(G,x,y)$ yields quantities of basic graph-theoretic
interest \cite{tutte3}-\cite{boll}.  We recall some definitions: a spanning
subgraph was defined at the beginning of the paper; a tree is a
connected graph with no cycles; a forest is a graph containing one or
more trees; and a spanning tree is a spanning subgraph that is a tree.  We
recall that the graphs $G$ that we consider are connected.  Then the number
of spanning trees of $G$, $N_{ST}(G)$, is
\beq
N_{ST}(G)=T(G,1,1) \ ,
\label{t11}
\eeq
the number of spanning forests of $G$, $N_{SF}(G)$, is
\beq
N_{SF}(G)=T(G,2,1) \ ,
\label{t21}
\eeq
the number of connected spanning subgraphs of $G$, $N_{CSSG}(G)$, is
\beq
N_{CSSG}(G)=T(G,1,2) \ ,
\label{T12}
\eeq
and the number of spanning subgraphs of $G$, $N_{SSG}(G)$, is
\beq
N_{SSG}(G)=T(G,2,2) \ .
\label{t22}
\eeq
From the duality relation (\ref{tuttedual}), one has, for planar graphs $G$ and
their planar duals $G^*$,
\beq
N_{ST}(G)=N_{ST}(G^*) \ , \quad N_{SSG}(G)=N_{SSG}(G^*)
\label{t11t11}
\eeq
and
\beq
N_{SF}(G)=N_{CSSG}(G^*) \ .
\label{t21t12}
\eeq
In previous works \cite{a}-\cite{ka} we have given resultant formulas for these
graphical quantities for the families of graphs considered therein.  It is
straightforward to do the same for the square-lattice strips considered here.
For spanning trees on lattice sections, see also \cite{wust,sw}.

\vfill
\eject

\clearpage

%
%

\clearpage

%
%

\clearpage

%
%

\clearpage

%
%

\clearpage
%
%

\clearpage
%
%

\clearpage
%
%

\clearpage
%
%

%
%
\clearpage

\clearpage
%
%
\clearpage

\clearpage
%
%
\clearpage
 
\clearpage
%
%
\clearpage


\clearpage
%
%
\clearpage

\clearpage
%
%
\clearpage

\clearpage
%
%
\clearpage

\vfill
\eject

\begin{thebibliography}{99}
\bibitem{rmis}{In the equation for $R_n$ given in
\cite{ss00} the lower limit of the summation should be 1 rather than 0.}
\bibitem{qge5}{For completeness, we recall that \cite{wurev,baxterbook} 
for $q > 4$, $\lim_{v \nearrow v_{p,FM}} E_r =  
(1+q^{-1/2})[1-D(q)\tanh(\mu/2)]$
and $\lim_{v \searrow v_{p,FM}} E_r = (1+q^{-1/2})[1+D(q)\tanh(\mu/2)]$ where
$\mu={\rm arccosh}(q^{1/2}/2)$ and $D(q) = \prod_{n=1}^\infty (\tanh(n
\mu))^2$.  For the Potts model on the infinite-length, finite-width strips,
this discontinuity is absent.}
\end{thebibliography}
\end{document}